\documentclass[11pt]{article}
\usepackage[hmargin={3.0cm,3.0cm},vmargin={3.0cm,3.0cm}]{geometry}

\usepackage{amsmath,amssymb}
\usepackage{amsfonts}
\usepackage{mathrsfs}
\usepackage{dsfont}

\usepackage{epic}
\usepackage{eepic}
\usepackage{graphicx}

\usepackage{pgf}
\usepackage{tikz}
\usetikzlibrary{arrows.meta}
\usetikzlibrary{decorations.markings}

\usepackage{amsmath}
\usepackage{mathtools}

\usepackage{bm}

\usepackage{empheq}

\usepackage{multirow}

\usepackage[skip=2pt,font=small]{caption}
\usepackage{hyperref}
\hypersetup{
colorlinks,
linkcolor={blue!60!black},
citecolor={blue!90!black},
urlcolor={blue!90!black}}

\usepackage{footnote}

\usepackage{cite}
\usepackage{enumerate}
\usepackage[curve]{xy}

\usepackage{titlesec}

\titleformat{\paragraph}
{\normalfont\normalsize\bfseries}{\theparagraph}{1em}{}
\titlespacing*{\paragraph}
{0pt}{3.25ex plus 1ex minus .2ex}{1.5ex plus .2ex}

\usepackage[titletoc,title]{appendix}

\numberwithin{equation}{section}

\tikzset{->-/.style={decoration={
  markings,
  mark=at position .60 
  with {\arrow{Latex[scale=1.1]}}},postaction={decorate}}}
\tikzset{->>-/.style={decoration={
  markings,
  mark=at position .56 
  with {\arrow{Latex[scale=1.1]}},
  mark=at position .64 
  with {\arrow{Latex[scale=1.1]}}},postaction={decorate}}}
\tikzset{->>>-/.style={decoration={
  markings,
  mark=at position .54 
  with {\arrow{Latex[scale=1.6,color=black]}}},postaction={decorate}}}
\tikzset{->>>>-/.style={decoration={
  markings,
  mark=at position .52 
  with {\arrow{Latex[scale=1.6,color=black]}},
  mark=at position .62 
  with {\arrow{Latex[scale=1.6,color=black]}}},postaction={decorate}}}
\tikzset{-<-/.style={decoration={
  markings,
  mark=at position .50 
  with {\arrow{Latex[reversed]}}},postaction={decorate}}}
\tikzset{-<<-/.style={decoration={
  markings,
  mark=at position .40 
  with {\arrow{Latex}}
  mark=at position .60 
  with {\arrow{Latex}}},postaction={decorate}}}
  
\tikzset{-|-/.style={decoration={
  markings,
  mark=at position .51 with {\arrow{Bar}}},postaction={decorate}}}
\tikzset{-||-/.style={decoration={
  markings,
  mark=at position .49 with {\arrow{Bar[sep=-5pt] Bar}}},postaction={decorate}}}
\tikzset{-!-/.style={decoration={
  markings,
  mark=at position .51 with {\arrow{Tee Barb[length=4pt]}}},postaction={decorate}}}
\tikzset{-!!-/.style={decoration={
  markings,
  mark=at position .51 with {\arrow{Tee Barb[sep=1pt,length=4pt] Tee Barb[length=4pt]}}},postaction={decorate}}}

\newcommand{\ds}{\displaystyle}

\renewcommand{\author}[1]{\large\rm #1\\ \bigskip}
\newcommand{\address}[1]{{\normalsize\it #1\\}\bigskip}
\renewcommand{\title}[1]{\bigskip\bigskip\Large\bf #1\bigskip\bigskip\\}

\newcommand{\Bigpsi}[3]{\phantom{\Psi}_2 \kern -.05em
\Psi_2\left(\genfrac{}{}{0pt}{}{#1}{#2}\biggl|#3\right)}

\newcommand{\bea}{\begin{eqnarray}}
\newcommand{\eea}{\end{eqnarray}}

\newcommand{\ii}{\mathsf{i}}
\newcommand{\Wh}{\hat{W}}
\newcommand{\oW}{\overline{W}}
\newcommand{\oWh}{\hat{\oW}}

\newcommand{\lag}{{\mathcal L}}
\newcommand{\ol}{\overline{\lag}}
\newcommand{\lam}{{\Lambda}}
\newcommand{\olam}{\overline{\lam}}
\newcommand{\lagh}{\hat{\lag}}
\newcommand{\olh}{\hat{\ol}}
\newcommand{\lamh}{\hat{\lam}}
\newcommand{\olamh}{\hat{\olamh}}

\newcommand{\iS}{{S}}

\newcommand{\bu}{{\mathbf u}}
\newcommand{\bv}{{\mathbf v}}
\newcommand{\bw}{{\mathbf w}}

\def\EXP{\textrm{{\large e}}}

\def\re{\mathop{\hbox{\rm Re}}\nolimits}
\def\im{\mathop{\hbox{\rm Im}}\nolimits}

\newcommand{\x}{{\boldsymbol{x}}}

\newcommand{\al}{{\bm{\alpha}}}
\newcommand{\bt}{{\bm{\beta}}}
\newcommand{\gm}{{\bm{\gamma}}}
\newcommand{\hal}{{\hat{\al}}}
\newcommand{\hbt}{{\hat{\bt}}}

\newcommand{\xv}{{x}}

\newcommand{\mpc}{\alpha}
\newcommand{\mqc}{\beta}
\newcommand{\mpa}{\alpha_1}
\newcommand{\mpb}{\alpha_2}
\newcommand{\mqa}{\beta_1}
\newcommand{\mqb}{\beta_2}
\newcommand{\mx}{x}
\newcommand{\mxa}{x_a}
\newcommand{\mxb}{x_b}
\newcommand{\mxc}{x_c}
\newcommand{\mxd}{x_d}
\newcommand{\dl}{\delta}

\newcommand{\dmp}{\dot{\alpha}}
\newcommand{\dmq}{\dot{\beta}}
\newcommand{\dmx}{\dot{\mx}}

\newcommand{\ccy}{y_0}
\newcommand{\ccya}{y_1}
\newcommand{\ccyb}{y_2}
\newcommand{\ccyc}{y_3}
\newcommand{\ccz}{z_0}
\newcommand{\ccza}{z_1}
\newcommand{\cczb}{z_2}
\newcommand{\cczc}{z_3}
\newcommand{\ccx}{x}
\newcommand{\cca}{x_a}
\newcommand{\ccb}{x_b}
\newcommand{\ccc}{x_c}
\newcommand{\ccd}{x_d}
\newcommand{\cce}{x_e}
\newcommand{\ccf}{x_f}
\newcommand{\ccpc}{\alpha}
\newcommand{\ccpa}{\alpha_1}
\newcommand{\ccpb}{\alpha_2}
\newcommand{\ccqc}{\beta}
\newcommand{\ccqa}{\beta_1}
\newcommand{\ccqb}{\beta_2}
\newcommand{\ccra}{\gamma_1}
\newcommand{\ccrb}{\gamma_2}
\newcommand{\ccpp}{{\al}}
\newcommand{\ccqq}{{\bt}}
\newcommand{\ccrr}{{\gm}}
\newcommand{\ccpph}{{\hal}}
\newcommand{\ccqqh}{{\hbt}}

\newcommand{\ow}{\overline{W}}
\newcommand{\oV}{\overline{V}}

\newcommand{\Log}{{\textrm{Log}\hspace{1pt}}}

\newcommand{\lie}{{\textrm{Li}_2}}

\newcommand{\spn}{{\sigma}}

\newcommand{\A}[7]{A(#1;#2,#3,#4,#5;#6,#7)}
\newcommand{\B}[7]{B(#1;#2,#3,#4,#5;#6,#7)}
\newcommand{\C}[7]{C(#1;#2,#3,#4,#5;#6,#7)}

\def\EXP{\textrm{{\large e}}}

\def\re{\mathop{\hbox{\rm Re}}\nolimits}
\def\im{\mathop{\hbox{\rm Im}}\nolimits}

\newcounter{app}
\newcounter{sapp}[app]

\begin{document}

\vglue 2cm

\begin{center}

\title{Interaction-round-a-face and consistency-around-a-face-centered-cube}
\author{Andrew P.~Kels}
\address{Scuola Internazionale Superiore di Studi Avanzati,\\ Via Bonomea 265, 34136 Trieste, Italy}

\end{center}

\begin{abstract}

There is a correspondence between integrable lattice models of statistical mechanics and discrete integrable equations which satisfy multidimensional consistency, where the latter may be found in a quasi-classical expansion of the former.  This paper extends this correspondence to interaction-round-a-face (IRF) models, resulting in a new formulation of the consistency-around-a-cube (CAC) integrability condition applicable to five-point equations in the square lattice. 
Multidimensional consistency for these equations is formulated as consistency-around-a-face-centered-cube (CAFCC), which namely involves satisfying an overdetermined system of fourteen five-point lattice equations for eight unknown variables on the face-centered cubic unit cell.  From the quasi-classical limit of IRF models, which are constructed from the continuous spin solutions of the star-triangle relations associated to the Adler-Bobenko-Suris (ABS) list, fifteen sets of equations are obtained which satisfy CAFCC. 

\end{abstract}

\tableofcontents




\section{Introduction}


The property of multidimensional consistency, or consistency-around-a-cube (CAC), is widely accepted as a definition for integrability of partial difference equations defined on faces of the square lattice ({\it e.g.}, \cite[Ch.~3]{hietarinta_joshi_nijhoff_2016} and references therein).  This property essentially implies that the equations may be consistently extended into two-dimensional sublattices of $n$-dimensional lattices depending on $n$ associated lattice parameters, which may be regarded as the discrete analogue of the existence of hierarchies of compatible equations that are found for integrable partial differential equations.  Perhaps the most interesting result following the emergence of CAC as an integrability condition was a classification of scalar CAC equations given by Adler, Bobenko, and Suris (ABS) \cite{ABS,ABS2}, now commonly referred to as the ABS list.

In recent works \cite{Bazhanov:2007mh,Bazhanov:2010kz,Bazhanov:2016ajm,Kels:2018xge}, it has been found how integrable lattice (or quad) equations in the ABS list arise as part of a more general integrable structure based on a special form of the Yang-Baxter equation (YBE) known as the star-triangle relation (STR).  Essentially, the ABS equations are equivalent to the equations for the saddle points in a quasi-classical expansion of the STR.
Through this connection the STRs themselves may be naturally interpreted as quantum counterparts (in a path-integral sense) of the equations in the ABS list, and the entire ABS list may be systematically generated through the degenerations and quasi-classical expansions of the STRs \cite{Kels:2018xge}.  

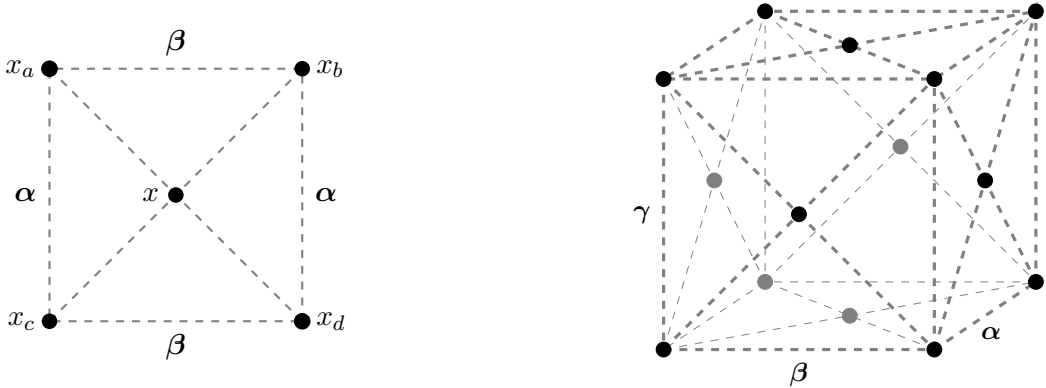
\begin{figure}[tbh]
\centering
\begin{tikzpicture}[scale=1.2]

\begin{scope}[scale=0.7]

\fill[white!] (1,1) circle (0.01pt)
node[left=1pt]{\color{black} $\al$};
\fill[white!] (5,1) circle (0.01pt)
node[right=1pt]{\color{black} $\al$};

\fill[white!] (3,3) circle (0.01pt)
node[above=1pt]{\color{black} $\bt$};
\fill[white!] (3,-1) circle (0.01pt)
node[below=1pt]{\color{black} $\bt$};

\draw[-,gray,thick,dashed] (5,-1)--(5,3)--(1,3)--(1,-1)--(5,-1);
\draw[-,gray,thick,dashed] (5,-1)--(1,3); \draw[-,gray,thick,dashed] (5,3)--(1,-1);

\fill (3,1) circle (3.5pt)
node[left=2.5pt]{\color{black} $\ccx$};
\fill (1,-1) circle (3.5pt)
node[left=1.5pt]{\color{black} $\ccc$};
\filldraw[fill=black,draw=black] (1,3) circle (3.5pt)
node[left=1.5pt]{\color{black} $\cca$};
\fill (5,3) circle (3.5pt)
node[right=1.5pt]{\color{black} $\ccb$};
\filldraw[fill=black,draw=black] (5,-1) circle (3.5pt)
node[right=1.5pt]{\color{black} $\ccd$};

\end{scope}



\begin{scope}[scale=0.75,xshift=370,yshift=-10]


\draw[-,gray,very thick,dashed] (2.5,4)--(-1.5,4)--(-3,3)--(1,3)--(2.5,4);
\draw[-,gray,very thick,dashed] (-3,-1)--(1,-1)--(2.5,0);\draw[-,gray,very thin,dashed] (2.5,0)--(-1.5,0)--(-3,-1);

\draw[-,gray,very thin,dashed] (-3,-1)--(2.5,0);\draw[-,gray,very thin,dashed] (-1.5,0)--(1,-1);
\draw[-,gray,very thick,dashed] (-3,3)--(-3,-1);\draw[-,gray,very thin,dashed] (-3,-1)--(-1.5,4);\draw[-,gray,very thin,dashed] (-3,3)--(-1.5,0)--(-1.5,4);
\draw[-,gray,very thin,dashed] (-1.5,0)--(2.5,4);\draw[-,gray,very thin,dashed] (-1.5,4)--(2.5,0);

\draw[-,gray,very thick,dashed] (-3,-1)--(1,3);\draw[-,gray,very thick,dashed] (-3,3)--(1,-1);
\draw[-,gray,very thick,dashed] (1,-1)--(2.5,4)--(2.5,0);\draw[-,gray,very thick,dashed] (1,-1)--(1,3)--(2.5,0);
\draw[-,gray,very thick,dashed] (-3,3)--(2.5,4);\draw[-,gray,very thick,dashed] (-1.5,4)--(1,3);

\filldraw[fill=black,draw=black] (1,-1) circle (3.1pt);
\filldraw[fill=gray,draw=gray] (-0.25,-0.5) circle (3.1pt);
\filldraw[fill=gray,draw=gray] (-1.5,0) circle (3.1pt);
\filldraw[fill=black,draw=black] (-3,-1) circle (3.1pt);
\filldraw[fill=black,draw=black] (2.5,0) circle (3.1pt);
\filldraw[fill=black,draw=black] (-3,3) circle (3.1pt);
\filldraw[fill=black,draw=black] (-1.5,4) circle (3.1pt);
\filldraw[fill=gray,draw=gray] (-2.25,1.5) circle (3.1pt);
\filldraw[fill=gray,draw=gray] (0.5,2) circle (3.1pt);
\filldraw[fill=black,draw=black] (2.5,4) circle (3.1pt);

\filldraw[fill=black,draw=black] (1,3) circle (3.1pt);
\filldraw[fill=black,draw=black] (-0.25,3.5) circle (3.1pt);
\filldraw[fill=black,draw=black] (-1.0,1) circle (3.1pt);
\filldraw[fill=black,draw=black] (1.75,1.5) circle (3.1pt);

\draw[white] (1.5,-0.75) circle (0.01pt) node[right=1pt]{\color{black}\small $\al$};
\draw[white] (-1,-1) circle (0.01pt) node[below=1pt]{\color{black}\small $\bt$};
\draw[white] (-3,1) circle (0.01pt) node[left=1pt]{\color{black}\small $\gm$};

\end{scope}

 \end{tikzpicture}
 
 \caption{Left: Variables and parameters of a five-point face-centered quad equation.  Right: The face-centered cube.  Equations on the face-centered cube will involve three different parameters $\al,\bt,\gm$.}
 \label{fig-intro}

\end{figure}

Motivated by these results, this paper considers the classical equations that arise from interaction-round-a-face (IRF) type models of statistical mechanics that are associated with the star-triangle relations.  The expression for the classical IRF Yang-Baxter equation will motivate a new formulation of multidimensional consistency, which is applicable to five-point lattice equations that come from the classical limit of the IRF Boltzmann weights.  Such five-point equations can be considered as equations defined on a face of the face-centered cube\footnote{Throughout this paper, {\it face-centered cube} refers to the face-centered cubic unit cell.}, which will be a central idea for their multidimensional consistency.  These equations will be referred to in this paper as {\it face-centered quad equations} to distinguish them from the regular quad equations which satisfy CAC.  The face-centered quad equation and the face-centered cube have a graphical interpretation shown in Figure \ref{fig-intro}.

The face-centered quad equations are expressed in terms of polynomials in five variables $x$, $x_a$, $x_b$, $x_c$, $x_d$, with linear dependence on the four corner variables $x_a$, $x_b$, $x_c$, $x_d$, respectively.  They also depend on the two-component parameters $\al=(\alpha_1,\alpha_2)$ and $\bt=(\beta_1,\beta_2)$, assigned to the edges.  If $n$ is the degree of $x$, such face-centered quad equations may be written in the form
\begin{align}\nonumber
\begin{split}
\sum_{i=0}^n P_i(x_a,x_b,x_c,x_d;\al,\bt)x^i=0,
\end{split}
\end{align}
where under a parameter specialisation, the coefficients $P_i(x_a,x_b,x_c,x_d;\al,\bt)$, $i=0,\ldots,n$, are expressions for affine-linear quad polynomials in the variables $x_a$, $x_b$, $x_c$, $x_d$.  

The above face-centered quad equations cannot satisfy a usual form of multidimensional consistency formulated as CAC. The main reason for this is the appearance of the face variable $x$ that has no counterpart for CAC, but also needs to be involved in determining the consistency of the equations.  To address this, an analogue of the property of CAC will be formulated as {\it consistency-around-a-face-centered-cube} (CAFCC).  For CAFCC, the six face-centered quad equations are defined on six faces of the face-centered cube, and the three parameters $\al$, $\bt$, and $\gm$, are associated to three orthogonal lattice directions, analogously to CAC.  However, in order to properly define an evolution around the face-centered cube, eight additional equations centered at corners are needed, which will provide the relations between the different face variables.  Then setting six initial variables on the face-centered cube, the CAFCC property is satisfied if the fourteen face-centered quad equations give consistent solutions for the remaining eight variables.  In comparison, CAC involves consistency of six quad equations on the cube for four unknown variables, so there is an increase in complexity in solving for CAFCC.

The two main results of this paper are the formulation of the CAFCC property described above, and the derivation of fifteen sets of face-centered quad equations which satisfy CAFCC, that come from new types of classical IRF Yang-Baxter equations associated to the star-triangle relations.   These face-centered quad equations are grouped as one of types-A, -B, or -C, where the type-A and -B equations are centered at faces of the face-centered cube, and the type-C equations are centered at corners of the face centered-cube.  The type-A equations possess the most symmetry, and satisfy CAFCC both on their own and in combination with type-B and -C equations.  The expressions for the type-A equations have previously appeared as discrete Toda-type (or Laplace-type) equations associated to type-Q equations in the ABS list \cite{AdlerPlanarGraphs,BobSurQuadGraphs,AdlerSurisQ4}, while some degenerate cases of the type-B equations may be identified with equations from the $H^6$ list of Boll \cite{BollThesis}.  For type-A and type-C equations, the type-Q and type-H ABS equations also respectively appear as a parameter specialisation of the coefficient $P_1(x_a,x_b,x_c,x_d;\al,\bt)$ of $x^1$.  Such connections to ABS equations is rather natural from the viewpoint of the connection between the  Yang-Baxter equation and multi-dimensional consistency \cite{Bazhanov:2016ajm,Kels:2018xge}, since both ABS equations and face-centered quad equations are constructed from the same Boltzmann weights/Lagrangian functions that satisfy the star-triangle relations.

The paper is organised as follows.  In Section \ref{sec:YBE}, it is shown how new expressions for classical IRF Yang-Baxter equations are constructed from the continuous spin solutions of the star-triangle relations.   In Section \ref{sec:FCC}, the face-centered quad equations and property of CAFCC are introduced.  Section \ref{sec:FCC} is intended to be self-contained, giving the formulation of the face-centered quad equations and CAFCC property without requiring knowledge of the connection to IRF equations that is given in Sections \ref{sec:YBE} and \ref{sec:CAFCCIRF}.  Section \ref{sec:FCClist} contains a list of face-centered quad equations which satisfy CAFCC, which also are given in their affine-linear form in Appendix \ref{app:afflin}.  In Section \ref{sec:CAFCCIRF}, it is shown case-by-case how the face-centered quad equations are derived from explicit solutions of the IRF equations of Section \ref{sec:YBE}.

\section{Star-triangle relations to classical IRF equations}\label{sec:YBE}

The star-triangle relation (STR) is a particular form of the Yang-Baxter equation that is satisfied for integrable Ising-type lattice models of statistical mechanics \cite{Baxter:1982zz,Bax02rip,PerkYBEs,Bazhanov:2016ajm}.  It is known that this relation also implies a Yang-Baxter equation for the model in either vertex or interaction-round-a-face (IRF) formulations \cite{Bazhanov:1992jqa,Baxter:1997tn,Bazhanov:2011mz}.  Such a construction will be used in this section to obtain new classical interaction-round-a-face (IRF) forms of the Yang-Baxter equation from continuous spin solutions of the STRs.  The expressions for the classical IRF equations will motivate the formulation of the new multidimensional consistency condition given in Section \ref{sec:FCC}.


The continuous spin solutions of the star-triangle relations and quasi-classical expansions at the rational level were previously considered in \cite{Bazhanov:2016ajm,Kels:2018xge}, and will be included in this section as examples.  The rational cases are chosen because they are the simplest cases that include examples of each of the three different types of star-triangle relation that will be used in this paper.

\subsection{Star-triangle relations: type 1}

The first type of star-triangle relation is given in terms of a pair of complex-valued Boltzmann weights, which are denoted here by $W_\theta(\spn_i,\spn_j)$ and $\oW_\theta(\spn_i,\spn_j)$.  Two different examples are respectively given by
\begin{align}
\label{BWex1a}
\begin{split}
W_{\theta}(\spn_i,\spn_j)&=\frac{\Gamma\bigl(\zeta-\theta+\ii(\spn_i+\spn_j)\bigr)\,\Gamma\bigl(\zeta-\theta+\ii(\spn_i-\spn_j)\bigr)}{\Gamma\bigl(\zeta+\theta+\ii(\spn_i+\spn_j)\bigr)\,\Gamma\bigl(\zeta+\theta+\ii(\spn_i-\spn_j)\bigr)}, \\[0.1cm]
\ow_{\theta}(\spn_i,\spn_j)&=\frac{\Gamma\bigl(\theta+\ii(\spn_i+\spn_j)\bigr)\,\Gamma\bigl(\theta+\ii(\spn_i-\spn_j)\bigr)\,\Gamma\bigl(\theta-\ii(\spn_i+\spn_j)\bigr)\,\Gamma\bigl(\theta-\ii(\spn_i-\spn_j)\bigr)}{2\pi\,\Gamma(2\theta)},
\end{split}
\end{align}
and
\begin{align}
\label{BWex1b}
\text{\scalebox{1.0}{$\ds W_{\theta}(\spn_i,\spn_j)=\frac{\Gamma\bigl(\zeta-\theta+\ii(\spn_i-\spn_j)\bigr)}{\Gamma\bigl(\zeta+\theta+\ii(\spn_i-\spn_j)\bigr)}, \quad
\ds\oW_{\theta}(\spn_i,\spn_j)=\frac{\Gamma\bigl(\theta+\ii(\spn_i-\spn_j)\bigr)\,\Gamma\bigl(\theta-\ii(\spn_i-\spn_j)\bigr)}{2\pi\,\Gamma(2\theta)},$}}
\end{align}
where $\Gamma(z)$ is the regular gamma function defined by
\begin{align}\label{ratgamdef}
\Gamma(z)=\int_{0}^\infty dtt^{z-1}\EXP^{-t}, \qquad \re(z)>0.
\end{align}
When the variables and parameters take values
\begin{align}\label{valsex}
\spn_i\in\mathbb{R},\qquad 0<\theta_i<\zeta,
\end{align}
where $\zeta$ is a positive parameter, the respective pairs of Boltzmann weights \eqref{BWex1a} and \eqref{BWex1b} satisfy the following form of the star-triangle relation \cite{Bazhanov:2016ajm,Kels:2018xge}
\begin{align}
\label{STR}
\begin{split}
\ds\int_\mathbb{R}\! d{\spn_d}\, \iS(\spn_d)\,\oW_{\theta_1}(\spn_a,\spn_d)\,W_{\theta_1+\theta_3}(\spn_b,\spn_d)\,\oW_{\theta_3}(\spn_d,\spn_c)\phantom{.} \\
\ds =W_{\theta_1}(\spn_b,\spn_c)\,\oW_{\theta_1+\theta_3}(\spn_a,\spn_c)\,W_{\theta_3}(\spn_b,\spn_a).
\end{split}
\end{align}
For the Boltzmann weights \eqref{BWex1b}, $\iS(\sigma)=1$, and for the Boltzmann weights \eqref{BWex1a},
\begin{align}\label{rats}
    \iS(\sigma)=\pi^{-1}\sigma_i\sin(2\pi\sigma_i).
\end{align}
The star-triangle relation \eqref{STR} can be analytically continued from \eqref{valsex} to more general complex values, as long as no poles of the integrand cross the real line (however the contour could also be deformed to allow for such a situation).  Up to a change of variables the equation \eqref{STR} for Boltzmann weights \eqref{BWex1a} and \eqref{BWex1b}, is respectively equivalent to hypergeometric integral formulas of Askey \cite{Askey1989} and Barnes's second lemma \cite{Barnes1910}.  

\subsubsection{Lagrangian functions and classical star-triangle relations}

For the quasi-classical expansion the variables and parameters are chosen as
\begin{align}\label{qclex}
    (\spn_a,\spn_b,\spn_c,\spn_d,\theta_1,\theta_3)=\hbar^{-1}(x_a,x_b,x_c,x_d,\alpha_1,\alpha_3),\qquad \hbar\to0.
\end{align}
Then the star-triangle relation \eqref{STR} is found to have an expansion in $\hbar$ of the form 
\begin{align}
\begin{split}
\ds\int_\mathbb{R}\! dx_d\, \exp\Bigl\{\hbar^{-1}\bigl(\ol_{\alpha_1}(x_a,x_d)+\lag_{\alpha_1+\alpha_3}(x_b,x_d)+\ol_{\alpha_3}(x_d,x_c)\bigr)+O(1)\Bigr\}\phantom{,} \\
\ds =\exp\Bigl\{\hbar^{-1}\bigl(\lag_{\alpha_1}(x_b,x_c)+\ol_{\alpha_1+\alpha_3}(x_a,x_c)+\lag_{\alpha_3}(x_b,x_a)\bigr)\Bigr\},
\end{split}
\end{align}
where the variables satisfy the three-leg saddle point equation
\begin{align}
\label{sym3leg1}
\frac{\partial}{\partial x_d}\bigl(\ol_{\alpha_1}(x_a,x_d)+\lag_{\alpha_1+\alpha_3}(x_b,x_d)+\ol_{\alpha_3}(x_d,x_c)\bigr)=0.
\end{align}
For the example of the Boltzmann weights \eqref{BWex1a},
\begin{align}
\label{lagex1a}
\begin{split}
\lag_{\alpha}(x_i,x_j)&=\gamma(x_i+x_j+\ii\alpha)+\gamma(x_i-x_j+\ii\alpha)-\gamma(x_i+x_j-\ii\alpha)-\gamma(x_i-x_j-\ii\alpha), \\
\ol_{\alpha}(x_i,x_j)&=\,\ds \gamma(x_i+x_j-\ii\alpha)+\gamma(-x_i+x_j-\ii\alpha) \\
&\hspace{0.15cm}+\gamma(x_i-x_j-\ii\alpha)+\gamma(-x_i-x_j-\ii\alpha) -\gamma(-2\ii\alpha),
\end{split}
\end{align}
and for the example of the Boltzmann weights \eqref{BWex1b},
\begin{align}
\label{lagex1b}
\begin{split}
\lag_{\alpha}(x_i,x_j)&=\gamma(x_i-x_j+\ii\alpha)-\gamma(x_i-x_j-\ii\alpha), \\
\ol_{\alpha}(x_i,x_j)&=\gamma(x_i-x_j-\ii\alpha)+\gamma(x_j-x_i-\ii\alpha)-\gamma(-2\ii\alpha),
\end{split}
\end{align}
where $\gamma(z)$ is defined in terms of the complex logarithm as
\begin{align}\label{gamlog}
    \gamma(z)=\ii z\Log(\ii z).
\end{align}
The function \eqref{gamlog} originates from the asymptotics of the gamma function \eqref{ratgamdef} given by the well-known Stirling formula.  The complex logarithm in \eqref{gamlog} (and throughout this paper) is to be taken with the principal branch.  

The Lagrangian functions \eqref{lagex1a} and \eqref{lagex1b} satisfy a classical form of the star-triangle relation \eqref{STR} given by
\begin{align}
\label{CSTR1}
\begin{split}
\ol_{\alpha_1}(x_a,x_d)+\lag_{\alpha_1+\alpha_3}(x_b,x_d)+\ol_{\alpha_3}(x_d,x_c)\phantom{,} \\
=\lag_{\alpha_1}(x_b,x_c)+\ol_{\alpha_1+\alpha_3}(x_a,x_c)+\lag_{\alpha_3}(x_b,x_a),
\end{split}
\end{align}
where the variables satisfy \eqref{sym3leg1}.  It is known \cite{Bazhanov:2016ajm,Kels:2018xge} that \eqref{sym3leg1} is equivalent modulo $2\pi\ii$ to an ABS equation \cite{ABS,ABS2} of the form
\begin{align}\label{sym3legABS}
    Q(y_a,y_b,y_c,y_d,\beta_1,\beta_3)=0,
\end{align}
where $Q$ is a multivariable polynomial that is linear in each of $y_a,y_b,y_c,y_d$.  The $y_a$, $y_b$, $y_c$, $y_d$, $\beta_1$, $\beta_3$, are respectively related to the $x_a$, $x_b$, $x_c$, $x_d$, $\alpha_1$, $\alpha_3$, by a change of variables.  For the rational solutions of the star-triangle relation, we have
\begin{align}\label{rattrans}
    \beta_1=-\ii\alpha_1,\qquad \beta_3=-\ii(\alpha_1+\alpha_3).
\end{align}
Then for the example of \eqref{lagex1a}, the polynomial $Q$ coming from \eqref{sym3leg1} is given by ($Q2$)
\begin{align}
    \label{Q2ex}
    Q=\beta_1(y_a-y_c)(y_b-y_d)-\beta(y_a-y_d)(y_b-y_c)+\beta_1\beta_3(\beta_1-\beta_3)(\!\!\sum_{I\in\{a,b,c,d\}}\!\!\!y_I-\beta_1^2-\beta_3^2+\beta_1\beta_3),
\end{align}
where $y_I=x_I^2$, for $I=a,b,c,d$, and for the example of \eqref{lagex1b}, the polynomial $Q$ coming from \eqref{sym3leg1} is given by ($Q1_{(\delta=1)}$)
\begin{align}
    \label{Q11ex}
    Q=\beta_1(y_a-y_c)(y_b-y_d)-\beta_3 (y_a-y_d)(y_b-y_c)+\beta_1\beta_3(\beta_1-\beta_3),
\end{align}
where $y_I=x_I$, for $I=a,b,c,d$.

Such a connection of the star-triangle relation of the form \eqref{STR} with integrable quad equations was first found for the Faddeev-Volkov model \cite{Bazhanov:2007mh} and Bazhanov-Sergeev master model \cite{Bazhanov:2010kz}, and is now known for the entire ABS list \cite{Bazhanov:2016ajm,Kels:2018xge}.  Classical star-triangle relations of the form \eqref{CSTR1} have also been derived for integrable quad equations independently of the quasi-classical expansion \cite{LobbNijhoff,BS09}.  The classical star-triangle relation \eqref{CSTR1} can be shown to hold in general through differentiation \cite{BS09}, because the derivatives with respect to its variables (and parameters) vanish on some subspace of the complex parameter space where the three-leg quad equation \eqref{sym3leg1} is satisfied, implying the equation \eqref{CSTR1} is equal to a constant.  This constant can be determined to be zero by evaluating \eqref{CSTR1} at a specific value of $x_a,x_b,x_c,x_d,\alpha_1,\alpha_3$.  

Due to the dependence on the complex logarithm, to determine the exact parameter space where the equations \eqref{CSTR1} and \eqref{sym3leg1} hold would require a complicated analysis of the arguments of the Lagrangian functions in \eqref{lagex1a} or \eqref{lagex1b}.  
However, for the purposes here it will be enough that for $x_I,\alpha_1,\alpha_3\in\mathbb{C}$ ($I\in\{a,b,c,d\}$), the classical star-triangle relation \eqref{CSTR1} holds on solutions of the quad equation \eqref{sym3legABS} up to a term of the form
\begin{align}\label{addfacs}
    2\pi\ii \sum_{I\in\{a,b,c,d\}}k_Ix_I+C(\alpha_1,\alpha_3),
\end{align}
for some $k_I\in\mathbb{Z}$ and $C$ constant with respect to the $x_I$.  This is because the equations of interest will be given in terms of exponentials of derivatives of equations related to \eqref{CSTR1}, which do not receive any contribution from terms of the form \eqref{addfacs}.  

For the examples of \eqref{lagex1a} and \eqref{lagex1b}, it can be seen directly that the classical star-triangle relation \eqref{CSTR1} holds up to terms of the form \eqref{addfacs}.  First, the equations \eqref{Q2ex} and \eqref{Q11ex} can be solved for $x_d$ and substituted into \eqref{CSTR1}, then by the repeated use of the formula 
\begin{align}\label{logident}
    \Log(xy)=\Log(x)+\Log(y)+2\pi\ii k,\qquad x,y\in\mathbb{C},
\end{align}
where $k$ is some integer, the products of the logarithms in the classical star-triangle relation \eqref{CSTR1} will cancel up to the additional terms of the form \eqref{addfacs}.  For the general case this follows from the method of derivatives used in \cite{BS09}, since the derivatives of \eqref{CSTR1} take the form of an additive three-leg ABS equation \eqref{sym3leg1}, which for $x_a,x_b,x_c,x_d,\alpha_1,\alpha_3\in\mathbb{C}$ and away from poles is satisfied on solutions of an ABS quad equation \eqref{sym3legABS} modulo $2\pi\ii$, in turn implying that the classical star-triangle relation \eqref{CSTR1} holds on solutions of \eqref{sym3legABS} modulo \eqref{addfacs}.

One useful consequence of allowing for terms of the form \eqref{addfacs}, is that the ordering of the variables in Lagrangian functions for \eqref{CSTR1} does not matter.  This usually already follows from the symmetries $\lag_\alpha(x_i,x_j)=\lag_\alpha(x_j,x_i)$ and $\ol_\alpha(x_i,x_j)=\ol_\alpha(x_j,x_i)$, which are satisfied by all cases, except \eqref{lagex1a} and \eqref{lagex1b}.  For these cases one has instead $\lag_\alpha(x_i,x_j)=\lag_\alpha(x_j,x_i)+2\pi\ii(k_1x_i+k_2x_j+k_3\ii\alpha)$, for some $k_1,k_2,k_3\in\mathbb{Z}$, and  thus exchanging the ordering of variables adds a term to \eqref{CSTR1} of the form of \eqref{addfacs}.

Another useful consequence of allowing for \eqref{addfacs} is that the Lagrangian functions \eqref{lagex1a} and \eqref{lagex1b} may respectively be replaced by the more convenient expressions
\begin{align}
\label{lagex1af}
\begin{split}
\lag_{\alpha}(x_i,x_j)=&\gamma(x_i+x_j+\ii\alpha)+\gamma(x_i-x_j+\ii\alpha)-\gamma(x_i+x_j-\ii\alpha)-\gamma(x_i-x_j-\ii\alpha), \\
\ol_{\alpha}(x_i,x_j)=&\,\ds \lag_{-\alpha}(x_i,x_j) -\gamma(-2\ii\alpha),
\end{split}
\end{align}
and 
\begin{align}
\label{lagex1bf}
\begin{split}
\lag_{\alpha}(x_i,x_j)&=\gamma(x_i-x_j+\ii\alpha)-\gamma(x_i-x_j-\ii\alpha), \\
\ol_{\alpha}(x_i,x_j)&=\lag_{-\alpha}(x_i,x_j)-\gamma(-2\ii\alpha).
\end{split}
\end{align}
With \eqref{lagex1af} or \eqref{lagex1bf}, the classical star-triangle relation \eqref{CSTR1} may now be more conveniently written in terms of a single Lagrangian function $\lag_\alpha$.  Comparing the expressions for Lagrangian functions in \eqref{lagex1a} and \eqref{lagex1af}, and \eqref{lagex1b} and \eqref{lagex1bf}, it is seen that this replacement only changes the classical star-triangle relation \eqref{CSTR1} by a term of the form \eqref{addfacs}.

\subsection{Star-triangle relations: type 2}

The second type of star-triangle relation involves a mixture of a pair of Boltzmann weights $W_{\theta}(\spn_i,\spn_j)$ and $\oW_{\theta}(\spn_i,\spn_j)$ which satisfy the first type of star-triangle relation \eqref{STR}, as well as another pair of Boltzmann weights which will be denoted by $V_{\theta}(\spn_i,\spn_j)$ and $\oV_{\theta}(\spn_i,\spn_j)$.  An example of the latter is given by
\begin{align}
\label{BWex2}
V_{\theta}(\spn_i,\spn_j)&\ds=\Gamma\bigl(\zeta-\theta+\ii(\spn_i+\spn_j)\bigr), \qquad \oV_{\theta}(\spn_i,\spn_j)\ds= V_{\zeta-\theta}(-\spn_i,-\spn_j).
\end{align}
Then for the values \eqref{valsex}, the Boltzmann weights \eqref{BWex2} together with the Boltzmann weights \eqref{BWex1b} collectively satisfy the following form of the star-triangle relation \cite{Kels:2018qzx}
\begin{align}
\label{ASTR}
\begin{split}
\ds\int_\mathbb{R} d{\spn_d}\,\oV_{\theta_1}(\spn_a,\spn_d)\,V_{\theta_1+\theta_3}(\spn_b,\spn_d)\,\oW_{\theta_3}(\spn_d,\spn_c)\phantom{.} \ds \\
\ds =V_{\theta_1}(\spn_b,\spn_c)\,\oV_{\theta_1+\theta_3}(\spn_a,\spn_c)\,W_{\theta_3}(\spn_b,\spn_a).
\end{split}
\end{align}
The star-triangle relation \eqref{ASTR} can be analytically continued from \eqref{valsex} to more general complex values, as long as no poles of the integrand cross the real line.  Up to a change of variables the integral \eqref{ASTR} with Boltzmann weights \eqref{BWex1b} and \eqref{BWex2} is equivalent to Barnes's first lemma \cite{Barnes1908}.

\subsubsection{Lagrangian functions and classical star-triangle relations}

For the quasi-classical expansion \eqref{qclex}, the star-triangle relation \eqref{ASTR} has the expansion in $\hbar$ of the form
\begin{align}\label{QCLASTR}
\begin{split}
\ds\int_\mathbb{R}\! dx_d\, \exp\Bigl\{\hbar^{-1}\bigl(\olam_{\alpha_1}(x_a,x_d)+\lam_{\alpha_1+\alpha_3}(x_b,x_d)+\ol_{\alpha_3}(x_d,x_c)\bigr)+O(1)\Bigr\}\phantom{.} \\
\ds =\exp\Bigl\{\hbar^{-1}\bigl(\lam_{\alpha_1}(x_b,x_c)+\olam_{\alpha_1+\alpha_3}(x_a,x_c)+\lag_{\alpha_3}(x_b,x_a)\bigr)\Bigr\},
\end{split}
\end{align}
where the variables satisfy the three-leg saddle point equation
\begin{align}
\label{mix3leg}
\frac{\partial}{\partial x_d}\bigl(\olam_{\alpha_1}(x_a,x_d)+\lam_{\alpha_1+\alpha_3}(x_b,x_d)+\ol_{\alpha_3}(x_d,x_c)\bigr)=0.
\end{align}
For the Boltzmann weights \eqref{BWex1b}, the Lagrangian functions $\lag_\alpha$ and $\ol_\alpha$ appeared in \eqref{lagex1b}, and for the Boltzmann weights \eqref{BWex2}, 
\begin{align}
\label{lagex2}
\Lambda_{\alpha}(x_i,x_j)=\gamma(x_i+x_j+\ii\alpha),\qquad \olam_{\alpha}(x_i,x_j)=\Lambda_{-\alpha}(-x_i,-x_j).
\end{align}
The Lagrangian functions \eqref{lagex1b} and \eqref{lagex2} satisfy a classical counterpart of \eqref{ASTR} given by
\begin{align}
\label{CASTR}
\begin{split}
\olam_{\alpha_1}(x_a,x_d)+\lam_{\alpha_1+\alpha_3}(x_b,x_d)+\ol_{\alpha_3}(x_d,x_c)\phantom{,} \\
=\lam_{\alpha_1}(x_b,x_c)+\olam_{\alpha_1+\alpha_3}(x_a,x_c)+\lag_{\alpha_3}(x_b,x_a),
\end{split}
\end{align}
where the variables satisfy \eqref{mix3leg}.  As for the previous case \eqref{CSTR1}, the classical star-triangle relation \eqref{CASTR} should be understood as being satisfied for $x_I,\alpha_1,\alpha_3\in\mathbb{C}$ ($I\in\{a,b,c,d\}$), up to a term of the form \eqref{addfacs}.

For the Lagrangian functions \eqref{lagex1b} and \eqref{lagex2}, the ABS equation \eqref{sym3legABS} that corresponds to \eqref{mix3leg} is given by ($H2_{(\varepsilon=0)}$)
\begin{align}
    Q= (y_a - y_b)(y_c-y_d) - (\beta_1-\beta_3)\Bigl(\beta_1+\beta_3 - \!\!\sum_{i\in\{a,b,c,d\}}\!\!\!y_i\Bigr),
\end{align}
where the parameters transform as \eqref{rattrans}, and $y_i=x_i$, for $i=a,b,c,d$.    
The Lagrangians \eqref{lagex2} also solve the classical star-triangle relation \eqref{CASTR} in combination with the more convenient form of the Lagrangians given in \eqref{lagex1bf}.  The ordering of variables in \eqref{CASTR} again does not matter, this time because the Lagrangian functions \eqref{lagex2} satisfy $\lam_\alpha(x_i,x_j)=\lam_\alpha(x_j,x_i)$.

\subsection{Star-triangle relations: type 3}

The third and final type of star-triangle relation is given in terms of two solutions of the second type of star-triangle relation \eqref{ASTR}, which are satisfied by a mixture of two pairs of Boltzmann weights which satisfy the first type star-triangle relation \eqref{STR}.  One pair of Boltzmann weights for \eqref{STR} will be denoted as usual by $W_\theta(\spn_i,\spn_j)$ and $\oW_\theta(\spn_i,\spn_j)$, and the other pair will be denoted by $\Wh_\theta(\spn_i,\spn_j)$ and $\oWh_\theta(\spn_i,\spn_j)$.

For example, $W_\theta(\spn_i,\spn_j)$ and $\oW_\theta(\spn_i,\spn_j)$ can be given by \eqref{BWex1b}, and $\Wh_\theta(\spn_i,\spn_j)$ and $\oWh_\theta(\spn_i,\spn_j)$ given by \eqref{BWex1a}.  Then two pairs of Boltzmann weights $V_{\theta}(\spn_i,\spn_j)$ and $\oV_{\theta}(\spn_i,\spn_j)$ are defined by  
\begin{align}
\label{BWex3a}
\ds V_{\theta}(\spn_i,\spn_j)=\Gamma\bigl(\zeta-\theta+\ii(\spn_i+\spn_j)\bigr)\,\Gamma\bigl(\zeta-\theta+\ii(\spn_i-\spn_j)\bigr), \; \ds \oV_{\theta}(\spn_i,\spn_j)=V_{\zeta-\theta}(-\spn_i,\spn_j),
\end{align}
and 
\begin{align}
\label{BWex3b}
\ds V_{\theta}(\spn_i,\spn_j)=\frac{\Gamma\bigl(\zeta-\theta+\ii(\spn_i+\spn_j)\bigr)}{\Gamma\bigl(\zeta+\theta+\ii(\spn_i-\spn_j)\bigr)}, \quad
\ds\oV_{\theta}(\spn_i,\spn_j)=\Gamma\bigl(\theta+\ii(\spn_i-\spn_j)\bigr)\,\Gamma\bigl(\theta-\ii(\spn_i+\spn_j)\bigr). 
\end{align}
Then for the values \eqref{valsex}, the star-triangle relation for the Boltzmann weights \eqref{BWex3a} is given by
\begin{align}
\label{ASTR1}
\begin{split}
\ds\int_{\mathbb{R}}\! d{\spn_d}\, \iS(\spn_d)\,\oV_{\theta_1}(\spn_a,\spn_d)\,V_{\theta_1+\theta_3}(\spn_b,\spn_d)\,\oWh_{\theta_3}(\spn_d,\spn_c)\phantom{.} \ds  \\
\ds =V_{\theta_1}(\spn_b,\spn_c)\,\oV_{\theta_1+\theta_3}(\spn_a,\spn_c)\,W_{\theta_3}(\spn_b,\spn_a),
\end{split}
\end{align}
where $\iS(\spn)$ is given in \eqref{rats}, and the star-triangle relation for the Boltzmann weights \eqref{BWex3b} is given by 
\begin{align}
\label{ASTR2}
\begin{split}
\ds\int_{\mathbb{R}}\! d{\spn_d}\, \oV_{\theta_1}(\spn_a,\spn_d)\,V_{\theta_1+\theta_3}(\spn_b,\spn_d)\,\oW_{\theta_3}(\spn_d,\spn_c)\phantom{.} \ds \\
\ds =V_{\theta_1}(\spn_b,\spn_c)\,\oV_{\theta_1+\theta_3}(\spn_a,\spn_c)\,\Wh_{\theta_3}(\spn_b,\spn_a),
\end{split}
\end{align}
The star-triangle relations \eqref{ASTR1} and \eqref{ASTR2} can respectively be analytically continued from \eqref{valsex} to more general complex values, as long as no poles of the respective integrands cross the real line.  In terms of hypergeometric integrals, the integral \eqref{ASTR1} is equivalent to the De Branges-Wilson integral \cite{DeBranges1972,Wilson1980}, and the integral \eqref{ASTR2} is equivalent to Barnes's second lemma \cite{Barnes1910} (with a different change of variables from the case of \eqref{STR} with \eqref{BWex1b}).

Unlike \eqref{ASTR}, the pair of Boltzmann weights $\oWh$ and $W$ in \eqref{ASTR1} together do not satisfy the star-triangle relation \eqref{STR}, and the pair of Boltzmann weights $\oW$ and $\Wh$ in \eqref{ASTR2} together also do not satisfy \eqref{STR}.  This is the main difference between the star-triangle relations for this case and the previous case \eqref{ASTR}.

\subsubsection{Lagrangian functions and classical star-triangle relations}

For the quasi-classical expansion \eqref{qclex}, the star-triangle relation \eqref{ASTR1} has an expansion in $\hbar$ of the form 
\begin{align} \label{QCLASTR1}
\begin{split}
\ds\int_\mathbb{R}\! dx_d\, \exp\Bigl\{\hbar^{-1}\bigl(\olam_{\alpha_1}(x_a,x_d)+\lam_{\alpha_1+\alpha_3}(x_b,x_d)+\olh_{\alpha_3}(x_d,x_c)\bigr)+O(1)\Bigr\}\phantom{.} \\
\ds =\exp\Bigr\{\hbar^{-1}\bigl(\lam_{\alpha_1}(x_b,x_c)+\olam_{\alpha_1+\alpha_3}(x_a,x_c)+\lag_{\alpha_3}(x_b,x_a)\bigr)\Bigr\},
\end{split}
\end{align}
where the variables satisfy the saddle point equation
\begin{align}
\label{asym3leg1}
\frac{\partial}{\partial x_d}\bigl(\olam_{\alpha_1}(x_a,x_d)+\lam_{\alpha_1+\alpha_3}(x_b,x_d)+\olh_{\alpha_3}(x_d,x_c)\bigr)=0.
\end{align}
For \eqref{QCLASTR1} and \eqref{asym3leg1},  $\olh$ is taken as $\ol$ from \eqref{lagex1a}, $\lag$ is taken from \eqref{lagex1b}, and 
\begin{align}
\label{lagex3a}
\Lambda_{\alpha}(x_i,x_j)=&\,\ds \gamma(x_i+x_j+\ii\alpha)+\gamma(x_i-x_j+\ii\alpha), \qquad
\olam_{\alpha}(x_i,x_j)=\Lambda_{-\alpha}(-x_i,x_j). 
\end{align}

Similarly, for the quasi-classical expansion \eqref{qclex}, the star-triangle relation  \eqref{ASTR2} has an expansion in $\hbar$ of the form 
\begin{align} \label{QCLASTR2}
\begin{split}
\ds\int_\mathbb{R}\! dx_d\, \exp\Bigl\{\hbar^{-1}\bigl(\olam_{\alpha_1}(x_a,x_d)+\lam_{\alpha_1+\alpha_3}(x_b,x_d)+\ol_{\alpha_3}(x_d,x_c)\bigr)+O(1)\Bigr\}\phantom{.} \\
\ds =\exp\Bigr\{\hbar^{-1}\bigl(\lam_{\alpha_1}(x_b,x_c)+\olam_{\alpha_1+\alpha_3}(x_a,x_c)+\lagh_{\alpha_3}(x_b,x_a)\bigr)\Bigr\},
\end{split}
\end{align}
where the variables satisfy the saddle point equation
\begin{align}
\label{asym3leg2t}
\frac{\partial}{\partial x_d}\bigl(\olam_{\alpha_1}(x_a,x_d)+\lam_{\alpha_1+\alpha_3}(x_b,x_d)+\ol_{\alpha_3}(x_d,x_c)\bigr)=0.
\end{align}
For \eqref{QCLASTR2} and \eqref{asym3leg2t}, $\lagh$ is taken as $\lag$ from \eqref{lagex1a}, $\ol$ is taken from \eqref{lagex1b}, and 
\begin{align}
\label{lagex3b}
\begin{split}
\Lambda_{\alpha}(x_i,x_j)=&\,\ds \gamma(x_i+x_j+\ii\alpha)-\gamma(x_i-x_j-\ii\alpha),\\ 
\olam_{\alpha}(x_i,x_j)=&\,\ds \gamma(x_i-x_j-\ii\alpha)+\gamma(-(x_i+x_j+\ii\alpha)).
\end{split}
\end{align}

The two sets of Lagrangian functions given in \eqref{lagex3a} and \eqref{lagex3b} satisfy the respective classical star-triangle relations of the form
\begin{align}
\label{CASTR1}
\begin{split}
\olam_{\alpha_1}(x_a,x_d)+\lam_{\alpha_1+\alpha_3}(x_b,x_d)+\olh_{\alpha_3}(x_d,x_c)\phantom{,} \\
=\lam_{\alpha_1}(x_b,x_c)+\olam_{\alpha_1+\alpha_3}(x_a,x_c)+\lag_{\alpha_3}(x_b,x_a),
\end{split}
\end{align}
and
\begin{align}
\label{CASTR2t}
\begin{split}
\olam_{\alpha_1}(x_a,x_d)+\lam_{\alpha_1+\alpha_3}(x_b,x_d)+\ol_{\alpha_3}(x_d,x_c)\phantom{,} \\
=\lam_{\alpha_1}(x_b,x_c)+\olam_{\alpha_1+\alpha_3}(x_a,x_c)+\lagh_{\alpha_3}(x_b,x_a),
\end{split}
\end{align}
where the variables for \eqref{CASTR1} satisfy \eqref{asym3leg1}, and the variables for \eqref{CASTR2t} satisfy \eqref{asym3leg2t}.   As for the previous cases of \eqref{CSTR1} and \eqref{CASTR}, the classical star-triangle relations \eqref{CASTR1} and \eqref{CASTR2t} should be understood as being satisfied for $x_I,\alpha_1,\alpha_3\in\mathbb{C}$ ($I\in\{a,b,c,d\}$), up to a term of the form \eqref{addfacs}.  Importantly, the ordering of variables does matter in \eqref{CASTR1} and \eqref{CASTR2t}, because for these cases there is no simple relation between $\lam_\alpha(x_i,x_j)$ and $\lam_{\alpha}(x_j,x_i)$.  

Both of the equations \eqref{asym3leg1} and \eqref{asym3leg2t} are equivalent to the ABS equation ($H3_{(\varepsilon=1)}$).  With the use of \eqref{rattrans}, the first equation \eqref{asym3leg1} gives
\begin{align}
    Q= (y_a-y_b)(y_c-y_d) +(\beta_1-\beta_3)(y_c+y_d-2y_ay_b+(\beta_1+\beta_3)(y_a+y_b)-\beta_1^2-\beta_3^2),
\end{align}
where $y_a=x_a$, $y_b=x_b$, $y_c=x_c^2$, and $y_d=x_d^2$. and the second equation \eqref{asym3leg2t} gives
\begin{align}
    Q= (y_a-y_b)(y_c-y_d)+(\beta_1-\beta_3)(y_a+y_b-2y_cy_d+(\beta_1+\beta_3)(y_c+y_d)-\beta_1^2-\beta_3^2),
\end{align}
where $y_a=x_a^2$, $y_b=x_b^2$, $y_c=x_c$, and $y_d=x_d$. 

The second expression for the classical star-triangle relation \eqref{CASTR2t}, can also be written in terms of the first set of Lagrangian functions \eqref{CASTR1}, as
\begin{align}
\label{CASTR2}
\begin{split}
\olam_{\alpha_1}(x_d,x_a)+\lam_{\alpha_1+\alpha_3}(x_d,x_b)+\ol_{\alpha_3}(x_c,x_d)\phantom{,} \\
=\lam_{\alpha_1}(x_c,x_b)+\olam_{\alpha_1+\alpha_3}(x_c,x_a)+\lagh_{\alpha_3}(x_a,x_b),
\end{split}
\end{align}
where
\begin{align}
\label{asym3leg2}
\frac{\partial}{\partial x_d}\bigl(\olam_{\alpha_1}(x_d,x_a)+\lam_{\alpha_1+\alpha_3}(x_d,x_b)+\ol_{\alpha_3}(x_c,x_d)\bigr)=0.
\end{align}
The classical star-triangle relation \eqref{CASTR2} with \eqref{lagex3a} only differs from the classical star-triangle relation \eqref{CASTR2t} with \eqref{lagex3b} by a term of the form \eqref{addfacs}.  The form of the classical star-triangle relation \eqref{CASTR2} will be more useful than the form \eqref{CASTR2t}, because the two star-triangle relations \eqref{CASTR1} and \eqref{CASTR2} are written in terms of the single Lagrangian function $\lam_\alpha$ given in \eqref{lagex3a}.  For a similar reason, it is also more convenient to use the Lagrangian functions \eqref{lagex1af} and \eqref{lagex1bf} in the classical star-triangle relations \eqref{CASTR1} and \eqref{CASTR2}, instead of the Lagrangian functions \eqref{lagex1a} and \eqref{lagex1b}.  This will be done in Section \ref{sec:CAFCCIRF}.

The classical star-triangle relations considered in this paper always take one of the forms given in \eqref{CSTR1}, \eqref{CASTR}, \eqref{CASTR1}, or \eqref{CASTR2}.  The star-triangle relations \eqref{CASTR1} and \eqref{CASTR2} can be considered as a general case, in the sense that setting  $\olh_\alpha=\ol_\alpha$ and $\lagh_\alpha=\lag_\alpha$ gives the second type of star-triangle relation \eqref{CASTR}, and setting $\olam_\alpha=\olh_\alpha=\ol_\alpha$ and $\lam_\alpha=\lagh_\alpha=\lag_\alpha$ gives the first type of star-triangle relation \eqref{CSTR1}.  Thus the remainder of this section will mainly focus on the IRF type equations that can be obtained from \eqref{CASTR1} and \eqref{CASTR2}, from which the IRF equations for the other types of star-triangle relations \eqref{CSTR1} and \eqref{CASTR} follow from similar, but simpler, constructions.

\begin{figure}[htb!]
\centering
\begin{tikzpicture}[scale=2.2]

\draw[-] (0,-0.5)--(0,0.5);
\filldraw[fill=black,draw=black] (0,-0.5) circle (1.0pt)
node[below=4pt]{\color{black} $i$};
\filldraw[fill=black,draw=black] (0,0.5) circle (1.0pt)
node[above=4pt]{\color{black}$j$};

\fill (0,-0.8) circle(0.01pt)
node[below=0.05pt]{\color{black} $ \lag_{\alpha}(\xv_i,\xv_j)$}
node[below=15.05pt]{\color{black} $ \lagh_{\alpha}(\xv_i,\xv_j)$};

\begin{scope}[xshift=55pt]
\draw[-|-] (0,-0.5)--(0,0.5);
\filldraw[fill=black,draw=black] (0,-0.5) circle (1.0pt)
node[below=4pt]{\color{black} $i$};
\filldraw[fill=black,draw=black] (0,0.5) circle (1.0pt)
node[above=4pt]{\color{black} $j$};


\fill (0,-0.8) circle(0.01pt)
node[below=0.05pt]{\color{black} $\ol_{\alpha}(\xv_i,\xv_j)$}
node[below=15.05pt]{\color{black} $\olh_{\alpha}(\xv_i,\xv_j)$};
\end{scope}

\begin{scope}[xshift=105pt]
\draw[-,double, thick] (0,-0.5)--(0,0.5);
\draw[->>>-,white] (0,-0.5)--(0,0.5);
\filldraw[fill=black,draw=black] (0,-0.5) circle (1.0pt)
node[below=4pt]{\color{black} $i$};
\filldraw[fill=black,draw=black] (0,0.5) circle (1.0pt)
node[above=4pt]{\color{black} $j$};


\fill (0,-0.9) circle(0.01pt)
node[below=0.05pt]{\color{black} $ \lam_{\alpha}(\xv_i,\xv_j)$};
\end{scope}

\begin{scope}[xshift=150pt]
\draw[-,double, thick] (0,-0.5)--(0,0.5);
\draw[->>>>-,white] (0,-0.5)--(0,0.5);
\filldraw[fill=black,draw=black] (0,-0.5) circle (1.0pt)
node[below=4pt]{\color{black} $i$};
\filldraw[fill=black,draw=black] (0,0.5) circle (1.0pt)
node[above=4pt]{\color{black} $j$};

\fill (0,-0.9) circle(0.01pt)
node[below=0.05pt]{\color{black} $\olam_{\alpha}(\xv_i,\xv_j)$};
\end{scope}

\end{tikzpicture}
\caption{The four types of edges associated to the different Lagrangian functions entering the classical star-triangle relations \eqref{CASTR1} and \eqref{CASTR2}.  The two types of directed edges on the right are used to indicate the ordering of variables for non-symmetric Lagrangian functions.}
\label{BWfig}
\end{figure}
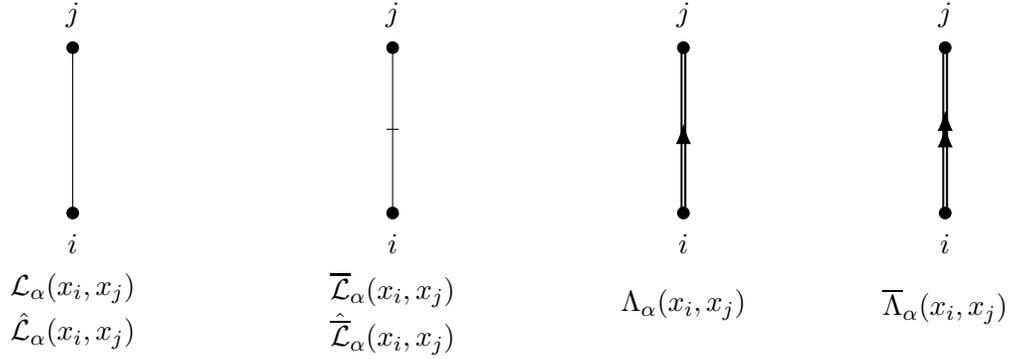

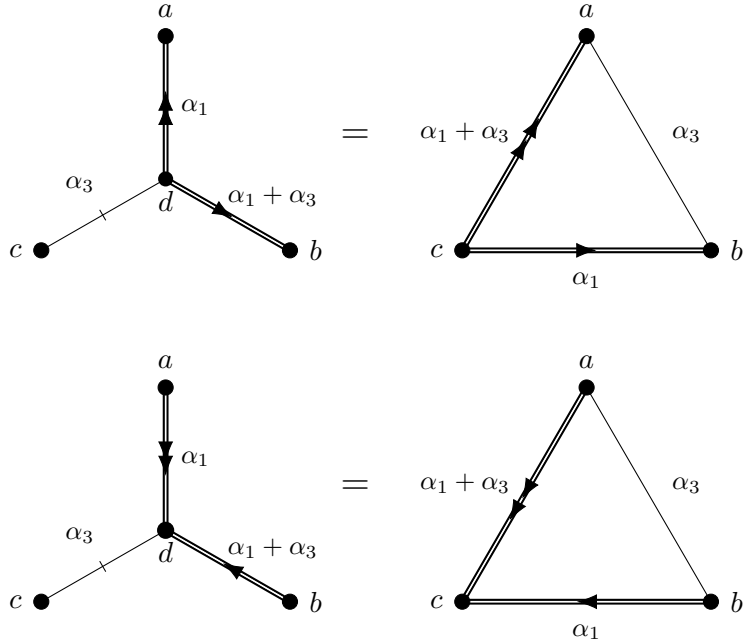
\begin{figure}[htb!]
\centering

\begin{tikzpicture}[scale=1.9]

\draw[white] (-2.0,0.5) circle (0.1pt)
node[right=2pt]{\color{black}\small $\alpha_1$};
\draw[white] (-2.6,-0.2) circle (0.1pt)
node[above=2pt]{\color{black}\small $\alpha_3$};
\draw[white] (-1.25,-0.3) circle (0.1pt)
node[above=2pt]{\color{black}\small $\alpha_1+\alpha_3$};
\draw[-|-] (-2,0)--(-2.87,-0.5);
\draw[-,double, thick] (-2,0)--(-2,1);
\draw[-,double, thick] (-2,0)--(-1.13,-0.5);
\draw[->>>>-,white] (-2,0)--(-2,1);
\draw[->>>-,white] (-2,0)--(-1.13,-0.5);
\fill (-2,0) circle (1.5pt)
node[below=0.5pt]{\color{black} $d$};
\filldraw[fill=black,draw=black] (-2,1) circle (1.5pt)
node[above=3.5pt] {\color{black} $a$};
\filldraw[fill=black,draw=black] (-2.87,-0.5) circle (1.5pt)
node[left=3.5pt] {\color{black} $c$};
\filldraw[fill=black,draw=black] (-1.13,-0.5) circle (1.5pt)
node[right=3.5pt] {\color{black} $b$};

\fill[white!] (-0.5,0.3) circle (0.01pt)
node[left=0.05pt] {\color{black}\Large $=$};

\begin{scope}[xshift=-30pt]

\draw[white] (2.0,-0.55) circle (0.1pt)
node[below=2pt]{\color{black}\small $\alpha_1$};
\draw[white] (2.7,0.15) circle (0.1pt)
node[above=2pt]{\color{black}\small $\alpha_3$};
\draw[white] (1.15,0.15) circle (0.1pt)
node[above=2pt]{\color{black}\small $\alpha_1+\alpha_3$};
\draw[-,double, thick] (1.13,-0.5)--(2,1);
\draw[-,double, thick] (2.87,-0.5)--(1.13,-0.5);
\draw[->>>>-,white] (1.13,-0.5)--(2,1);
\draw[->>>-,white] (1.13,-0.5)--(2.87,-0.5);
\draw[-] (2.87,-0.5)--(2,1);
\filldraw[fill=black,draw=black] (2,1) circle (1.5pt)
node[above=3.5pt]{\color{black} $a$};
\filldraw[fill=black,draw=black] (1.13,-0.5) circle (1.5pt)
node[left=3.5pt]{\color{black} $c$};
\filldraw[fill=black,draw=black] (2.87,-0.5) circle (1.5pt)
node[right=3.5pt]{\color{black} $b$};

\end{scope}


\begin{scope}[yshift=-70pt]

\draw[white] (-2.0,0.5) circle (0.1pt)
node[right=2pt]{\color{black}\small $\alpha_1$};
\draw[white] (-2.6,-0.2) circle (0.1pt)
node[above=2pt]{\color{black}\small $\alpha_3$};
\draw[white] (-1.25,-0.3) circle (0.1pt)
node[above=2pt]{\color{black}\small $\alpha_1+\alpha_3$};
\draw[-|-] (-2,0)--(-2.87,-0.5);
\draw[-,double, thick] (-2,1)--(-2,0);
\draw[-,double, thick] (-1.13,-0.5)--(-2,0);
\draw[->>>>-,white] (-2,1)--(-2,0);
\draw[->>>-,white] (-1.13,-0.5)--(-2,0);
\filldraw[fill=black,draw=black, thick] (-2,0) circle (1.5pt)
node[below=0.5pt]{\color{black} $d$};
\filldraw[fill=black,draw=black] (-2,1) circle (1.5pt)
node[above=3.5pt] {\color{black} $a$};
\filldraw[fill=black,draw=black] (-2.87,-0.5) circle (1.5pt)
node[left=3.5pt] {\color{black} $c$};
\filldraw[fill=black,draw=black] (-1.13,-0.5) circle (1.5pt)
node[right=3.5pt] {\color{black} $b$};

\fill[white!] (-0.5,0.3) circle (0.01pt)
node[left=0.05pt] {\color{black}\Large $=$};

\begin{scope}[xshift=-30pt]

\draw[white] (2.0,-0.55) circle (0.1pt)
node[below=2pt]{\color{black}\small $\alpha_1$};
\draw[white] (2.7,0.15) circle (0.1pt)
node[above=2pt]{\color{black}\small $\alpha_3$};
\draw[white] (1.15,0.15) circle (0.1pt)
node[above=2pt]{\color{black}\small $\alpha_1+\alpha_3$};
\draw[-,double, thick] (2,1)--(1.13,-0.5);
\draw[-,double, thick] (2.87,-0.5)--(1.13,-0.5);
\draw[->>>>-,white] (2,1)--(1.13,-0.5);
\draw[->>>-,white] (2.87,-0.5)--(1.13,-0.5);
\draw[-] (2.87,-0.5)--(2,1);
\filldraw[fill=black,draw=black] (2,1) circle (1.5pt)
node[above=3.5pt]{\color{black} $a$};
\filldraw[fill=black,draw=black] (1.13,-0.5) circle (1.5pt)
node[left=3.5pt]{\color{black} $c$};
\filldraw[fill=black,draw=black] (2.87,-0.5) circle (1.5pt)
node[right=3.5pt]{\color{black} $b$};

\end{scope}

\end{scope}

\end{tikzpicture}

\caption{The classical star-triangle relations \eqref{CASTR1} (top) and \eqref{CASTR2} (bottom) in terms of the edges of Figure \ref{BWfig}.  Both of the above diagrams are equivalent for the classical star-triangle relation \eqref{CASTR}, because the Lagrangians associated to the directed edges would be symmetric. The diagram for the classical star-triangle relation \eqref{CSTR1} would only involve the two types of edges on the left of Figure \ref{BWfig}.}
\label{YBErapidityfig1}
\end{figure}

Finally, if the edges are associated to the different types of Lagrangian functions as shown in Figure \ref{BWfig}, the classical star-triangle relations \eqref{CASTR1} and \eqref{CASTR2} have the convenient graphical representation shown in Figure \ref{YBErapidityfig1}.  Such a graphical representation will be useful for understanding the more complicated equations that arise in the following subsections.

\subsection{Classical star-star relations}

The above star-triangle relations can be used to construct another equation called the star-star relation \cite{Baxter:1986df,Bazhanov:1992jqa,Baxter:1997tn,Bazhanov:2011mz}.  Rather than the parameters $\alpha_1$ and $\alpha_3$ that appeared for the star-triangle relations, it will be more convenient to work with the following two-component parameters
\begin{align}
\bu=(u_1,u_2),\qquad \bv=(v_1,v_2).
\end{align}
The parameters entering the Lagrangian functions for the star-star relation will then be given in terms of differences of the components of $\bu$ and $\bv$.  Let $\lam^{(1)}_{\bu\bv}$,  $\lam^{(2)}_{\bu\bv}$, $\lamh^{(1)}_{\bu\bv}$, and $\lamh^{(2)}_{\bu\bv}$, respectively denote the following sums of Lagrangian functions from the third type of the classical star-triangle relations \eqref{CASTR1} and \eqref{CASTR2}:
\begin{align}\label{cssrfunctions}
\begin{split}
 \text{\scalebox{0.98}{$\lam^{(1)}_{\bu\bv}\!\left(\!x_e\Big|\,\arraycolsep=0.7pt\begin{array}{ll}x_a & x_b \\  x_c &x_d\end{array}\right)$}}
=& \text{\scalebox{0.98}{ $\lag_{u_2-v_1}(x_a,x_e)+\olam_{u_2-v_2}(x_e,x_b)+\ol_{u_1-v_1}(x_c,x_e)+\lam_{u_1-v_2}(x_e,x_d),$}} \\
 \text{\scalebox{0.98}{ $\lam^{(2)}_{\bu\bv}\!\left(\!x_f\Big|\,\arraycolsep=0.7pt\begin{array}{ll}x_a & x_b \\  x_c &x_d\end{array}\right)$}}
=& \text{\scalebox{0.98}{ $\lam_{u_1-v_2}(x_a,x_f)+\olh_{u_1-v_1}(x_f,x_b)+\olam_{u_2-v_2}(x_c,x_f)+\lagh_{u_2-v_1}(x_f,x_d),$}} \\  
\text{\scalebox{0.98}{ $\lamh^{(1)}_{\bu\bv}\!\left(\!x_e\Big|\,\arraycolsep=0.7pt\begin{array}{ll}x_a & x_b \\  x_c &x_d\end{array}\right)$}}
=& \text{\scalebox{0.98}{ $\olam_{v_1-u_2}(x_e,x_a)+\lam_{v_2-u_2}(x_e,x_b)+\ol_{u_1-v_1}(x_c,x_e)+\lag_{u_1-v_2}(x_d,x_e),$}} \\
 \text{\scalebox{0.98}{ $\lamh^{(2)}_{\bu\bv}\!\left(\!x_f\Big|\,\arraycolsep=0.7pt\begin{array}{ll}x_a & x_b \\  x_c &x_d\end{array}\right)$}}
=& \text{\scalebox{0.98}{ $\lagh_{u_1-v_2}(x_f,x_a)+\olh_{u_1-v_1}(x_f,x_b)+\lam_{v_2-u_2}(x_c,x_f)+\olam_{v_1-u_2}(x_d,x_f).$}}
\end{split}
\end{align}
Then two expressions for the classical star-star relations are given by
\begin{align}
\label{CSSR1}
\begin{split}
\lam_{v_1-v_2}(x_a,x_b)+\lag_{u_1-u_2}(x_a,x_c)+\lam^{(1)}_{\bu\bv}\!\left(\!x_e\Big|\,\arraycolsep=0.7pt\begin{array}{ll}x_a & x_b \\  x_c &x_d\end{array}\right)\!\phantom{,}   \\
=\lam_{v_1-v_2}(x_c,x_d)+\lag_{u_1-u_2}(x_b,x_d)+\lam^{(2)}_{\bu\bv}\!\left(\!x_f\Big|\,\arraycolsep=0.7pt\begin{array}{ll}x_a & x_b \\  x_c &x_d\end{array}\right)\!,
\end{split}
\end{align}
and
\begin{align}
\label{CSSR2}
\begin{split}
\olam_{u_1-u_2}(x_d,x_b)+\lag_{v_2-v_1}(x_c,x_d)+\lamh^{(1)}_{\bu\bv}\!\left(\!x_e\Big|\,\arraycolsep=0.7pt\begin{array}{ll}x_a & x_b \\  x_c &x_d\end{array}\right)\!\phantom{,} \\
=\olam_{u_1-u_2}(x_c,x_a)+\lag_{v_2-v_1}(x_a,x_b)+\lamh^{(2)}_{\bu\bv}\!\left(\!x_f\Big|\,\arraycolsep=0.7pt\begin{array}{ll}x_a & x_b \\  x_c &x_d\end{array}\right)\!,
\end{split}
\end{align}
where the variables of \eqref{CSSR1} and \eqref{CSSR2} satisfy the respective pairs of equations for the derivatives
\begin{align}
\frac{\partial}{\partial x_e}\lam^{(1)}_{\bu\bv}\!\left(\!x_e\Big|\,\arraycolsep=0.7pt\begin{array}{ll}x_a & x_b \\  x_c &x_d\end{array}\right)=0,\qquad
\frac{\partial}{\partial x_f}\lam^{(2)}_{\bu\bv}\!\left(\!x_f\Big|\,\arraycolsep=0.7pt\begin{array}{ll}x_a & x_b \\  x_c &x_d\end{array}\right)=0,
\end{align}
and 
\begin{align}
\frac{\partial}{\partial x_e}\lamh^{(1)}_{\bu\bv}\!\left(\!x_e\Big|\,\arraycolsep=0.7pt\begin{array}{ll}x_a & x_b \\  x_c &x_d\end{array}\right)=0,\qquad \frac{\partial}{\partial x_f}\lamh^{(2)}_{\bu\bv}\!\left(\!x_f\Big|\,\arraycolsep=0.7pt\begin{array}{ll}x_a & x_b \\  x_c &x_d\end{array}\right)=0.
\end{align}
As for the classical star-triangle relations which are satisfied up to \eqref{addfacs}, for $x_I,u_1,u_2,v_1,v_2\in\mathbb{C}$ ($I\in\{a,b,c,d,e,f\}$) and away from poles, the classical star-star relations should be regarded as equations being satisfied up to terms of the form $2\pi\ii\sum_{I\in\{a,b,c,d,e,f\}}k_Ix_I+C(\bu,\bv)$, for some $k_I\in\mathbb{Z}$, and $C$ constant with respect to the $x_I$.

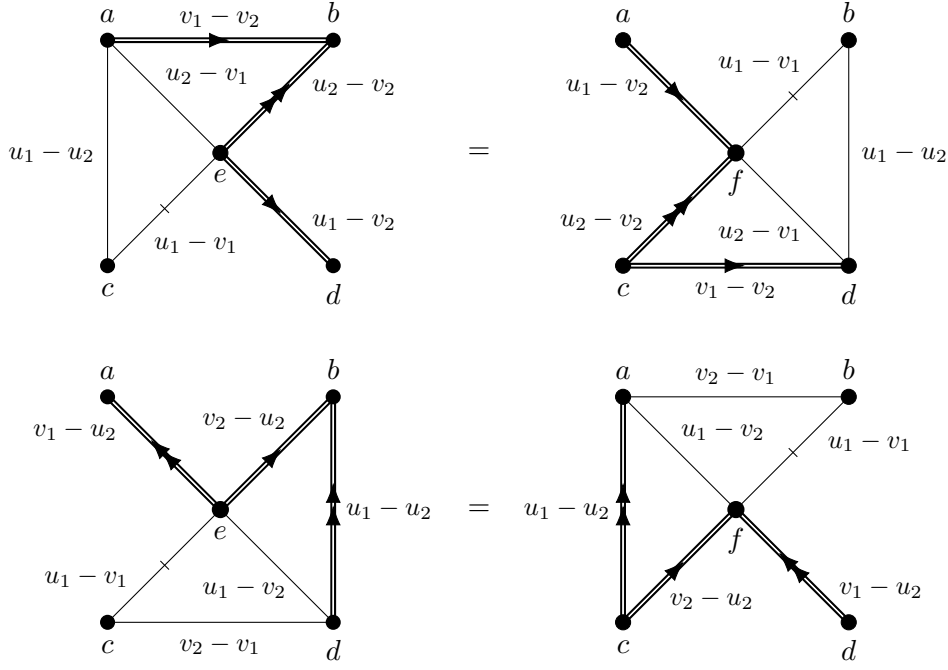
\begin{figure}[htb!]
\centering
\begin{tikzpicture}[scale=1.5]


\draw[white] (0,1) circle (0.01pt)
node[above=1pt]{\color{black}\small $v_1-v_2$};
\draw[white] (-1,0) circle (0.01pt)
node[left=1pt]{\color{black}\small $u_1-u_2$};
\draw[white] (-0.6,0.7) circle (0.01pt)
node[right=1pt]{\color{black}\small $u_2-v_1$};
\draw[white] (0.7,0.6) circle (0.01pt)
node[right=1pt]{\color{black}\small $u_2-v_2$};
\draw[white] (0.7,-0.6) circle (0.01pt)
node[right=1pt]{\color{black}\small $u_1-v_2$};
\draw[white] (-0.7,-0.8) circle (0.01pt)
node[right=1pt]{\color{black}\small $u_1-v_1$};

\draw[-] (-1,1)--(0,0);
\draw[-,double,thick] (0,0)--(1,-1);
\draw[->>>-,white] (0,0)--(1,-1);

\draw[-|-] (-1,-1)--(0,0);
\draw[-,double,thick] (0,0)--(1,1);
\draw[->>>>-,white] (0,0)--(1,1);

\draw[-,double,thick] (-1,1)--(1,1);
\draw[->>>-,white] (-1,1)--(1,1);
\draw[-] (-1,-1)--(-1,1);

\fill[black!] (-1,-1) circle (1.9pt)
node[below=3pt]{\color{black} $c$};
\fill[black!] (1,-1) circle (1.9pt)
node[below=3pt]{\color{black} $d$};
\fill[black!] (-1,1) circle (1.9pt)
node[above=3pt]{\color{black} $a$};
\fill[black!] (1,1) circle (1.9pt)
node[above=3pt]{\color{black} $b$};

\filldraw[fill=black!,draw=black!] (0,0) circle (1.9pt)
node[below=2.5pt]{\color{black} $e$};

\draw[white!] (2.1,0.0) circle (0.01pt)
node[right=0.1pt]{\color{black}$=$};

\begin{scope}[xshift=130pt]


\draw[white] (0,-1) circle (0.01pt)
node[below=1pt]{\color{black}\small $v_1-v_2$};
\draw[white] (1,0) circle (0.01pt)
node[right=1pt]{\color{black}\small $u_1-u_2$};
\draw[white] (-0.65,0.6) circle (0.01pt)
node[left=1pt]{\color{black}\small $u_1-v_2$};
\draw[white] (0.7,0.8) circle (0.01pt)
node[left=1pt]{\color{black}\small $u_1-v_1$};
\draw[white] (0.7,-0.7) circle (0.01pt)
node[left=1pt]{\color{black}\small $u_2-v_1$};
\draw[white] (-0.7,-0.6) circle (0.01pt)
node[left=1pt]{\color{black}\small $u_2-v_2$};

\draw[-,double,thick] (-1,1)--(0,0);
\draw[->>>-,white] (-1,1)--(0,0);
\draw[-] (0,0)--(1,-1);

\draw[-,double,thick] (-1,-1)--(0,0);
\draw[->>>>-,white] (-1,-1)--(0,0);
\draw[-|-] (0,0)--(1,1);

\draw[-] (1,1)--(1,-1);
\draw[-,double,thick] (-1,-1)--(1,-1);
\draw[->>>-,white] (-1,-1)--(1,-1);

\filldraw[fill=black!,draw=black!] (-1,-1) circle (1.9pt)
node[below=3pt]{\color{black} $c$};
\filldraw[fill=black!,draw=black!] (1,-1) circle (1.9pt)
node[below=3pt]{\color{black} $d$};
\filldraw[fill=black!,draw=black!] (-1,1) circle (1.9pt)
node[above=3pt]{\color{black} $a$};
\filldraw[fill=black!,draw=black!] (1,1) circle (1.9pt)
node[above=3pt]{\color{black} $b$};

\filldraw[fill=black!,draw=black!,thick] (0,0) circle (1.9pt)
node[below=2.5pt]{\color{black} $f$};

\end{scope}

\begin{scope}[yshift=-90pt]


\draw[white] (1.0,0) circle (0.01pt)
node[right=1pt]{\color{black}\small $u_1-u_2$};
\draw[white] (0,-1) circle (0.01pt)
node[below=1pt]{\color{black}\small $v_2-v_1$};
\draw[white] (-0.7,-0.6) circle (0.01pt)
node[left=1pt]{\color{black}\small $u_1-v_1$};
\draw[white] (-0.8,0.7) circle (0.01pt)
node[left=1pt]{\color{black}\small $v_1-u_2$};
\draw[white] (0.7,0.8) circle (0.01pt)
node[left=1pt]{\color{black}\small $v_2-u_2$};
\draw[white] (0.7,-0.7) circle (0.01pt)
node[left=1pt]{\color{black}\small $u_1-v_2$};

\draw[-|-] (-1,-1)--(0,0);
\draw[-,double,thick] (0,0)--(1,1);
\draw[->>>-,white] (0,0)--(1,1);

\draw[-] (1,-1)--(0,0);
\draw[-,double,thick] (0,0)--(-1,1);
\draw[->>>>-,white] (0,0)--(-1,1);

\draw[-] (-1,-1)--(1,-1);
\draw[-,double,thick] (1,-1)--(1,1);
\draw[->>>>-,white] (1,-1)--(1,1);

\fill[black!] (-1,-1) circle (1.9pt)
node[below=3pt]{\color{black} $c$};
\fill[black!] (1,-1) circle (1.9pt)
node[below=3pt]{\color{black} $d$};
\fill[black!] (-1,1) circle (1.9pt)
node[above=3pt]{\color{black} $a$};
\fill[black!] (1,1) circle (1.9pt)
node[above=3pt]{\color{black} $b$};

\filldraw[fill=black!,draw=black!,thick] (0,0) circle (1.9pt)
node[below=2.5pt]{\color{black} $e$};

\draw[white!] (2.1,0.0) circle (0.01pt)
node[right=0.1pt]{\color{black}$=$};

\begin{scope}[xshift=130pt]


\draw[white] (-1.0,0) circle (0.01pt)
node[left=1pt]{\color{black}\small $u_1-u_2$};
\draw[white] (0,1) circle (0.01pt)
node[above=1pt]{\color{black}\small $v_2-v_1$};
\draw[white] (-0.7,-0.8) circle (0.01pt)
node[right=1pt]{\color{black}\small $v_2-u_2$};
\draw[white] (-0.6,0.7) circle (0.01pt)
node[right=1pt]{\color{black}\small $u_1-v_2$};
\draw[white] (0.7,0.6) circle (0.01pt)
node[right=1pt]{\color{black}\small $u_1-v_1$};
\draw[white] (0.8,-0.7) circle (0.01pt)
node[right=1pt]{\color{black}\small $v_1-u_2$};

\draw[-,double,thick] (-1,-1)--(0,0);
\draw[->>>-,white] (-1,-1)--(0,0);
\draw[-|-] (0,0)--(1,1);

\draw[-,double,thick] (1,-1)--(0,0);
\draw[->>>>-,white] (1,-1)--(0,0);
\draw[-] (0,0)--(-1,1);

\draw[-,double,thick] (-1,-1)--(-1,1);
\draw[->>>>-,white] (-1,-1)--(-1,1);
\draw[-] (-1,1)--(1,1);

\filldraw[fill=black!,draw=black!] (-1,-1) circle (1.9pt)
node[below=3pt]{\color{black} $c$};
\filldraw[fill=black!,draw=black!] (1,-1) circle (1.9pt)
node[below=3pt]{\color{black} $d$};
\filldraw[fill=black!,draw=black!] (-1,1) circle (1.9pt)
node[above=3pt]{\color{black} $a$};
\filldraw[fill=black!,draw=black!] (1,1) circle (1.9pt)
node[above=3pt]{\color{black} $b$};

\filldraw[fill=black!,draw=black!,thick] (0,0) circle (1.9pt)
node[below=3pt]{\color{black} $f$};

\end{scope}

\end{scope}

\end{tikzpicture}

\caption{Expressions for classical star-star relations \eqref{CSSR1} (top) and \eqref{CSSR2} (bottom).}
\label{ssfig1}
\end{figure}

The classical star-star relations \eqref{CSSR1} and \eqref{CSSR2} are pictured in Figure \ref{ssfig1}, according to the assignment of edges to Lagrangian functions of Figure \ref{BWfig}.  The classical star-star relations \eqref{CSSR1} and \eqref{CSSR2} can respectively be shown to hold through a sequence of four transformations with the appropriate star-triangle relations from \eqref{CSTR1}, \eqref{CASTR1}, and \eqref{CASTR2}.

The equations \eqref{CSSR1} and \eqref{CSSR2} are the most complicated forms of star-star relation which involve each of the different Lagrangian functions in Figure \ref{YBErapidityfig1}.  Similarly to the star-triangle relations, there are also star-star relations that are given in terms of just two Lagrangian functions $\lag_\alpha$ and $\ol_\alpha$.  Recall that the classical star-triangle relation \eqref{CSTR1} can be found from the following substitutions in the classical star-triangle relations \eqref{CASTR1} and \eqref{CASTR2}: 
\begin{align}\label{ssrsub}
\lam\to \lag,\qquad \lagh \to \lag,\qquad \olam\to \ol,\qquad \olh\to \ol.
\end{align}
Then making the substitutions \eqref{ssrsub} in \eqref{CSSR1} results in the classical star-star relation of the form 
\begin{align}
\label{CSSR}
\begin{split}
\lag_{v_1-v_2}(x_a,x_b)+\lag_{u_1-u_2}(x_a,x_c)+\lag^{(1)}_{\bu\bv}\!\left(\!x_e\Big|\,\arraycolsep=0.7pt\begin{array}{ll}x_a & x_b \\  x_c &x_d\end{array}\right)\!\phantom{,}   \\
=\lag_{v_1-v_2}(x_c,x_d)+\lag_{u_1-u_2}(x_b,x_d)+\lag^{(2)}_{\bu\bv}\!\left(\!x_f\Big|\,\arraycolsep=0.7pt\begin{array}{ll}x_a & x_b \\  x_c &x_d\end{array}\right)\!,
\end{split}
\end{align}
where 
\begin{align}
\begin{split}
\text{\scalebox{1.0}{$\lag^{(1)}_{\bu\bv}\!\left(\!x_e\Big|\,\arraycolsep=0.7pt\begin{array}{ll}x_a & x_b \\  x_c &x_d\end{array}\right)
=$}}&\text{\scalebox{1.0}{$\,\lag_{u_2-v_1}(x_a,x_e)+\ol_{u_2-v_2}(x_e,x_b)+\ol_{u_1-v_1}(x_c,x_e)+\lag_{u_1-v_2}(x_e,x_d),$}} \\
\text{\scalebox{1.0}{$\lag^{(2)}_{\bu\bv}\!\left(\!x_f\Big|\,\arraycolsep=0.7pt\begin{array}{ll}x_a & x_b \\  x_c &x_d\end{array}\right)
=$}}&\text{\scalebox{1.0}{$\,\lag_{u_1-v_2}(x_c,x_d)+\ol_{u_1-v_1}(x_f,x_b)+\ol_{u_2-v_2}(x_c,x_f)+\lag_{u_2-v_1}(x_f,x_d),$}}
\end{split}
\end{align}
and the variables of \eqref{CSSR} satisfy the pair of equations for the derivatives
\begin{align}
\frac{\partial}{\partial x_e}\lag^{(1)}_{\bu\bv}\!\left(\!x_e\Big|\,\arraycolsep=0.7pt\begin{array}{ll}x_a & x_b \\  x_c &x_d\end{array}\right)=0,\qquad
\frac{\partial}{\partial x_f}\lag^{(2)}_{\bu\bv}\!\left(\!x_f\Big|\,\arraycolsep=0.7pt\begin{array}{ll}x_a & x_b \\  x_c &x_d\end{array}\right)=0.
\end{align}
Equation \eqref{CSSR} has a graphical representation similar to that shown in the top diagram of Figure \ref{ssfig1}, but involving only the two types of edges on the right of Figure \ref{BWfig}.  Similarly to \eqref{CSSR1} and \eqref{CSSR2}, the classical star-star relation \eqref{CSSR} can be shown to hold through a sequence of four transformations involving just the classical star-triangle relation \eqref{CSTR1}.  There is also another star-star relation that is obtained from the substitution \eqref{ssrsub} into \eqref{CSSR2}, which can be shown in a similar way.

\subsection{Classical interaction-round-a-face equations}\label{sec:CIRFYBEs}

\begin{figure}[htb!]
\centering
\begin{tikzpicture}[scale=0.8]

\draw[-|-] (5,3)--(3,1);\draw[-|-] (3,1)--(1,-1);
\draw[-] (5,-1)--(3,1);\draw[-] (3,1)--(1,3);
\fill (1.8,0) circle (0.1pt)
node[left=0.5pt]{\color{black}\small $u_1-v_1$};
\fill (4.2,0) circle (0.1pt)
node[right=0.5pt]{\color{black}\small $u_1-v_2$};
\fill (4.2,2) circle (0.1pt)
node[right=0.5pt]{\color{black}\small $u_2-v_2$};
\fill (1.8,2) circle (0.1pt)
node[left=0.5pt]{\color{black}\small $u_2-v_1$};
\fill (3,1) circle (3.5pt)
node[left=4.5pt]{\color{black} $e$};
\fill (1,-1) circle (3.5pt)
node[left=3.5pt]{\color{black} $c$};
\filldraw[fill=black,draw=black] (1,3) circle (3.5pt)
node[left=3.5pt]{\color{black} $a$};
\fill (5,3) circle (3.5pt)
node[right=3.5pt]{\color{black} $b$};
\filldraw[fill=black,draw=black] (5,-1) circle (3.5pt)
node[right=3.5pt]{\color{black} $d$};

\begin{scope}[xshift=250pt]

\draw[-,double,thick] (5,3)--(3,1);\draw[-,double,thick] (1,-1)--(3,1);
\draw[-,double,thick] (5,-1)--(3,1);\draw[-,double,thick] (1,3)--(3,1);
\draw[->>>>-,white] (5,3)--(3,1);\draw[->>>>-,white] (1,-1)--(3,1);
\draw[->>>-,white] (5,-1)--(3,1);\draw[->>>-,white] (1,3)--(3,1);
\fill (1.8,0) circle (0.1pt)
node[left=0.5pt]{\color{black}\small $u_1-v_1$};
\fill (4.2,0) circle (0.1pt)
node[right=0.5pt]{\color{black}\small $u_1-v_2$};
\fill (4.2,2) circle (0.1pt)
node[right=0.5pt]{\color{black}\small $u_2-v_2$};
\fill (1.8,2) circle (0.1pt)
node[left=0.5pt]{\color{black}\small $u_2-v_1$};
\filldraw[fill=black,draw=black,thick] (3,1) circle (3.5pt)
node[left=4.5pt]{\color{black}\small $e$};
\fill (1,-1) circle (3.5pt)
node[left=3.5pt]{\color{black} $c$};
\filldraw[fill=black,draw=black] (1,3) circle (3.5pt)
node[left=3.5pt]{\color{black} $a$};
\fill (5,3) circle (3.5pt)
node[right=3.5pt]{\color{black} $b$};
\filldraw[fill=black,draw=black] (5,-1) circle (3.5pt)
node[right=3.5pt]{\color{black} $d$};


\end{scope}

\end{tikzpicture}
\caption{The IRF Lagrangian functions \eqref{IRFlag1} or \eqref{IRFlag2} (left), and \eqref{IRFlam1} (right).  The IRF Lagrangian function \eqref{IRFlam2} would correspond to the diagram on the right with each directed edge having the reverse orientation.}
\label{fig-IRFBW}
\end{figure}
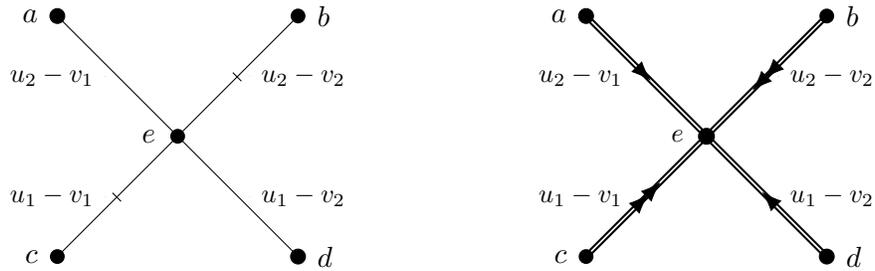

The star-star relations imply an interaction-round-a-face (IRF) form of the Yang-Baxter equation \cite{Bazhanov:1992jqa,Baxter:1997tn,Bazhanov:2011mz}.  These equations will involve the three types of two-component parameters
\begin{align}\label{32comps}
\bu=(u_1,u_2),\qquad \bv=(v_1,v_2),\qquad \bw=(w_1,w_2).
\end{align}
Let $\lag_{\bu\bv}$,  $\lagh_{\bu\bv}$, $\lam_{\bu\bv}$, and $\lamh_{\bu\bv}$, respectively denote the following sums of Lagrangian functions from the third type of the classical star-triangle relations \eqref{CASTR1} and \eqref{CASTR2}:
\begin{align}\label{IRFlag1}
 \text{\scalebox{1.0}{$\lag_{\bu\bv}\!\left(\!x_e\Big|\,\arraycolsep=0.7pt\begin{array}{ll}x_a & x_b \\  x_c & x_d\end{array}\right)\!=$}}
& \text{\scalebox{1.0}{$\,\lag_{u_2-v_1}(x_a,x_e)+\ol_{u_2-v_2}(x_b,x_e)+\ol_{u_1-v_1}(x_c,x_e)+\lag_{u_1-v_2}(x_d,x_e),$}}
\\[-0.0cm]
\label{IRFlag2}
 \text{\scalebox{1.0}{$\lagh_{\bu\bv}\!\left(\!x_e\Big|\,\arraycolsep=0.7pt\begin{array}{ll}x_a & x_b \\  x_c & x_d\end{array}\right)\!=$}}
& \text{\scalebox{1.0}{$\,\lagh_{u_2-v_1}(x_e,x_a)+\olh_{u_2-v_2}(x_e,x_b)+\olh_{u_1-v_1}(x_e,x_c)+\lagh_{u_1-v_2}(x_e,x_d),$}}
\\[-0.0cm]
\label{IRFlam1}
 \text{\scalebox{1.0}{$\lam_{\bu\bv}\!\left(\!x_e\Big|\,\arraycolsep=0.7pt\begin{array}{ll}x_a & x_b \\ x_c & x_d\end{array}\right)\!=$}}
 & \text{\scalebox{1.0}{$\,\lam_{u_2-v_1}(x_a,x_e)+\olam_{u_2-v_2}(x_b,x_e)+\olam_{u_1-v_1}(x_c,x_e)+\lam_{u_1-v_2}(x_d,x_e),$}}
\\[-0.0cm]
\label{IRFlam2}
 \text{\scalebox{1.0}{$\lamh_{\bu\bv}\!\left(\!x_e\Big|\,\arraycolsep=0.7pt\begin{array}{ll}x_a & x_b \\  x_c & x_d\end{array}\right)\!=$}}
 & \text{\scalebox{1.0}{$\,\lam_{u_2-v_1}(x_e,x_a)+\olam_{u_2-v_2}(x_e,x_b)+\olam_{u_1-v_1}(x_e,x_c)+\lam_{u_1-v_2}(x_e,x_d).$}}
\end{align}
In this paper, the expressions of the form \eqref{IRFlag1}--\eqref{IRFlam2} will be referred to as IRF Lagrangian functions, since they arise from the leading asymptotics of IRF Boltzmann weights \cite{Kels:2018qzx}.  In terms of the assignment of edges to Lagrangian functions of Figure \ref{BWfig}, the expressions \eqref{IRFlag1}--\eqref{IRFlam1} are pictured in Figure \ref{fig-IRFBW}.

\begin{figure}[htb!]
\centering
\begin{tikzpicture}[scale=1.25]

\begin{scope}[xshift=-170pt]



\draw[-,gray,dashed] (-2,0)--(-1,-1.73)--(1,-1.73)--(2,0)--(1,1.73)--(-1,1.73)--(-2,0)--(0,0);
\draw[-,gray,dashed] (1,1.73)--(0,0)--(1,-1.73);

\draw[-,double,thick] (2,0)--(1,0);
\draw[-,double,thick] (0,0)--(1,0);
\draw[-,double,thick] (1,1.73)--(1,0);
\draw[-,double,thick] (1,-1.73)--(1,0);
\draw[->>>-,white] (2,0)--(1,0);
\draw[->>>-,white] (0,0)--(1,0);
\draw[->>>>-,white] (1,1.73)--(1,0);
\draw[->>>>-,white] (1,-1.73)--(1,0);

\draw[-,double,thick] (1,1.73)--(-0.5,0.87);
\draw[-,double,thick] (-2,0)--(-0.5,0.87);
\draw[-,double,thick] (-1,1.73)--(-0.5,0.87);
\draw[-,double,thick] (0,0)--(-0.5,0.87);
\draw[->>>-,white] (1,1.73)--(-0.5,0.87);
\draw[->>>-,white] (-2,0)--(-0.5,0.87);
\draw[->>>>-,white] (-1,1.73)--(-0.5,0.87);
\draw[->>>>-,white] (0,0)--(-0.5,0.87);

\draw[-] (1,-1.73)--(-0.5,-0.87);
\draw[-] (-2,0)--(-0.5,-0.87);
\draw[-|-] (-1,-1.73)--(-0.5,-0.87);
\draw[-|-] (0,0)--(-0.5,-0.87);

\draw[-] (-2,-0.0)--(0,-0.0);
\draw[-] (-1,-1.73)--(1,-1.73);


\filldraw[fill=black,draw=black] (-1,-1.73) circle (2.0pt)
node[below=3pt]{ $a$};
\filldraw[fill=black,draw=black] (1,-1.73) circle (2.0pt)
node[below=3pt]{ $b$};
\filldraw[fill=black,draw=black] (2,0) circle (2.0pt)
node[right=3pt]{ $c$};
\filldraw[fill=black,draw=black] (1,1.73) circle (2.0pt)
node[above=3pt]{ $d$};
\filldraw[fill=black,draw=black] (-1,1.73) circle (2.0pt)
node[above=3pt]{ $e$};
\filldraw[fill=black,draw=black] (-2,0) circle (2.0pt)
node[left=3pt]{ $f$};
 \filldraw[fill=black,draw=black] (0,0) circle (2.0pt)
 node[below=8pt]{\small $h$};
\filldraw[fill=black,draw=black,thick] (-0.5,0.87) circle (2.0pt);
\filldraw[fill=white,draw=white,thick] (-0.4,0.87) circle (0.01pt)
 node[above=8pt]{\small $k$};
\filldraw[fill=black,draw=black] (-0.5,-0.87) circle (2.0pt)
 node[below=8pt]{\small $i$};
\filldraw[fill=black,draw=black,thick] (1,0) circle (2.0pt);
\filldraw[fill=white,draw=white,thick] (1.2,0) circle (0.01pt)
 node[below=4pt]{\small $j$};

\end{scope}

\draw[-,gray,dashed] (2,0)--(1,1.73)--(-1,1.73)--(-2,0)--(-1,-1.73)--(1,-1.73)--(2,0)--(0,0);
\draw[-,gray,dashed] (-1,1.73)--(0,0)--(-1,-1.73);

\draw[-,double,thick] (-2,0)--(-1,0);
\draw[-,double,thick] (0,0)--(-1,0);
\draw[-,double,thick] (-1,-1.73)--(-1,0);
\draw[-,double,thick] (-1,1.73)--(-1,0);
\draw[->>>-,white] (-2,0)--(-1,0);
\draw[->>>-,white] (0,0)--(-1,0);
\draw[->>>>-,white] (-1,-1.73)--(-1,0);
\draw[->>>>-,white] (-1,1.73)--(-1,0);

\draw[-,double,thick] (-1,-1.73)--(0.5,-0.87);
\draw[-,double,thick] (2,0)--(0.5,-0.87);
\draw[-,double,thick] (1,-1.73)--(0.5,-0.87);
\draw[-,double,thick] (0,0)--(0.5,-0.87);
\draw[->>>-,,white] (-1,-1.73)--(0.5,-0.87);
\draw[->>>-,white] (2,0)--(0.5,-0.87);
\draw[->>>>-,white] (1,-1.73)--(0.5,-0.87);
\draw[->>>>-,white] (0,0)--(0.5,-0.87);

\draw[-] (-1,1.73)--(0.5,0.87);
\draw[-] (2,0)--(0.5,0.87);
\draw[-|-] (1,1.73)--(0.5,0.87);
\draw[-|-] (0,0)--(0.5,0.87);

\draw[-] (0,0)--(2,0.0);
\draw[-] (-1,1.73)--(1,1.73);


\filldraw[fill=black,draw=black] (-1,-1.73) circle (2.0pt)
node[below=3pt]{ $a$};
\filldraw[fill=black,draw=black] (1,-1.73) circle (2.0pt)
node[below=3pt]{ $b$};
\filldraw[fill=black,draw=black] (2,0) circle (2.0pt)
node[right=3pt]{ $c$};
\filldraw[fill=black,draw=black] (1,1.73) circle (2.0pt)
node[above=3pt]{ $d$};
\filldraw[fill=black,draw=black] (-1,1.73) circle (2.0pt)
node[above=3pt]{ $e$};
\filldraw[fill=black,draw=black] (-2,0) circle (2.0pt)
node[left=3pt]{ $f$};
 \filldraw[fill=black,draw=black] (0,0) circle (2.0pt)
 node[above=8pt]{\small $h'$};
\filldraw[fill=black,draw=black] (0.5,0.87) circle (2.0pt)
 node[above=8pt]{\small $i'$};
\filldraw[fill=black,draw=black,thick] (0.5,-0.87) circle (2.0pt);
\filldraw[fill=white,draw=black,thick] (0.3,-0.87) circle (0.01pt)
 node[below=8pt]{\small $k'$};
\filldraw[fill=black,draw=black,thick] (-1,0) circle (2.0pt);
\filldraw[fill=white,draw=white,thick] (-1.2,0) circle (0.01pt)
 node[below=4pt]{\small $j'$};

\draw[white!] (-2.7,0) circle (0.01pt)
node[left=3pt]{\color{black} $=$};

\end{tikzpicture}
\caption{The classical Yang-Baxter equation \eqref{CAYBE1} in terms of IRF Lagrangian functions of Figure \ref{fig-IRFBW}.}
\label{SSRYBE}
\end{figure}
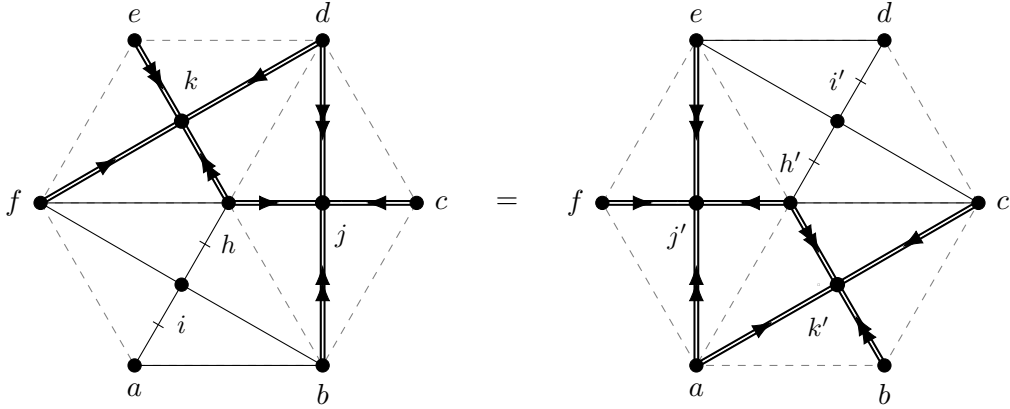

Define also
\begin{align}\label{IRFolag1}
\begin{split}
\ol_{\bu\bv}\!\left(\!x_e\Big|\,\arraycolsep=0.7pt\begin{array}{ll}x_a & x_b \\  x_c & x_d\end{array}\right)=&\, \lag_{v_1-v_2}(x_a,x_b)+\lag_{v_2-v_1}(x_c,x_d) +\lag_{\bu\bv}\!\left(\!x_e\Big|\,\arraycolsep=0.7pt\begin{array}{ll}x_a & x_b \\  x_c & x_d\end{array}\right)\!,
\end{split}
\end{align}
and 
\begin{align}
\label{IRFolag2}
\begin{split}
\olh_{\bu\bv}\!\left(\!x_e\Big|\,\arraycolsep=0.7pt\begin{array}{ll}x_a & x_b \\  x_c & x_d\end{array}\right)=&\, \lagh_{v_1-v_2}(x_a,x_b)+\lagh_{v_2-v_1}(x_c,x_d) 
+\lagh_{\bu\bv}\!\left(\!x_e\Big|\,\arraycolsep=0.7pt\begin{array}{ll}x_a & x_b \\  x_c & x_d\end{array}\right)\!.
\end{split}
\end{align}
Then the first expression for the classical IRF YBE is given by  (see also Figure \ref{SSRYBE})
\begin{align}
\label{CAYBE1}
\begin{split} 
\ol_{\bu\bv}\!\left(\!x_i\Big|\,\arraycolsep=0.7pt\begin{array}{ll}x_f & x_h \\ x_a & x_b\end{array}\right) +
\lam_{\bu\bw}\!\left(\!x_j\Big|\,\arraycolsep=0.7pt\begin{array}{ll}x_h & x_d \\ x_b & x_c\end{array}\right) +
\lam_{\bv\bw}\!\left(\!x_k\Big|\,\arraycolsep=0.7pt\begin{array}{ll}x_f & x_e \\ x_h & x_d\end{array}\right)\!\phantom{,} \\ = 
\lam_{\bv\bw}\!\left(\!x'_k\Big|\,\arraycolsep=0.7pt\begin{array}{ll}x_a & x'_h \\ x_b & x_c\end{array}\right) +
\lam_{\bu\bw}\!\left(\!x'_j\Big|\,\arraycolsep=0.7pt\begin{array}{ll}x_f & x_e \\ x_a & x'_h\end{array}\right) +
\ol_{\bu\bv}\!\left(\!x'_i\Big|\,\arraycolsep=0.7pt\begin{array}{ll}x_e & x_d \\ x'_h & x_c\end{array}\right)\!,
 \end{split}
\end{align}
and the second expression for the classical IRF YBE is given by
\begin{align}
\label{CAYBE2}
\begin{split} 
\olh_{\bu\bv}\!\left(\!x_i\Big|\,\arraycolsep=0.7pt\begin{array}{ll}x_f & x_h \\ x_a & x_b\end{array}\right) +
\lamh_{\bu\bw}\!\left(\!x_j\Big|\,\arraycolsep=0.7pt\begin{array}{ll}x_h & x_d \\ x_b & x_c\end{array}\right) +
\lamh_{\bv\bw}\!\left(\!x_k\Big|\,\arraycolsep=0.7pt\begin{array}{ll}x_f & x_e \\ x_h & x_d\end{array}\right)\!\phantom{,} \\ =  
\lamh_{\bv\bw}\!\left(\!x'_k\Big|\,\arraycolsep=0.7pt\begin{array}{ll}x_a & x'_h \\ x_b & x_c\end{array}\right) +
\lamh_{\bu\bw}\!\left(\!x'_j\Big|\,\arraycolsep=0.7pt\begin{array}{ll}x_f & x_e \\ x_a & x'_h\end{array}\right) +
\olh_{\bu\bv}\!\left(\!x'_i\Big|\,\arraycolsep=0.7pt\begin{array}{ll}x_e & x_d \\ x'_h & x_c\end{array}\right)\!.
\end{split}
\end{align}

For \eqref{CAYBE1}, the variables on the left and right hand sides satisfy the following eight equations for the derivatives
\begin{align}\label{CIRFeqmos}
\begin{gathered}
\frac{\partial}{\partial x_i}\lag_{\bu\bv}\!\left(\!x_i\Big|\,\arraycolsep=0.7pt\begin{array}{ll}x_f & x_h \\ x_a & x_b\end{array}\right)=0, \quad
\frac{\partial}{\partial x_j}\lam_{\bu\bw}\!\left(\!x_j\Big|\,\arraycolsep=0.7pt\begin{array}{ll}x_h & x_d \\ x_b & x_c\end{array}\right)=0, \quad 
\frac{\partial}{\partial x_k}\lam_{\bv\bw}\!\left(\!x_k\Big|\,\arraycolsep=0.7pt\begin{array}{ll}x_f & x_e \\ x_h & x_d\end{array}\right)=0,
\\
\frac{\partial}{\partial x'_i}\lag_{\bu\bv}\!\left(\!x'_i\Big|\,\arraycolsep=0.7pt\begin{array}{ll}x_e & x_d \\ x'_h & x_c\end{array}\right)=0, \quad
\frac{\partial}{\partial x'_j}\lam_{\bu\bw}\!\left(\!x'_j\Big|\,\arraycolsep=0.7pt\begin{array}{ll}x_f & x_e \\ x_a & x'_h\end{array}\right)=0, \quad
\frac{\partial}{\partial x'_k}\lam_{\bv\bw}\!\left(\!x'_k\Big|\,\arraycolsep=0.7pt\begin{array}{ll}x_a & x'_h \\ x_b & x_c\end{array}\right)=0,
\\
\frac{\partial}{\partial x_h}\lam^{(1)}_{\theta_1\theta_2}\!\left(\!x_h\Big|\,\arraycolsep=0.7pt\begin{array}{ll}x_k & x_j \\ x_f & x_i\end{array}\right)=0, \qquad
\frac{\partial}{\partial x'_h}\lam^{(1)}_{\phi_1\phi_2}\!\left(\!x'_h\Big|\,\arraycolsep=0.7pt\begin{array}{ll}x'_k & x'_j \\ x_c & x'_i\end{array}\right)=0,
\end{gathered}
\end{align}
where in the last two equations $\theta_1=(u_2,v_1)$, $\theta_2=(v_2,w_1)$, and $\phi_1=(v_2,u_1)$, $\phi_2=(w_2,v_1)$.  The eight equations \eqref{CIRFeqmos} essentially mean that the derivatives of the classical IRF YBE \eqref{CAYBE1} with respect to the eight variables
 \begin{align}\label{interiorvars}
 x_h,x_i,x_j,x_k,\qquad x'_h,x'_i,x'_j,x'_k,
 \end{align}
at interior vertices should vanish. There is a similar set of eight equations to be satisfied for \eqref{CAYBE2}.  

  Similarly to the classical star-triangle and star-star relations from which they are derived, for $x_I,u_1,u_2,v_1,v_2,w_1,w_2\in\mathbb{C}$ ($I\in\{a,b,c,d,e,f,h,h',i,i',j,j',k,k'\})$ and away from poles, the classical IRF YBE's \eqref{CAYBE1} and \eqref{CAYBE2} should be understood as being satisfied up to terms of the form
\begin{align}
\label{addfacs2}
2\pi\ii\sum_{I\in\{a,b,c,d,e,f,h,h',i,i',j,j',k,k'\}}k_Ix_I+C(\bu,\bv,\bw),
\end{align}
for some $k_I\in\mathbb{Z}$ and $C$ constant with respect to the $x_I$.

\begin{figure}[htb!]
\centering
\begin{tikzpicture}[scale=0.95]

\begin{scope}[xshift=-180pt]

\draw[-,gray,dashed] (-2,0)--(-1,-1.73)--(1,-1.73)--(2,0)--(1,1.73)--(-1,1.73)--(-2,0)--(0,0);
\draw[-,gray,dashed] (1,1.73)--(0,0)--(1,-1.73);

\draw[-,double,thick] (2,0)--(1,0);
\draw[-,double,thick] (0,0)--(1,0);
\draw[-,double,thick] (1,1.73)--(1,0);
\draw[-,double,thick] (1,-1.73)--(1,0);
\draw[->>>-,white] (2,0)--(1,0);
\draw[->>>-,white] (0,0)--(1,0);
\draw[->>>>-,white] (1,1.73)--(1,0);
\draw[->>>>-,white] (1,-1.73)--(1,0);

\draw[-,double,thick] (1,1.73)--(-0.5,0.87);
\draw[-,double,thick] (-2,0)--(-0.5,0.87);
\draw[-,double,thick] (-1,1.73)--(-0.5,0.87);
\draw[-,double,thick] (0,0)--(-0.5,0.87);
\draw[->>>-,white] (1,1.73)--(-0.5,0.87);
\draw[->>>-,white] (-2,0)--(-0.5,0.87);
\draw[->>>>-,white] (-1,1.73)--(-0.5,0.87);
\draw[->>>>-,white] (0,0)--(-0.5,0.87);

\draw[-] (1,-1.73)--(-0.5,-0.87);
\draw[-] (-2,0)--(-0.5,-0.87);
\draw[-|-] (-1,-1.73)--(-0.5,-0.87);
\draw[-|-] (0,0)--(-0.5,-0.87);

\draw[-] (-2,-0.0)--(0,-0.0);
\draw[-] (-1,-1.73)--(1,-1.73);

\filldraw[fill=black,draw=black] (-1,-1.73) circle (2.5pt)
node[below=3pt]{ $a$};
\filldraw[fill=black,draw=black] (1,-1.73) circle (2.5pt)
node[below=3pt]{ $b$};
\filldraw[fill=black,draw=black] (2,0) circle (2.5pt)
node[right=3pt]{ $c$};
\filldraw[fill=black,draw=black] (1,1.73) circle (2.5pt)
node[above=3pt]{ $d$};
\filldraw[fill=black,draw=black] (-1,1.73) circle (2.5pt)
node[above=3pt]{ $e$};
\filldraw[fill=black,draw=black] (-2,0) circle (2.5pt)
node[left=3pt]{ $f$};
 \filldraw[fill=black,draw=black] (0,0) circle (2.5pt);
\filldraw[fill=black,draw=black] (-0.5,0.87) circle (2.5pt);
\filldraw[fill=black,draw=black] (-0.5,-0.87) circle (2.5pt);
\filldraw[fill=black,draw=black] (1,0) circle (2.5pt);

\end{scope}

\draw[-,gray,dashed] (2,0)--(1,1.73)--(-1,1.73)--(-2,0)--(-1,-1.73)--(1,-1.73)--(2,0)--(0,0);
\draw[-,gray,dashed] (-1,1.73)--(0,0)--(-1,-1.73);

\draw[-,double,thick] (-2,0)--(-1,0);
\draw[-,double,thick] (0,0)--(-1,0);
\draw[-,double,thick] (-1,-1.73)--(-1,0);
\draw[-,double,thick] (-1,1.73)--(-1,0);
\draw[->>>-,white] (-2,0)--(-1,0);
\draw[->>>-,white] (0,0)--(-1,0);
\draw[->>>>-,white] (-1,-1.73)--(-1,0);
\draw[->>>>-,white] (-1,1.73)--(-1,0);

\draw[-,double,thick] (-1,-1.73)--(0.5,-0.87);
\draw[-,double,thick] (2,0)--(0.5,-0.87);
\draw[-,double,thick] (1,-1.73)--(0.5,-0.87);
\draw[-,double,thick] (0,0)--(0.5,-0.87);
\draw[->>>-,,white] (-1,-1.73)--(0.5,-0.87);
\draw[->>>-,white] (2,0)--(0.5,-0.87);
\draw[->>>>-,white] (1,-1.73)--(0.5,-0.87);
\draw[->>>>-,white] (0,0)--(0.5,-0.87);

\draw[-] (-1,1.73)--(0.5,0.87);
\draw[-] (2,0)--(0.5,0.87);
\draw[-|-] (1,1.73)--(0.5,0.87);
\draw[-|-] (0,0)--(0.5,0.87);

\draw[-] (0,0)--(2,0.0);
\draw[-] (-1,1.73)--(1,1.73);

\filldraw[fill=black,draw=black] (-1,-1.73) circle (2.5pt)
node[below=3pt]{ $a$};
\filldraw[fill=black,draw=black] (1,-1.73) circle (2.5pt)
node[below=3pt]{ $b$};
\filldraw[fill=black,draw=black] (2,0) circle (2.5pt)
node[right=3pt]{ $c$};
\filldraw[fill=black,draw=black] (1,1.73) circle (2.5pt)
node[above=3pt]{ $d$};
\filldraw[fill=black,draw=black] (-1,1.73) circle (2.5pt)
node[above=3pt]{ $e$};
\filldraw[fill=black,draw=black] (-2,0) circle (2.5pt)
node[left=3pt]{ $f$};
 \filldraw[fill=black,draw=black] (0,0) circle (2.5pt);
\filldraw[fill=black,draw=black] (0.5,0.87) circle (2.5pt);
\filldraw[fill=black,draw=black] (0.5,-0.87) circle (2.5pt);
\filldraw[fill=black,draw=black] (-1,0) circle (2.5pt);

\draw[<->,thick] (-8.5,-2)--(-8.5,-3);
\draw[<->,thick] (2.0,-2)--(2.0,-3);

\begin{scope}[xshift=-13,yshift=-50,scale=0.8]
\draw[<->,thick] (-7.5,-7)--(-6.5,-8);
\end{scope}
\begin{scope}[xshift=-30,yshift=-50,scale=0.8]
\draw[<->,thick] (1.3,-7)--(0.3,-8);
\end{scope}

\draw[white!] (-2.9,0) circle (0.01pt)
node[left=3pt]{\color{black} $=$};

\begin{scope}[yshift=-140pt]

\draw[-,gray,dashed] (2,0)--(1,1.73)--(-1,1.73)--(-2,0)--(-1,-1.73)--(1,-1.73)--(2,0)--(0,0);
\draw[-,gray,dashed] (-1,1.73)--(0,0)--(-1,-1.73);


\draw[-,double,thick] (-2,0)--(-1,0);
\draw[-,double,thick] (-1,-1.73)--(-1,0);
\draw[-,double,thick] (-1,1.73)--(-1,0);
\draw[->>>-,white] (-2,0)--(-1,0);
\draw[->>>>-,white] (-1,-1.73)--(-1,0);
\draw[->>>>-,white] (-1,1.73)--(-1,0);
\draw[-] (-1,0)--(1,0);

\draw[-,double,thick] (-1,-1.73)--(0.5,-0.87);
\draw[-,double,thick] (1,-1.73)--(0.5,-0.87);
\draw[->>>-,white] (-1,-1.73)--(0.5,-0.87);
\draw[->>>>-,white] (1,-1.73)--(0.5,-0.87);
\draw[-|-] (0.5,-0.87)--(1,-0);
\draw[-|-] (0.5,-0.87)--(1,-0);
\draw[-] (0.5,-0.87)--(-1,0);

\draw[-,double,thick] (0.5,0.87)--(-1,0);
\draw[-,double,thick] (0.5,0.87)--(1,0);
\draw[->>>-,white] (0.5,0.87)--(-1,0);
\draw[->>>>-,white] (0.5,0.87)--(1,0);
\draw[-] (-1,1.73)--(0.5,0.87);
\draw[-|-] (1,1.73)--(0.5,0.87);

\draw[-,double,thick] (2,0)--(1,0);
\draw[->>>-,white] (2,0)--(1,0);

\draw[-] (-1,1.73)--(1,1.73);

\filldraw[fill=black,draw=black] (-1,-1.73) circle (2.5pt)
node[below=3pt]{ $a$};
\filldraw[fill=black,draw=black] (1,-1.73) circle (2.5pt)
node[below=3pt]{ $b$};
\filldraw[fill=black,draw=black] (2,0) circle (2.5pt)
node[right=3pt]{ $c$};
\filldraw[fill=black,draw=black] (1,1.73) circle (2.5pt)
node[above=3pt]{ $d$};
\filldraw[fill=black,draw=black] (-1,1.73) circle (2.5pt)
node[above=3pt]{ $e$};
\filldraw[fill=black,draw=black] (-2,0) circle (2.5pt)
node[left=3pt]{ $f$};
 \filldraw[fill=black,draw=black] (1,0) circle (2.5pt);
\filldraw[fill=black,draw=black] (0.5,0.87) circle (2.5pt);
\filldraw[fill=black,draw=black] (0.5,-0.87) circle (2.5pt);
\filldraw[fill=black,draw=black] (-1,0) circle (2.5pt);


\end{scope}

\begin{scope}[xshift=-180pt,yshift=-140pt]

\draw[-,gray,dashed] (-2,0)--(-1,-1.73)--(1,-1.73)--(2,0)--(1,1.73)--(-1,1.73)--(-2,0)--(0,0);
\draw[-,gray,dashed] (1,1.73)--(0,0)--(1,-1.73);


\draw[-,double,thick] (2,0)--(1,0);
\draw[-,double,thick] (1,1.73)--(1,0);
\draw[-,double,thick] (1,-1.73)--(1,0);
\draw[->>>-,white] (2,0)--(1,0);
\draw[->>>>-,white] (1,1.73)--(1,0);
\draw[->>>>-,white] (1,-1.73)--(1,0);
\draw[-] (-1,0)--(1,0);

\draw[-,double,thick] (1,1.73)--(-0.5,0.87);
\draw[-,double,thick] (-1,1.73)--(-0.5,0.87);
\draw[->>>-,white] (1,1.73)--(-0.5,0.87);
\draw[->>>>-,white] (-1,1.73)--(-0.5,0.87);
\draw[-] (-0.5,0.87)--(1,0);
\draw[-|-] (-0.5,0.87)--(-1,0);

\draw[-,double,thick] (-0.5,-0.87)--(1,0);
\draw[-,double,thick] (-1,0)--(-0.5,-0.87);
\draw[->>>-,white] (-0.5,-0.87)--(1,0);
\draw[->>>>-,white] (-0.5,-0.87)--(-1,0);
\draw[-] (1,-1.73)--(-0.5,-0.87);
\draw[-|-] (-1,-1.73)--(-0.5,-0.87);

\draw[-,double,thick] (-2,0)--(-1,0);
\draw[->>>-,white] (-2,0)--(-1,0);

\draw[-] (-1,-1.73)--(1,-1.73);


\filldraw[fill=black,draw=black] (-1,-1.73) circle (2.5pt)
node[below=3pt]{ $a$};
\filldraw[fill=black,draw=black] (1,-1.73) circle (2.5pt)
node[below=3pt]{ $b$};
\filldraw[fill=black,draw=black] (2,0) circle (2.5pt)
node[right=3pt]{ $c$};
\filldraw[fill=black,draw=black] (1,1.73) circle (2.5pt)
node[above=3pt]{ $d$};
\filldraw[fill=black,draw=black] (-1,1.73) circle (2.5pt)
node[above=3pt]{ $e$};
\filldraw[fill=black,draw=black] (-2,0) circle (2.5pt)
node[left=3pt]{ $f$};
 \filldraw[fill=black,draw=black] (-1,0) circle (2.5pt);
\filldraw[fill=black,draw=black] (-0.5,0.87) circle (2.5pt);
\filldraw[fill=black,draw=black] (-0.5,-0.87) circle (2.5pt);
\filldraw[fill=black,draw=black] (1,0) circle (2.5pt);

\end{scope}

\begin{scope}[xshift=-90pt,yshift=-270pt]

\draw[-,gray,dashed] (2,0)--(1,1.73)--(-1,1.73)--(-2,0)--(-1,-1.73)--(1,-1.73)--(2,0);

\draw[-,double,thick] (2,0)--(1,0);
\draw[-,double,thick] (-1,1.73)--(-1,0);
\draw[-,double,thick] (1,-1.73)--(1,0);
\draw[->>>-,white] (2,0)--(1,0);
\draw[->>>>-,white] (-1,1.73)--(-1,0);
\draw[->>>>-,white] (1,-1.73)--(1,0);

\draw[-,double,thick] (-0,0.87)--(1,0);
\draw[-,double,thick] (-0,0.87)--(-1,0);
\draw[->>>>-,white] (-0,0.87)--(1,0);
\draw[->>>-,white] (-0,0.87)--(-1,0);
\draw[-|-] (1,1.73)--(-0,0.87);
\draw[-] (-1,1.73)--(-0,0.87);

\draw[-,double,thick] (-0,-0.87)--(1,0);
\draw[-,double,thick] (-0,-0.87)--(-1,0);
\draw[->>>-,white] (-0,-0.87)--(1,0);
\draw[->>>>-,white] (-0,-0.87)--(-1,0);
\draw[-] (1,-1.73)--(-0,-0.87);
\draw[-|-] (-1,-1.73)--(-0,-0.87);

\draw[-,double,thick] (-2,0)--(-1,0);
\draw[->>>-,white] (-2,0)--(-1,0);

\draw[-] (-1,1.73)--(1,1.73);

\draw[-] (-1,-1.73)--(1,-1.73);

\filldraw[fill=black,draw=black] (-1,-1.73) circle (2.5pt)
node[below=3pt]{ $a$};
\filldraw[fill=black,draw=black] (1,-1.73) circle (2.5pt)
node[below=3pt]{ $b$};
\filldraw[fill=black,draw=black] (2,0) circle (2.5pt)
node[right=3pt]{ $c$};
\filldraw[fill=black,draw=black] (1,1.73) circle (2.5pt)
node[above=3pt]{ $d$};
\filldraw[fill=black,draw=black] (-1,1.73) circle (2.5pt)
node[above=3pt]{ $e$};
\filldraw[fill=black,draw=black] (-2,0) circle (2.5pt)
node[left=3pt]{ $f$};
 \filldraw[fill=black,draw=black] (1,0) circle (2.5pt);
\filldraw[fill=black,draw=black] (-0.0,0.87) circle (2.5pt);
\filldraw[fill=black,draw=black] (0.0,-0.87) circle (2.5pt);
\filldraw[fill=black,draw=black] (-1,0) circle (2.5pt);

\end{scope}

\end{tikzpicture}
\caption{Deformations of the classical Yang-Baxter equation of Figure \ref{SSRYBE} with the use of the classical star-star relations of Figure \ref{ssfig1}.}
\label{SSRYBEdef}
\end{figure}
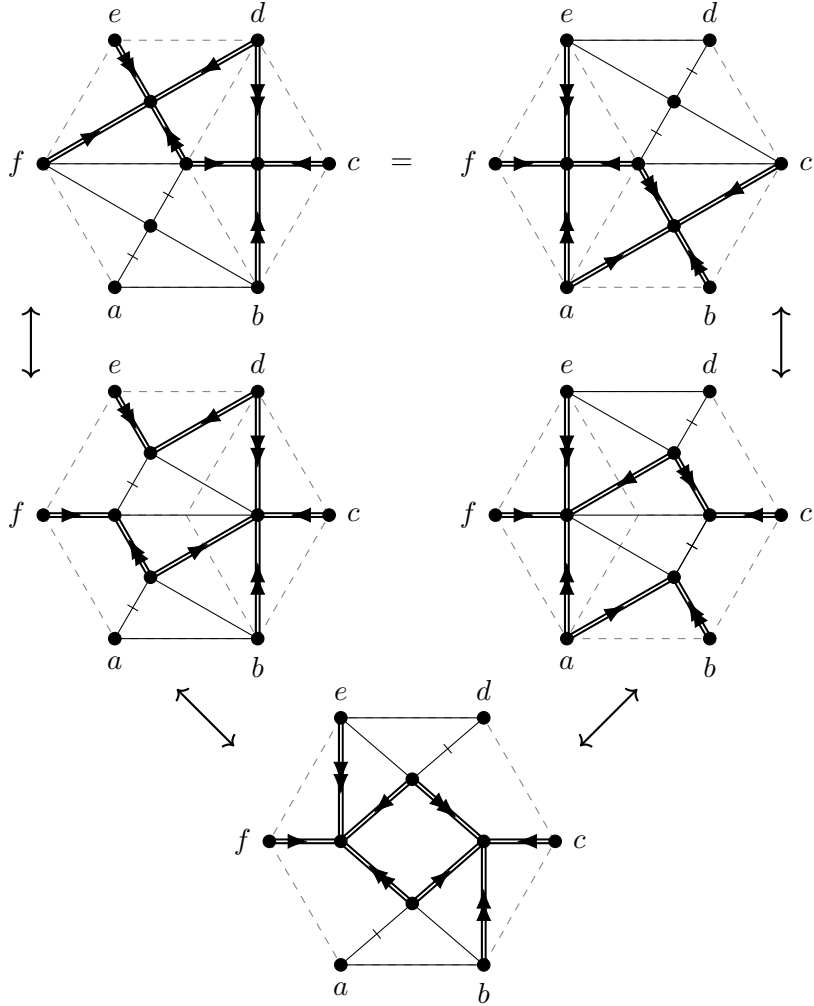

The first expression  for the classical IRF YBE \eqref{CAYBE1} has the graphical representation shown in Figure \ref{SSRYBE},  
and may be derived with the use of the star-star relations \eqref{CSSR1} and \eqref{CSSR2} through the sequence of deformations pictured in Figure \ref{SSRYBEdef}.  The second expression for the classical IRF YBE \eqref{CAYBE2} has the same graphical representation shown in Figures \ref{SSRYBE} and \ref{SSRYBEdef}, but with all directed edges having the reverse orientation. 

The IRF Lagrangian functions \eqref{IRFlag1} and \eqref{IRFolag1} also satisfy the classical IRF YBE (up to terms of the form \eqref{addfacs2})
\begin{align}
\label{CYBE1}
\begin{split} 
\ol_{\bu\bv}\!\left(\!x_i\Big|\,\arraycolsep=0.7pt\begin{array}{ll}x_f & x_h \\ x_a & x_b\end{array}\right) +
\lag_{\bu\bw}\!\left(\!x_j\Big|\,\arraycolsep=0.7pt\begin{array}{ll}x_h & x_d \\ x_b & x_c\end{array}\right) +
\lag_{\bv\bw}\!\left(\!x_k\Big|\,\arraycolsep=0.7pt\begin{array}{ll}x_f & x_e \\ x_h & x_d\end{array}\right)\!\phantom{,} \\ = 
\lag_{\bv\bw}\!\left(\!x'_k\Big|\,\arraycolsep=0.7pt\begin{array}{ll}x_a & x'_h \\ x_b & x_c\end{array}\right) +
\lag_{\bu\bw}\!\left(\!x'_j\Big|\,\arraycolsep=0.7pt\begin{array}{ll}x_f & x_e \\ x_a & x'_h\end{array}\right) +
\ol_{\bu\bv}\!\left(\!x'_i\Big|\,\arraycolsep=0.7pt\begin{array}{ll}x_e & x_d \\ x'_h & x_c\end{array}\right)\!.
\end{split}
\end{align}
As was the case for \eqref{CAYBE1} and \eqref{CAYBE2}, the derivatives of \eqref{CYBE1} with respect to each of the eight variables \eqref{interiorvars} must vanish.  The equation \eqref{CYBE1}  follows from the classical star-star relations \eqref{CSSR1} and \eqref{CSSR2}, with the same sequence of deformations shown in Figure \ref{SSRYBEdef}.  A similar equation to \eqref{CYBE1} is also satisfied by the IRF Lagrangian functions \eqref{IRFlag2} and \eqref{IRFolag2}.

For the cases involving just two Lagrangian functions $\lag_\alpha$ and $\ol_\alpha$, equations of the form \eqref{CIRFeqmos} have previously appeared in the integrable systems literature as discrete Laplace-type (or Toda-type) equations \cite{AdlerPlanarGraphs,BobSurQuadGraphs,AdlerSurisQ4}.  Thus in this paper, any equation (not neccessarily in terms of just $\lag_\alpha$ and $\ol_\alpha$) which is obtained from the derivative with respect to a variable of a classical IRF YBE will also sometimes be referred to as a discrete Laplace-type equation.  However, the derivatives of the classical IRF YBEs \eqref{CAYBE1} and \eqref{CAYBE2} involve a mixture of up to four different Lagrangian functions $\lag_\alpha$, $\ol_\alpha$, $\lam_\alpha$, and $\olam_\alpha$, which have not previously been considered in the context of discrete integrability.  For the classical IRF YBE \eqref{CAYBE1}, the fourteen discrete Laplace-type equations obtained as derivatives with respect to its fourteen variables are given in Appendix \ref{app:FCCequations}. These equations will be utilised in the following sections in the derivation of the new multidimensional consistency property and associated difference equations.




\section{Face-centered quad equations and CAFCC}\label{sec:FCC}

A main goal of this paper is to show how the classical IRF equations of Section \ref{sec:YBE} may be reinterpreted in terms of lattice equations which are multidimensionally consistent.  For this purpose the new concept of a face-centered quad equation is introduced, which is an evolution equation that is defined on a vertex and its four nearest neighbours in the square lattice.  The multidimensional consistency for face-centered quad equations is proposed as consistency-around-a-face-centered-cube (CAFCC), which involves satisfying an overdetermined system of fourteen face-centered quad equations for eight unknown variables on the face-centered cube.  The CAFCC condition for the face-centered quad equations may be regarded as an analogue of the consistency-around-the-cube (CAC) integrability condition for the usual quad equations \cite{nijhoffwalker,BobSurQuadGraphs,ABS}.


\subsection{Face-centered quad equations}
\label{sec:Mcomp}

\begin{figure}[tbh]
\centering

\begin{tikzpicture}[scale=0.7]

\fill[white!] (1,1) circle (0.01pt)
node[left=3pt]{\color{black} $\al$};
\fill[white!] (5,1) circle (0.01pt)
node[right=3pt]{\color{black} $\al$};

\fill[white!] (3,3) circle (0.01pt)
node[above=3pt]{\color{black} $\bt$};
\fill[white!] (3,-1) circle (0.01pt)
node[below=3pt]{\color{black} $\bt$};

\draw[-,gray,very thin,dashed] (5,-1)--(5,3)--(1,3)--(1,-1)--(5,-1);
\draw[-,gray,very thin,dashed] (5,-1)--(1,3); \draw[-,gray,very thin,dashed] (5,3)--(1,-1);

\fill (3,1) circle (3.5pt)
node[left=2.5pt]{\color{black} $\ccx$};
\fill (1,-1) circle (3.5pt)
node[left=1.5pt]{\color{black} $\ccc$};
\filldraw[fill=black,draw=black] (1,3) circle (3.5pt)
node[left=1.5pt]{\color{black} $\cca$};
\fill (5,3) circle (3.5pt)
node[right=1.5pt]{\color{black} $\ccb$};
\filldraw[fill=black,draw=black] (5,-1) circle (3.5pt)
node[right=1.5pt]{\color{black} $\ccd$};

\end{tikzpicture}

\caption{Variables and parameters on the vertices and edges of a face of the face-centered cube.}  
\label{fig-face}
\end{figure}
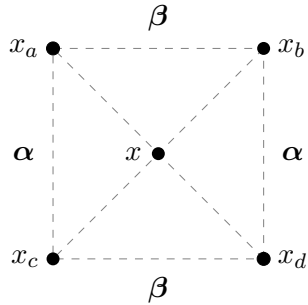

Consider a face of the face-centered cube, as is shown in Figure \ref{fig-face}, where the four variables $\cca,\ccb,\ccc,\ccd$, are associated to the four corner vertices, and the variable $\ccx$ is associated to the face vertex.\footnote{Note that in comparison to a previous work \cite{Kels:2018qzx}, the face variable $\ccx$ will be treated here as a variable on the same level as the corner variables $\cca,\ccb,\ccc,\ccd$, rather than as a parameter.} The two-component parameters  
\begin{align}
\label{pardefs}
\ccpp=(\ccpa,\ccpb),\qquad\ccqq=(\ccqa,\ccqb),
\end{align}
are associated to the edges of the face, where two opposite edges that are parallel should be assigned the same  parameters.

A face-centered quad equation associated to Figure \ref{fig-face}, will be denoted by
\begin{align}\label{afflin}
\A{\ccx}{\cca}{\ccb}{\ccc}{\ccd}{\ccpp}{\ccqq}=0,
\end{align}
where $A$ is a polynomial in the five variables $\ccx$, $\cca$, $\ccb$, $\ccc$, $\ccd$, with linear dependence on the four corner variables $\cca$, $\ccb$, $\ccc$, $\ccd$, but with no restriction on the degree of the face variable $x$.  Such an equation may be written in the following form
\begin{align}\label{quadseries}
\begin{split}
\A{\ccx}{\cca}{\ccb}{\ccc}{\ccd}{\ccpp}{\ccqq}=\sum_{i=0}^n P_i(x_a,x_b,x_c,x_d;\al,\bt)x^i,
\end{split}
\end{align}
where $n$ is the degree of $x$.  Under a parameter specialisation, the $P_i(x_a,x_b,x_c,x_d;\al,\bt)$, $i=0,\ldots,n$, are the usual expressions for affine-linear quad polynomials in the variables $x_a$, $x_b$, $x_c$, $x_d$.

The face-centered quad equations \eqref{afflin} that will be considered here, typically may also be expressed in an equivalent form (this expression does not generally follow from \eqref{quadseries})
\begin{align}
\label{4leg}
\begin{split}
\frac{
a(x;x_a;\alpha_2,\beta_1)a(x;x_d;\alpha_1,\beta_2)}{a(x;x_b;\alpha_2,\beta_2)a(x;x_c;\alpha_1,\beta_1)}=1,
\end{split}
\end{align}
 where $
     a(x;y;\ccpc,\ccqc)$
is a rational linear function of the corner variable $y$, 
and satisfies
\begin{align}\label{fcinvrel}
a(x;y;\ccpc,\ccqc)a(x;y;\ccqc,\ccpc)=1.
\end{align}
The face-centered quad equation \eqref{afflin} in the form \eqref{quadseries}, is then typically recovered by multiplying both sides of \eqref{4leg} by the denominator, bringing all terms to one side, and simplifying the resulting expression.  The expression \eqref{4leg} may be regarded as a four-leg form of a face-centered quad equation \eqref{afflin}, in analogy to the three-leg forms that are found for the regular integrable quad equations \cite{ABS}.  The expression \eqref{4leg} naturally arises through the connection (given in Section \ref{sec:CAFCCIRF}) to the IRF equations that were derived in the previous section.

The form of the equation \eqref{4leg} along with the symmetry \eqref{fcinvrel} implies that the following symmetries are satisfied for \eqref{afflin}
\begin{align}
\label{quadsym2}
\begin{split}
-\A{\ccx}{\cca}{\ccb}{\ccc}{\ccd}{\ccpp}{\ccqq}=&\A{\ccx}{\ccd}{\ccb}{\ccc}{\cca}{\ccqq}{\ccpp} \\
=&\A{\ccx}{\ccc}{\ccd}{\cca}{\ccb}{\ccpph}{\ccqq} \\
=&\A{\ccx}{\ccb}{\cca}{\ccd}{\ccc}{\ccpp}{\ccqqh},
\end{split}
\end{align}
where $\ccpph$ and $\ccqqh$ respectively represent $\ccpp$ and $\ccqq$ with their components exchanged, {\it i.e.}
\begin{align}
\ccpph=(\ccpb,\ccpa),\qquad\ccqqh=(\ccqb,\ccqa).
\end{align}
The equations \eqref{quadsym2} may be regarded as the analogues of the square symmetries that are satisfied by the regular integrable quad equations.

\begin{figure}[htb]
\centering
\begin{tikzpicture}[scale=0.55]

\draw[-,gray,very thin,dashed] (0,0)--(-1,-1);
\draw[-,gray,very thin,dashed] (0,0)--(-1,1);
\draw[-,gray,very thin,dashed] (0,0)--(1,1);
\draw[-,gray,very thin,dashed] (0,0)--(1,-1);

\draw[-,gray,very thin,dashed] (-2,-2)--(-3,-3);
\draw[-,gray,very thin,dashed] (-2,-2)--(-3,-1);
\draw[-,gray,very thin,dashed] (-2,-2)--(-1,-1);
\draw[-,gray,very thin,dashed] (-2,-2)--(-1,-3);

\draw[-,gray,very thin,dashed] (-2,0)--(-3,-1);
\draw[-,gray,very thin,dashed] (-2,0)--(-3,1);
\draw[-,gray,very thin,dashed] (-2,0)--(-1,1);
\draw[-,gray,very thin,dashed] (-2,0)--(-1,-1);

\draw[-,gray,very thin,dashed] (-2,2)--(-3,1);
\draw[-,gray,very thin,dashed] (-2,2)--(-3,3);
\draw[-,gray,very thin,dashed] (-2,2)--(-1,3);
\draw[-,gray,very thin,dashed] (-2,2)--(-1,1);

\draw[-,gray,very thin,dashed] (0,2)--(-1,1);
\draw[-,gray,very thin,dashed] (0,2)--(-1,3);
\draw[-,gray,very thin,dashed] (0,2)--(1,3);
\draw[-,gray,very thin,dashed] (0,2)--(1,1);

\draw[-,gray,very thin,dashed] (2,2)--(1,1);
\draw[-,gray,very thin,dashed] (2,2)--(1,3);
\draw[-,gray,very thin,dashed] (2,2)--(3,3);
\draw[-,gray,very thin,dashed] (2,2)--(3,1);

\draw[-,gray,very thin,dashed] (2,0)--(1,-1);
\draw[-,gray,very thin,dashed] (2,0)--(1,1);
\draw[-,gray,very thin,dashed] (2,0)--(3,1);
\draw[-,gray,very thin,dashed] (2,0)--(3,-1);

\draw[-,gray,very thin,dashed] (2,-2)--(1,-3);
\draw[-,gray,very thin,dashed] (2,-2)--(1,-1);
\draw[-,gray,very thin,dashed] (2,-2)--(3,-1);
\draw[-,gray,very thin,dashed] (2,-2)--(3,-3);

\draw[-,gray,very thin,dashed] (0,-2)--(-1,-3);
\draw[-,gray,very thin,dashed] (0,-2)--(-1,-1);
\draw[-,gray,very thin,dashed] (0,-2)--(1,-1);
\draw[-,gray,very thin,dashed] (0,-2)--(1,-3);

\draw[-,gray,very thin,dashed] (-4,-4)--(-3,-3)--(-4,-2)--(-3,-1)--(-4,0)--(-3,1)--(-4,2)--(-3,3)--(-4,4);
\draw[-,gray,very thin,dashed]          (-3,-3)--(-2,-4)--(-1,-3)--(0,-4)--(1,-3)--(2,-4)--(3,-3)--(4,-4);
\draw[-,gray,very thin,dashed] (4,4)--(3,3)--(4,2)--(3,1)--(4,0)--(3,-1)--(4,-2)--(3,-3);
\draw[-,gray,very thin,dashed]        (3,3)--(2,4)--(1,3)--(0,4)--(-1,3)--(-2,4)--(-3,3);


\foreach \x in {-2,0,...,4}{
\fill[black] (-2,\x) circle (3.5pt);
\fill[black] (\x,-2) circle (3.5pt);}

\foreach \x in {-1,1,3}{
\fill[black] (-1,\x) circle (3.5pt);
\fill[black] (\x,-1) circle (3.5pt);}

\foreach \x in {0,2,4}{
\filldraw[fill=white,draw=black] (0,\x) circle (3.5pt);
\filldraw[fill=white,draw=black] (\x,0) circle (3.5pt);}

\foreach \x in {1,3}{
\filldraw[fill=white,draw=black] (1,\x) circle (3.5pt);
\filldraw[fill=white,draw=black] (\x,1) circle (3.5pt);}

\foreach \x in {2,4}{
\filldraw[fill=white,draw=black] (2,\x) circle (3.5pt);
\filldraw[fill=white,draw=black] (\x,2) circle (3.5pt);}

\filldraw[fill=white,draw=black] (3,3) circle (3.5pt);
\filldraw[fill=white,draw=black] (4,4) circle (3.5pt);

\begin{scope}[xshift=400]

\draw[-,gray,very thin,dashed] (0,0)--(-1,-1);
\draw[-,gray,very thin,dashed] (0,0)--(-1,1);
\draw[-,gray,very thin,dashed] (0,0)--(1,1);
\draw[-,gray,very thin,dashed] (0,0)--(1,-1);

\draw[-,gray,very thin,dashed] (-2,-2)--(-3,-3);
\draw[-,gray,very thin,dashed] (-2,-2)--(-3,-1);
\draw[-,gray,very thin,dashed] (-2,-2)--(-1,-1);
\draw[-,gray,very thin,dashed] (-2,-2)--(-1,-3);

\draw[-,gray,very thin,dashed] (-2,0)--(-3,-1);
\draw[-,gray,very thin,dashed] (-2,0)--(-3,1);
\draw[-,gray,very thin,dashed] (-2,0)--(-1,1);
\draw[-,gray,very thin,dashed] (-2,0)--(-1,-1);

\draw[-,gray,very thin,dashed] (-2,2)--(-3,1);
\draw[-,gray,very thin,dashed] (-2,2)--(-3,3);
\draw[-,gray,very thin,dashed] (-2,2)--(-1,3);
\draw[-,gray,very thin,dashed] (-2,2)--(-1,1);

\draw[-,gray,very thin,dashed] (0,2)--(-1,1);
\draw[-,gray,very thin,dashed] (0,2)--(-1,3);
\draw[-,gray,very thin,dashed] (0,2)--(1,3);
\draw[-,gray,very thin,dashed] (0,2)--(1,1);

\draw[-,gray,very thin,dashed] (2,2)--(1,1);
\draw[-,gray,very thin,dashed] (2,2)--(1,3);
\draw[-,gray,very thin,dashed] (2,2)--(3,3);
\draw[-,gray,very thin,dashed] (2,2)--(3,1);

\draw[-,gray,very thin,dashed] (2,0)--(1,-1);
\draw[-,gray,very thin,dashed] (2,0)--(1,1);
\draw[-,gray,very thin,dashed] (2,0)--(3,1);
\draw[-,gray,very thin,dashed] (2,0)--(3,-1);

\draw[-,gray,very thin,dashed] (2,-2)--(1,-3);
\draw[-,gray,very thin,dashed] (2,-2)--(1,-1);
\draw[-,gray,very thin,dashed] (2,-2)--(3,-1);
\draw[-,gray,very thin,dashed] (2,-2)--(3,-3);

\draw[-,gray,very thin,dashed] (0,-2)--(-1,-3);
\draw[-,gray,very thin,dashed] (0,-2)--(-1,-1);
\draw[-,gray,very thin,dashed] (0,-2)--(1,-1);
\draw[-,gray,very thin,dashed] (0,-2)--(1,-3);

\draw[-,gray,very thin,dashed] (-4,2)--(-3,3)--(-4,4);\draw[-,gray,very thin,dashed] (2,-4)--(3,-3)--(4,-4);

\draw[-,gray,very thin,dashed] (-4,-4)--(-3,-3)--(-4,-2)--(-3,-1)--(-4,0)--(-3,1)--(-4,2);
\draw[-,gray,very thin,dashed]          (-3,-3)--(-2,-4)--(-1,-3)--(0,-4)--(1,-3)--(2,-4);
\draw[-,gray,very thin,dashed] (4,4)--(3,3)--(4,2)--(3,1)--(4,0)--(3,-1)--(4,-2)--(3,-3);
\draw[-,gray,very thin,dashed]        (3,3)--(2,4)--(1,3)--(0,4)--(-1,3)--(-2,4)--(-3,3);

\foreach \x in {-4,...,4}{
\fill[black] (\x,-\x) circle (3.5pt);}

\foreach \x in {-3,...,3}{
\fill[black] (\x-1,-\x-1) circle (3.5pt);}

\foreach \x in {-3,...,3}{
\filldraw[fill=white,draw=black] (-\x+1,\x+1) circle (3.5pt);}

\foreach \x in {-2,...,2}{
\filldraw[fill=white,draw=black] (-\x+2,\x+2) circle (3.5pt);}

\foreach \x in {-1,...,1}{
\filldraw[fill=white,draw=black] (-\x+3,\x+3) circle (3.5pt);}

\foreach \x in {0,...,0}{
\filldraw[fill=white,draw=black] (-\x+4,\x+4) circle (3.5pt);}

\end{scope}

\end{tikzpicture}
\caption{The corner-type (left) and staircase-type (right) initial conditions on the rotated square lattice for the face-centered quad equation \eqref{afflin} of Figure \ref{fig-face}.  Filled vertices are the initial values and the unfilled vertices are values that may be uniquely determined through evolution of the equation.}
\label{fig-initial}
\end{figure}
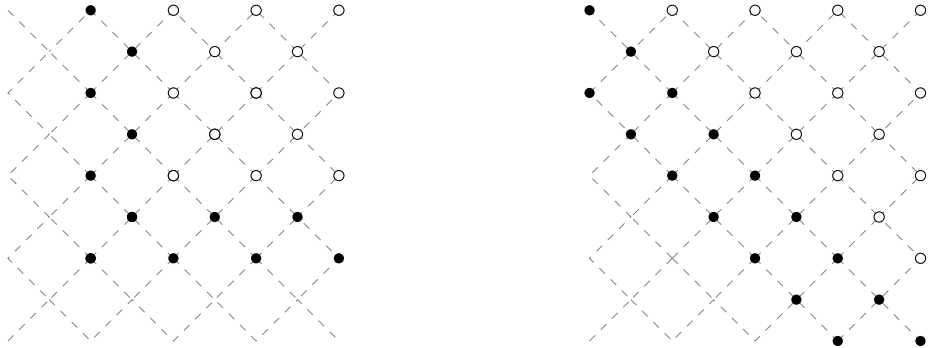

Due to the affine-linearity in the corner variables, under appropriate initial conditions the face-centered quad equations should have a unique evolution in the square lattice.  For regular quad equations which satisfy CAC, two typical initial conditions on the square lattice are the corner-type initial condition, and the staircase-type initial condition.  For the face-centered quad equations \eqref{afflin}, the analogues of these initial conditions on the rotated square lattice are shown in Figure \ref{fig-initial}.  The evolutions are unique because the variables of the square lattice are always determined by solving for the linear corner variables, and not the face variable.

Two more types of face-centered quad equations that are different from \eqref{afflin} will also be used, denoted here respectively by
\begin{align}\label{typebc}
    \B{\ccx}{\cca}{\ccb}{\ccc}{\ccd}{\ccpp}{\ccqq}=0,\qquad \C{\ccx}{\cca}{\ccb}{\ccc}{\ccd}{\ccpp}{\ccqq}=0.
\end{align}
Then the face-centered quad equation \eqref{afflin} will be denoted as type-A, while the equations in \eqref{typebc} will be denoted respectively as type-B and type-C. Similarly to the type-A equation \eqref{afflin}, the type-B and type-C equations \eqref{typebc} are polynomials in the five variables $\ccx$, $\cca$, $\ccb$, $\ccc$, $\ccd$, with linear dependence on the four corner variables $\cca$, $\ccb$, $\ccc$, $\ccd$, and hence may be written in the same form as \eqref{quadseries}.  

The equations \eqref{typebc} will also be found to have the following equivalent four-leg forms
\begin{align}
\label{4legb}
\frac{b(x;x_a,\alpha_2,\beta_1)b(x;x_d,\alpha_1,\beta_2)}{b(x;x_b,\alpha_2,\beta_2)b(x;x_c,\alpha_1,\beta_1)}=1, \\
\label{4legc}
\frac{a(x;x_a,\alpha_2,\beta_1)c(x;x_d,\alpha_1,\beta_2)}{a(x;x_b,\alpha_2,\beta_2)c(x;x_c,\alpha_1,\beta_1)}=1,
\end{align}
for type-B, and type-C respectively.  Here $a(x;y;\ccpc,\ccqc)$ is the same function from \eqref{4leg} that is associated to a type-A equation \eqref{afflin}.  The two new edge functions $b(x;y;\ccpc,\ccqc)$ and $c(x;y;\ccpc,\ccqc)$, are again rational linear functions of $y$, but do not satisfy the same reflection symmetry \eqref{fcinvrel} that was satisfied by $a(x;y;\ccpc,\ccqc)$.  Thus the expression \eqref{4legb} implies just the two symmetries for type-B equations
\begin{align}
\label{quadsymb}
-\B{\ccx}{\cca}{\ccb}{\ccc}{\ccd}{\ccpp}{\ccqq}=\B{\ccx}{\ccc}{\ccd}{\cca}{\ccb}{\ccpph}{\ccqq} =\B{\ccx}{\ccb}{\cca}{\ccd}{\ccc}{\ccpp}{\ccqqh},
\end{align}
while the expression \eqref{4legc} implies the single symmetry for type-C equations
\begin{align}
\label{quadsymc}
-\C{\ccx}{\cca}{\ccb}{\ccc}{\ccd}{\ccpp}{\ccqq}=\C{\ccx}{\ccb}{\cca}{\ccd}{\ccc}{\ccpp}{\ccqqh}.
\end{align}

\begin{figure}[tbh]
\centering
\begin{tikzpicture}[scale=0.74]

\draw[-,gray,very thin,dashed] (5,-1)--(5,3)--(1,3)--(1,-1)--(5,-1);
\draw[-] (5,3)--(1,-1);
\draw[-] (5,-1)--(1,3);
\fill (0.8,-0.0) circle (0.1pt)
node[left=0.5pt]{\color{black}\small $\ccpa$};
\fill (5.2,-0.0) circle (0.1pt)
node[right=0.5pt]{\color{black}\small $\ccpa$};
\fill (4,-1.2) circle (0.1pt)
node[below=0.5pt]{\color{black}\small $\ccqb$};
\fill (4,3.2) circle (0.1pt)
node[above=0.5pt]{\color{black}\small $\ccqb$};
\fill (0.8,2.0) circle (0.1pt)
node[left=0.5pt]{\color{black}\small $\ccpb$};
\fill (5.2,2.0) circle (0.1pt)
node[right=0.5pt]{\color{black}\small $\ccpb$};
\fill (2,-1.2) circle (0.1pt)
node[below=0.5pt]{\color{black}\small $\ccqa$};
\fill (2,3.2) circle (0.1pt)
node[above=0.5pt]{\color{black}\small $\ccqa$};
\fill (3,1) circle (4.1pt)
node[left=2.5pt]{\color{black} $\ccx$};
\fill (1,-1) circle (4.1pt)
node[left=1.5pt]{\color{black} $\ccc$};
\filldraw[fill=black,draw=black] (1,3) circle (4.1pt)
node[left=1.5pt]{\color{black} $\cca$};
\fill (5,3) circle (4.1pt)
node[right=1.5pt]{\color{black} $\ccb$};
\filldraw[fill=black,draw=black] (5,-1) circle (4.1pt)
node[right=1.5pt]{\color{black} $\ccd$};

\draw[-,dotted,thick] (2,-1.2)--(2,3.2);\draw[-,dotted,thick] (4,-1.2)--(4,3.2);
\draw[-,dotted,thick] (0.8,-0.0)--(5.2,-0.0);\draw[-,dotted,thick] (0.8,2.0)--(5.2,2.0);

\fill (3,-2) circle (0.01pt)
node[below=0.5pt]{\color{black}\small $\A{\ccx}{\cca}{\ccb}{\ccc}{\ccd}{\ccpp}{\ccqq} $};


\begin{scope}[xshift=210pt]

\draw[-,gray,very thin,dashed] (5,-1)--(5,3)--(1,3)--(1,-1)--(5,-1);
\draw[-,double,thick] (5,3)--(1,-1);
\draw[-,double,thick] (5,-1)--(1,3);
\fill (0.8,-0.0) circle (0.1pt)
node[left=0.5pt]{\color{black}\small $\ccpa$};
\fill (5.2,-0.0) circle (0.1pt)
node[right=0.5pt]{\color{black}\small $\ccpa$};
\fill (4,-1.2) circle (0.1pt)
node[below=0.5pt]{\color{black}\small $\ccqb$};
\fill (4,3.2) circle (0.1pt)
node[above=0.5pt]{\color{black}\small $\ccqb$};
\fill (0.8,2.0) circle (0.1pt)
node[left=0.5pt]{\color{black}\small $\ccpb$};
\fill (5.2,2.0) circle (0.1pt)
node[right=0.5pt]{\color{black}\small $\ccpb$};
\fill (2,-1.2) circle (0.1pt)
node[below=0.5pt]{\color{black}\small $\ccqa$};
\fill (2,3.2) circle (0.1pt)
node[above=0.5pt]{\color{black}\small $\ccqa$};
\fill (3,1) circle (4.1pt)
node[left=2.5pt]{\color{black} $\ccx$};
\fill (1,-1) circle (4.1pt)
node[left=1.5pt]{\color{black} $\ccc$};
\filldraw[fill=black,draw=black] (1,3) circle (4.1pt)
node[left=1.5pt]{\color{black} $\cca$};
\fill (5,3) circle (4.1pt)
node[right=1.5pt]{\color{black} $\ccb$};
\filldraw[fill=black,draw=black] (5,-1) circle (4.1pt)
node[right=1.5pt]{\color{black} $\ccd$};

\draw[-,dotted,thick] (2,-1.2)--(2,3.2);\draw[-,dotted,thick] (4,-1.2)--(4,3.2);
\draw[-,dotted,thick] (0.8,-0.0)--(5.2,-0.0);\draw[-,dotted,thick] (0.8,2.0)--(5.2,2.0);

\fill (3,-2) circle (0.01pt)
node[below=0.5pt]{\color{black}\small $\B{\ccx}{\cca}{\ccb}{\ccc}{\ccd}{\ccpp}{\ccqq} $};

\end{scope}

\begin{scope}[xshift=420pt]

\draw[-,gray,very thin,dashed] (5,-1)--(5,3)--(1,3)--(1,-1)--(5,-1);
\draw[-] (5,3)--(3,1)--(1,3);
\draw[-,double,thick] (1,-1)--(3,1)--(5,-1);
\fill (0.8,-0.0) circle (0.1pt)
node[left=0.5pt]{\color{black}\small $\ccpa$};
\fill (5.2,-0.0) circle (0.1pt)
node[right=0.5pt]{\color{black}\small $\ccpa$};
\fill (4,-1.2) circle (0.1pt)
node[below=0.5pt]{\color{black}\small $\ccqb$};
\fill (4,3.2) circle (0.1pt)
node[above=0.5pt]{\color{black}\small $\ccqb$};
\fill (0.8,2.0) circle (0.1pt)
node[left=0.5pt]{\color{black}\small $\ccpb$};
\fill (5.2,2.0) circle (0.1pt)
node[right=0.5pt]{\color{black}\small $\ccpb$};
\fill (2,-1.2) circle (0.1pt)
node[below=0.5pt]{\color{black}\small $\ccqa$};
\fill (2,3.2) circle (0.1pt)
node[above=0.5pt]{\color{black}\small $\ccqa$};
\fill (3,1) circle (4.1pt)
node[left=2.5pt]{\color{black} $\ccx$};
\fill (1,-1) circle (4.1pt)
node[left=1.5pt]{\color{black} $\ccc$};
\filldraw[fill=black,draw=black] (1,3) circle (4.1pt)
node[left=1.5pt]{\color{black} $\cca$};
\fill (5,3) circle (4.1pt)
node[right=1.5pt]{\color{black} $\ccb$};
\filldraw[fill=black,draw=black] (5,-1) circle (4.1pt)
node[right=1.5pt]{\color{black} $\ccd$};

\draw[-,dotted,thick] (2,-1.2)--(2,3.2);\draw[-,dotted,thick] (4,-1.2)--(4,3.2);
\draw[-,dotted,thick] (0.8,-0.0)--(5.2,-0.0);\draw[-,dotted,thick] (0.8,2.0)--(5.2,2.0);

\fill (3,-2) circle (0.01pt)
node[below=0.5pt]{\color{black}\small $\C{\ccx}{\cca}{\ccb}{\ccc}{\ccd}{\ccpp}{\ccqq} $};

\end{scope}

\end{tikzpicture}
\caption{Graphical representation of four-leg forms \eqref{4leg}, \eqref{4legb}, and \eqref{4legc}, for the face-centered quad equations of type-A, type-B, and type-C respectively.  Type-C equations share the same types of single-line solid edges (and edge functions $a(x;y;\alpha,\beta)$) associated to type-A equations, and double-line solid edges (but not the same edge functions) associated to type-B equations.}
\label{3fig4quad}
\end{figure}
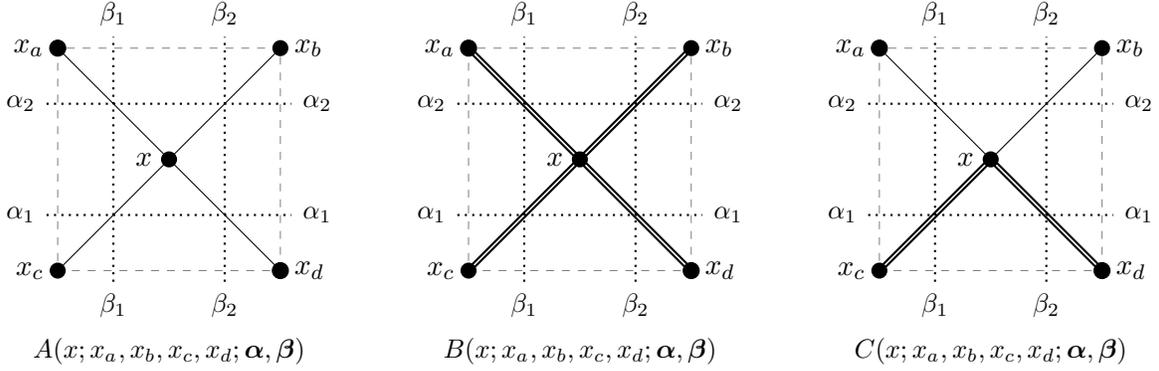

The four-leg expressions for the face-centered quad equations in \eqref{4leg}, \eqref{4legb}, and \eqref{4legc}, may respectively be associated to the faces of the face-centered cube shown in Figure \ref{3fig4quad}.  The edge function $a(x;y;\ccpc,\ccqc)$ for type-A equations is associated to single-line edges, and the edge function $b(x;y;\ccpc,\ccqc)$ for type-B equations is associated to double-line edges.  The two edge functions for type-C equations are associated to the same single- and double-line edges as type-A and type-B equations respectively.  The reason for this is that the type-C equations will be seen to effectively be defined from the combination of type-A and -B equations that meet at a corner vertex on the face-centered cube. 


\subsection{Consistency-around-a-face-centered-cube}\label{sec:CAFCC}

A multidimensional consistency property can be formulated for the face-centered quad equations introduced above, whereby under an appropriate initial condition they are required to have a consistent evolution around the vertices and edges of the face-centered cube.  This property will be referred to as consistency-around-a-face-centered-cube (CAFCC), and may be regarded as an analogue of the consistency-around-a-cube (CAC) property that is used as an integrability condition for the usual integrable quad equations.

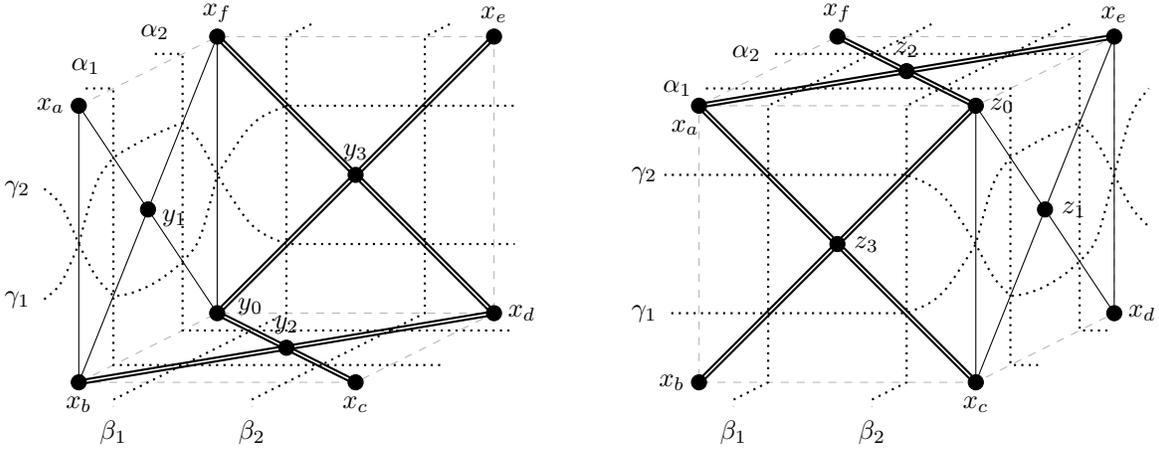
\begin{figure}[hbt!]
\centering
\begin{tikzpicture}

\begin{scope}[scale=0.92]

\draw[-,gray!60!white,very thin,dashed] (-3,3)--(-1,4)--(3,4)--(3,0);
\draw[-,gray!60!white,very thin,dashed] (-3,-1)--(-1,0)--(3,0)--(1,-1)--(-3,-1);

\draw[-,double,thick] (-3,-1)--(3,0);\draw[-,double,thick] (-1,0)--(1,-1);
\draw[-] (-3,-1)--(-1,4);\draw[-] (-3,3)--(-1,0);
\draw[-,double,thick] (-1,0)--(3,4);\draw[-,double,thick] (-1,4)--(3,0);
\draw[-] (-3.0,-1)--(-3.0,3);\draw[-] (-1.0,0)--(-1.0,4);
\draw[-,black,thick,dotted] (-1.9,3.75)--(-1.5,3.75)--(-1.5,-0.25)--(3.25,-0.25);
\draw[-,black,thick,dotted] (-2.5,-1.25)--(0,0)--(0,4)--(0.4,4.2);
\draw[-,black,thick,dotted] (-2.9,3.25)--(-2.5,3.25)--(-2.5,-0.75)--(2.25,-0.75);
\draw[-,black,thick,dotted] (-0.5,-1.25)--(2,0)--(2,4)--(2.4,4.2);
\draw[-,black,thick,dotted] (-3.5,0.2) .. controls (-3.25,0.4) ..(-3,1) .. controls (-2.75,1.95) .. (-2.5,2.25) .. controls (-2.2,2.45) and (-1.8,2.65) .. (-1.5,2.75) .. controls (-1.25,2.5) .. (-1,2) .. controls (-0.5,1.2) .. (0,1) -- (3.3,1); 
\draw[-,black,thick,dotted] (-3.5,1.8) .. controls (-3.25,1.6) .. (-3,1) .. controls (-2.75,0.5) .. (-2.5,0.25) .. controls (-2.2,0.25) and (-1.8,0.45) .. (-1.5,0.75) .. controls (-1.25,1) .. (-1,2) .. controls (-0.5,2.8) .. (0,3) -- (3.3,3); 

\filldraw[fill=black,draw=black] (1,-1) circle (3.1pt)
node[below=1.5pt]{\small $\ccc$};
\filldraw[fill=black,draw=black] (0,-0.5) circle (3.1pt)
node[above=1.5pt]{\small $\ccyb$};
\filldraw[fill=black,draw=black] (-1,0) circle (3.1pt);
\fill[black!] (-1,0.1) circle (0.01pt)
node[right=4pt]{\color{black}\small $\ccy$};
\filldraw[fill=black,draw=black] (-3,-1) circle (3.1pt)
node[below=1.5pt]{\small $\ccb$};
\filldraw[fill=black,draw=black] (3,0) circle (3.1pt)
node[right=1.5pt]{\color{black}\small $\ccd$};
\filldraw[fill=black,draw=black] (-3,3) circle (3.1pt)
node[left=1.5pt]{\color{black}\small $\cca$};
\filldraw[fill=black,draw=black] (-1,4) circle (3.1pt)
node[above=1.5pt]{\small $\ccf$};
\filldraw[fill=black,draw=black] (-2,1.5) circle (3.1pt);
\fill[black!] (-2,1.37) circle (0.01pt)
node[right=1.5pt]{\color{black}\small $\ccya$};
\filldraw[fill=black,draw=black] (1,2) circle (3.1pt)
node[above=1.5pt]{\small $\ccyc$};
\filldraw[fill=black,draw=black] (3,4) circle (3.1pt)
node[above=1.5pt]{\small $\cce$};

\draw[black] (-1.9,3.75) circle (0.01pt)
node[above=1.5pt]{\color{black}\small $\ccpb$};
\draw[black] (-2.9,3.25) circle (0.01pt)
node[above=1.5pt]{\color{black}\small $\ccpa$};
\draw[black] (-2.5,-1.35) circle (0.01pt)
node[below=1.5pt]{\color{black}\small $\ccqa$};
\draw[black] (-0.5,-1.35) circle (0.01pt)
node[below=1.5pt]{\color{black}\small $\ccqb$};
\draw[black] (-3.5,1.8) circle (0.01pt)
node[left=1.5pt]{\color{black}\small $\ccrb$};
\draw[black] (-3.5,0.2) circle (0.01pt)
node[left=1.5pt]{\color{black}\small $\ccra$};

\begin{scope}[xshift=255,yshift=0,rotate=0]

\draw[-,gray!60!white,very thin,dashed] (-3,3)--(-1,4)--(3,4)--(1,3)--(-3,3);
\draw[-,gray!60!white,very thin,dashed] (3,4)--(3,0)--(1,-1)--(-3,-1)--(-3,3);

\draw[-,double,thick] (-3,-1)--(1,3);\draw[-,double,thick] (-3,3)--(1,-1);
\draw[-] (1,-1)--(3,4);\draw[-] (1,3)--(3,0);
\draw[-,double,thick] (-3,3)--(3,4);\draw[-,double,thick] (-1,4)--(1,3);
\draw[-] (1.0,-1)--(1.0,3);\draw[-] (3.0,0)--(3.0,4);
\draw[-,black,thick,dotted] (-2.9,3.25)--(1.5,3.25)--(1.5,-0.75)--(1.9,-0.75);
\draw[-,black,thick,dotted] (-0.5,-1.25)--(0,-1)--(0,3)--(2,4)--(2.4,4.2);
\draw[-,black,thick,dotted] (-3.4,0)--(-2,0)--(0,0) .. controls (0.5,0.2) .. (1,1) .. controls (1.25,1.95) .. (1.5,2.25) .. controls (1.8,2.45) and (2.2,2.65) .. (2.5,2.75) .. controls (2.75,2.5) .. (3,2) .. controls (3.25,1.5) .. (3.5,1.25);
\draw[-,black,thick,dotted] (-1.9,3.75)--(2.5,3.75)--(2.5,-0.25)--(2.9,-0.25);
\draw[-,black,thick,dotted] (-2.5,-1.25)--(-2,-1)--(-2,3)--(0,4)--(0.4,4.2);
\draw[-,black,thick,dotted] (-3.5,2)--(-2,2)--(0,2) .. controls (0.5,1.8) .. (1,1) .. controls (1.25,0.5) .. (1.5,0.25) .. controls (1.8,0.25) and (2.2,0.45) .. (2.5,0.75) .. controls (2.75,1) .. (3,2) .. controls (3.25,2.95) .. (3.5,3.25);

\filldraw[fill=black,draw=black] (-3,3) circle (3.1pt);
\fill[black!] (-3.2,3) circle (0.01pt)
node[below=1.5pt]{\color{black}\small $\cca$};
\filldraw[fill=black,draw=black] (1,3) circle (3.1pt)
node[right=1.5pt]{\small $\ccz$};
\filldraw[fill=black,draw=black] (0,3.5) circle (3.1pt)
node[above=1.5pt]{\small $\cczb$};
\filldraw[fill=black,draw=black] (-1,4) circle (3.1pt)
node[above=1.5pt]{\small $\ccf$};
\filldraw[fill=black,draw=black] (3,4) circle (3.1pt)
node[above=1.5pt]{\small $\cce$};
\filldraw[fill=black,draw=black] (3,0) circle (3.1pt)
node[right=1.5pt]{\color{black}\small $\ccd$};
\filldraw[fill=black,draw=black] (-3,-1) circle (3.1pt)
node[left=1.5pt]{\small $\ccb$};
\filldraw[fill=black,draw=black] (-1,1) circle (3.1pt)
node[right=2.5pt]{\small $\cczc$};
\filldraw[fill=black,draw=black] (1,-1) circle (3.1pt)
node[below=1.5pt]{\small $\ccc$};
\filldraw[fill=black,draw=black] (2,1.5) circle (3.1pt)
node[right=2.5pt]{\small $\ccza$};

\draw[black] (-1.9,3.75) circle (0.01pt)
node[left=1.5pt]{\color{black}\small $\ccpb$};
\draw[black] (-2.9,3.25) circle (0.01pt)
node[left=1.5pt]{\color{black}\small $\ccpa$};
\draw[black] (-2.5,-1.35) circle (0.01pt)
node[below=1.5pt]{\color{black}\small $\ccqa$};
\draw[black] (-0.5,-1.35) circle (0.01pt)
node[below=1.5pt]{\color{black}\small $\ccqb$};
\draw[black] (-3.4,2) circle (0.01pt)
node[left=1.5pt]{\color{black}\small $\ccrb$};
\draw[black] (-3.4,0) circle (0.01pt)
node[left=1.5pt]{\color{black}\small $\ccra$};
\end{scope}

\end{scope}

\end{tikzpicture}
\caption{The six type-A and type-B equations \eqref{6face}, and eight type-C equations \eqref{8corner}, centered at the fourteen vertices of the face-centered cube.  Note that the edges on the right hand side appear with reverse orientation with respect to the type-C equations \eqref{8corner} pictured in Figure \ref{3fig4quad}. }
\label{CAFCC5}
\end{figure}

The CAFCC property has a convenient graphical representation in terms of the four-leg expressions that are pictured in Figure \ref{3fig4quad}.  The two relevant halves of the face-centered cube are pictured in Figure \ref{CAFCC5}, where the following six type-A and -B equations from Figure \ref{3fig4quad}
\begin{align}\label{6face}
\begin{array}{rrr}
\ds\A{\ccya}{\ccf}{\ccy}{\cca}{\ccb}{\ccpp}{\ccrr}=0,&\hspace{0.07cm} \B{\ccyb}{\ccy}{\ccd}{\ccb}{\ccc}{\ccpp}{\ccqq}=0,&\hspace{0.07cm} \B{\ccyc}{\ccf}{\cce}{\ccy}{\ccd}{\ccrr}{\ccqq}=0, \\[0.1cm]
\ds\A{\ccza}{\cce}{\ccd}{\ccz}{\ccc}{\ccpp}{\ccrr}=0,&\hspace{0.07cm} \B{\cczb}{\ccf}{\cce}{\cca}{\ccz}{\ccpp}{\ccqq}=0,&\hspace{0.07cm} \B{\cczc}{\cca}{\ccz}{\ccb}{\ccc}{\ccrr}{\ccqq}=0,
\end{array}
\end{align}
are centered at the six face vertices $\ccya,\ccyb,\ccyc,\ccza,\cczb,\cczc$, respectively.   The face-centered cube is drawn this way to more clearly visualise the parameter dependencies on the edges, and this also more clearly shows the connection to the IRF form of Yang-Baxter equation of Figure \ref{SSRYBE} from which CAFCC is derived.

The six equations \eqref{6face} may be regarded as the analogues of the six equations on the faces of a cube for CAC. However, they are not enough to describe an evolution around a face-centered cube because the face variables $\ccya,\ccyb,\ccyc,\ccza,\cczb,\cczc$, are not shared by any two of the equations in \eqref{6face}.  This may be addressed by also taking into account the following eight equations of type-C
\begin{align}\label{8corner}
\begin{array}{rr}
\text{\scalebox{1.0}{$\C{\ccy}{\ccya}{\ccf}{\ccyb}{\ccyc}{(\ccqa,\ccrb)}{(\ccpb,\ccra)}=0,$}}&
\text{\scalebox{1.0}{$\C{\ccz}{\ccza}{\ccc}{\cczb}{\cczc}{(\ccqb,\ccra)}{(\ccpa,\ccrb)}=0,$}} \\[0.1cm]
\text{\scalebox{1.0}{$\C{\cca}{\ccya}{\ccb}{\cczb}{\cczc}{(\ccqa,\ccra)}{(\ccpa,\ccrb)}=0,$}}&
\text{\scalebox{1.0}{$\C{\ccb}{\ccya}{\cca}{\ccyb}{\cczc}{(\ccqa,\ccrb)}{(\ccpa,\ccra)}=0,$}} \\[0.1cm]
\text{\scalebox{1.0}{$\C{\ccc}{\ccza}{\ccz}{\ccyb}{\cczc}{(\ccqb,\ccrb)}{(\ccpa,\ccra)}=0,$}}&
\text{\scalebox{1.0}{$\C{\ccd}{\ccza}{\cce}{\ccyb}{\ccyc}{(\ccqb,\ccrb)}{(\ccpb,\ccra)}=0,$}} \\[0.1cm]
\text{\scalebox{1.0}{$\C{\cce}{\ccza}{\ccd}{\cczb}{\ccyc}{(\ccqb,\ccra)}{(\ccpb,\ccrb)}=0,$}}&
\text{\scalebox{1.0}{$\C{\ccf}{\ccya}{\ccy}{\cczb}{\ccyc}{(\ccqa,\ccra)}{(\ccpb,\ccrb)}=0,$}}
\end{array}
\end{align}
which appear in Figure \ref{CAFCC5} centered at the eight corner vertices $\ccy,\ccz,\cca,\ccb,\ccc,\ccd,\cce,\ccf$, respectively (note that the edges of the type-C equations on the right hand side of Figure \ref{CAFCC5}, are pictured with a reversed orientation with respect to the edges of the same type-C equation on the left hand side of Figure \ref{CAFCC5}). The type-C equations in \eqref{8corner} always share one edge with a type-A equation and two edges with type-B equations from \eqref{6face}.  The remaining edge for the type-C equation is one of the four vertical edges which connect the four pairs of corner vertices $(x_a,x_b)$, $(x_f,y_0)$, $(x_c,x_d)$, $(z_0,x_c)$, respectively, and these edges are only associated to the type-C equations.  These vertical edges have the effect of exchanging the components of the parameter $\ccrr$ for type-A and type-B equations on the neighbouring faces.  

\subsubsection{CAFCC algorithm}

CAFCC for the fourteen equations \eqref{6face} and \eqref{8corner} on the two halves of the face-centered cube of Figure \ref{CAFCC5}, can be formulated as follows.

In Figure \ref{CAFCC5}, there are six components of the parameters $\ccpp$, $\ccqq$, $\gm$, 
\begin{align}
\al=(\ccpa,\ccpb),\qquad\bt=(\ccqa,\ccqb),\qquad\gm=(\ccra,\ccrb),
\end{align}
which take some fixed values.  Also the following six variables
\begin{align}
\ccy,\ccya,\ccyb,\ccyc,\ccb,\ccd,
\end{align}
are chosen to take some fixed initial values.  There remain a total of eight undetermined variables associated to the vertices of the face-centered cube, and the fourteen equations \eqref{6face} and \eqref{8corner} to be satisfied.  

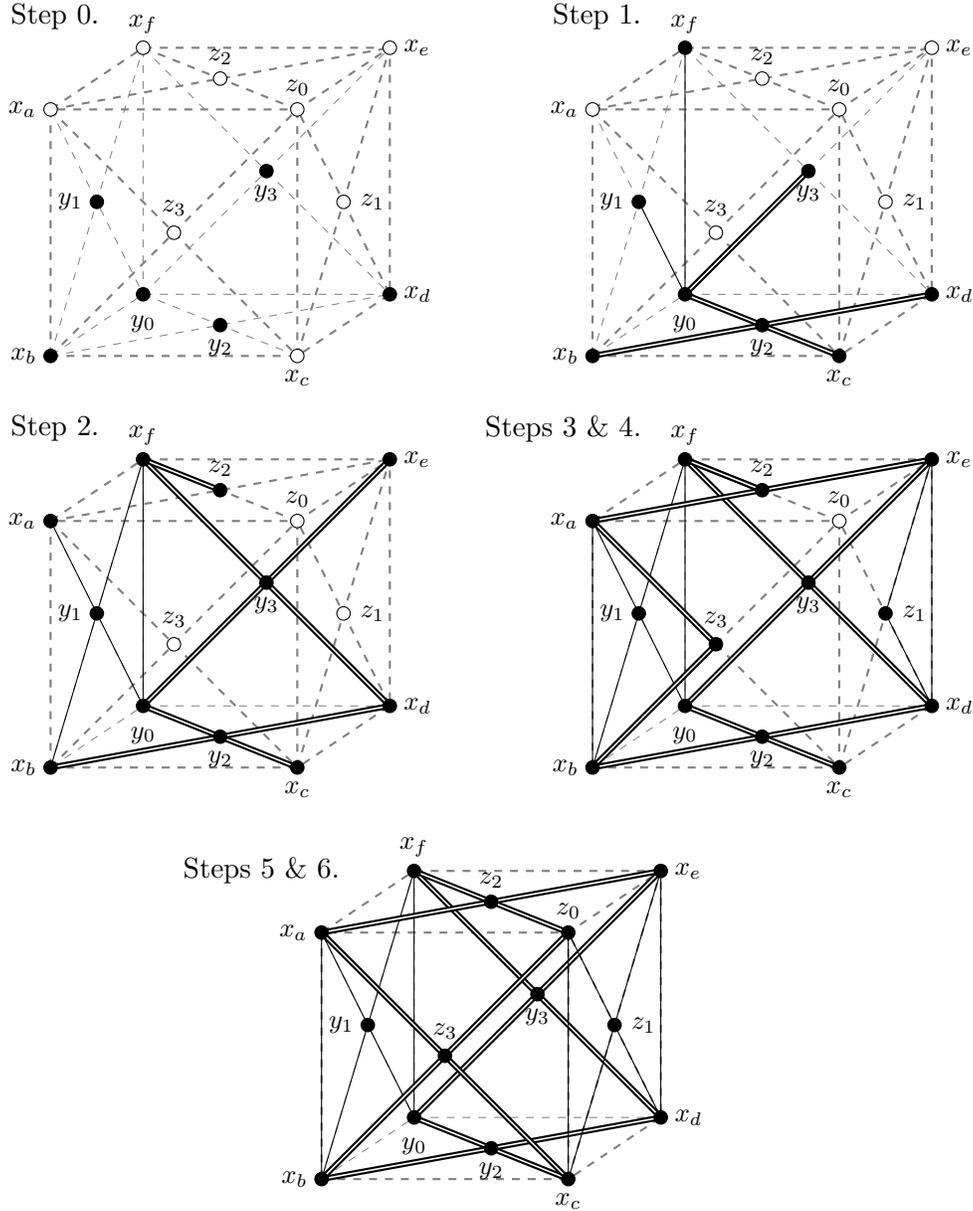
\begin{figure}[hbt!]
\centering
\begin{tikzpicture}[scale=0.82]

\begin{scope}[xshift=0,yshift=0]


\draw[-,gray,thick,dashed] (2.5,4)--(-1.5,4)--(-3,3)--(1,3)--(2.5,4);
\draw[-,gray,thick,dashed] (-3,-1)--(1,-1)--(2.5,0);\draw[-,gray,very thin,dashed] (2.5,0)--(-1.5,0)--(-3,-1);

\draw[-,gray,very thin,dashed] (-3,-1)--(2.5,0);\draw[-,gray,very thin,dashed] (-1.5,0)--(1,-1);
\draw[-,gray,thick,dashed] (-3,3)--(-3,-1);\draw[-,gray,very thin,dashed] (-3,-1)--(-1.5,4);\draw[-,gray,very thin,dashed] (-3,3)--(-1.5,0)--(-1.5,4);
\draw[-,gray,very thin,dashed] (-1.5,0)--(2.5,4);\draw[-,gray,very thin,dashed] (-1.5,4)--(2.5,0);

\draw[-,gray,thick,dashed] (-3,-1)--(1,3);\draw[-,gray,thick,dashed] (-3,3)--(1,-1);
\draw[-,gray,thick,dashed] (1,-1)--(2.5,4)--(2.5,0);\draw[-,gray,thick,dashed] (1,-1)--(1,3)--(2.5,0);
\draw[-,gray,thick,dashed] (-3,3)--(2.5,4);\draw[-,gray,thick,dashed] (-1.5,4)--(1,3);

\filldraw[fill=white,draw=black] (1,-1) circle (3.1pt) node[below=1.5pt]{\small $\ccc$};
\filldraw[fill=black,draw=black] (-0.25,-0.5) circle (3.1pt) node[below=1.5pt]{\small $\ccyb$};
\filldraw[fill=black,draw=black] (-1.5,0) circle (3.1pt) node[below=4pt]{\color{black}\small $\ccy$};
\filldraw[fill=black,draw=black] (-3,-1) circle (3.1pt) node[left=1.5pt]{\small $\ccb$};
\filldraw[fill=black,draw=black] (2.5,0) circle (3.1pt) node[right=1.5pt]{\color{black}\small $\ccd$};
\filldraw[fill=white,draw=black] (-3,3) circle (3.1pt) node[left=1.5pt]{\color{black}\small $\cca$};
\filldraw[fill=white,draw=black] (-1.5,4) circle (3.1pt) node[above=1.5pt]{\small $\ccf$};
\filldraw[fill=black,draw=black] (-2.25,1.5) circle (3.1pt) node[left=1.5pt]{\color{black}\small $\ccya$};
\filldraw[fill=black,draw=black] (0.5,2) circle (3.1pt) node[below=1.5pt]{\small $\ccyc$};
\filldraw[fill=white,draw=black] (2.5,4) circle (3.1pt) node[right=1.5pt]{\small $\cce$};

\filldraw[fill=white,draw=black] (1,3) circle (3.1pt) node[above=1.5pt]{\small $\ccz$};
\filldraw[fill=white,draw=black] (-0.25,3.5) circle (3.1pt) node[above=1.5pt]{\small $\cczb$};
\filldraw[fill=white,draw=black] (-1.0,1) circle (3.1pt) node[above=2.5pt]{\small $\cczc$};
\filldraw[fill=white,draw=black] (1.75,1.5) circle (3.1pt) node[right=2.5pt]{\small $\ccza$};

\fill[white] (-2,4.5) circle (0.01pt) node[left=1pt]{\color{black} Step 0.};

\end{scope}

\begin{scope}[xshift=250,yshift=0]


\draw[-,gray,thick,dashed] (2.5,4)--(-1.5,4)--(-3,3)--(1,3)--(2.5,4);
\draw[-,gray,thick,dashed] (-3,-1)--(1,-1)--(2.5,0);\draw[-,gray,very thin,dashed] (2.5,0)--(-1.5,0)--(-3,-1);

\draw[-,gray,very thin,dashed] (-3,-1)--(2.5,0);\draw[-,gray,very thin,dashed] (-1.5,0)--(1,-1);
\draw[-,gray,thick,dashed] (-3,3)--(-3,-1);\draw[-,gray,very thin,dashed] (-3,-1)--(-1.5,4);\draw[-,gray,very thin,dashed] (-3,3)--(-1.5,0)--(-1.5,4);
\draw[-,gray,very thin,dashed] (-1.5,0)--(2.5,4);\draw[-,gray,very thin,dashed] (-1.5,4)--(2.5,0);

\draw[-,gray,thick,dashed] (-3,-1)--(1,3);\draw[-,gray,thick,dashed] (-3,3)--(1,-1);
\draw[-,gray,thick,dashed] (1,-1)--(2.5,4)--(2.5,0);\draw[-,gray,thick,dashed] (1,-1)--(1,3)--(2.5,0);
\draw[-,gray,thick,dashed] (-3,3)--(2.5,4);\draw[-,gray,thick,dashed] (-1.5,4)--(1,3);

\draw[-,double,thick] (-3,-1)--(2.5,0);\draw[-,double,thick] (1,-1)--(-1.5,0);
\draw[-] (-2.25,1.5)--(-1.5,0)--(-1.5,4);
\draw[-,double,thick](-1.5,0)--(0.5,2);

\filldraw[fill=black,draw=black] (1,-1) circle (3.1pt) node[below=1.5pt]{\small $\ccc$};
\filldraw[fill=black,draw=black] (-0.25,-0.5) circle (3.1pt) node[below=1.5pt]{\small $\ccyb$};
\filldraw[fill=black,draw=black] (-1.5,0) circle (3.1pt) node[below=4pt]{\color{black}\small $\ccy$};
\filldraw[fill=black,draw=black] (-3,-1) circle (3.1pt) node[left=1.5pt]{\small $\ccb$};
\filldraw[fill=black,draw=black] (2.5,0) circle (3.1pt) node[right=1.5pt]{\color{black}\small $\ccd$};
\filldraw[fill=white,draw=black] (-3,3) circle (3.1pt) node[left=1.5pt]{\color{black}\small $\cca$};
\filldraw[fill=black,draw=black] (-1.5,4) circle (3.1pt) node[above=1.5pt]{\small $\ccf$};
\filldraw[fill=black,draw=black] (-2.25,1.5) circle (3.1pt) node[left=1.5pt]{\color{black}\small $\ccya$};
\filldraw[fill=black,draw=black] (0.5,2) circle (3.1pt) node[below=1.5pt]{\small $\ccyc$};
\filldraw[fill=white,draw=black] (2.5,4) circle (3.1pt) node[right=1.5pt]{\small $\cce$};

\filldraw[fill=white,draw=black] (1,3) circle (3.1pt) node[above=1.5pt]{\small $\ccz$};
\filldraw[fill=white,draw=black] (-0.25,3.5) circle (3.1pt) node[above=1.5pt]{\small $\cczb$};
\filldraw[fill=white,draw=black] (-1.0,1) circle (3.1pt) node[above=2.5pt]{\small $\cczc$};
\filldraw[fill=white,draw=black] (1.75,1.5) circle (3.1pt) node[right=2.5pt]{\small $\ccza$};

\fill[white] (-2,4.5) circle (0.01pt) node[left=1pt]{\color{black} Step 1.};

\end{scope}

\begin{scope}[xshift=0,yshift=-190]


\draw[-,gray,thick,dashed] (2.5,4)--(-1.5,4)--(-3,3)--(1,3)--(2.5,4);
\draw[-,gray,thick,dashed] (-3,-1)--(1,-1)--(2.5,0);\draw[-,gray,very thin,dashed] (2.5,0)--(-1.5,0)--(-3,-1);

\draw[-,gray,very thin,dashed] (-3,-1)--(2.5,0);\draw[-,gray,very thin,dashed] (-1.5,0)--(1,-1);
\draw[-,gray,thick,dashed] (-3,3)--(-3,-1);\draw[-,gray,very thin,dashed] (-3,-1)--(-1.5,4);\draw[-,gray,very thin,dashed] (-3,3)--(-1.5,0)--(-1.5,4);
\draw[-,gray,very thin,dashed] (-1.5,0)--(2.5,4);\draw[-,gray,very thin,dashed] (-1.5,4)--(2.5,0);

\draw[-,gray,thick,dashed] (-3,-1)--(1,3);\draw[-,gray,thick,dashed] (-3,3)--(1,-1);
\draw[-,gray,thick,dashed] (1,-1)--(2.5,4)--(2.5,0);\draw[-,gray,thick,dashed] (1,-1)--(1,3)--(2.5,0);
\draw[-,gray,thick,dashed] (-3,3)--(2.5,4);\draw[-,gray,thick,dashed] (-1.5,4)--(1,3);

\draw[-,double,thick] (-3,-1)--(2.5,0);\draw[-,double,thick] (1,-1)--(-1.5,0);
\draw[-] (-3,-1)--(-1.5,4);\draw[-] (-3,3)--(-1.5,0)--(-1.5,4);
\draw[-,double,thick] (-1.5,0)--(2.5,4);\draw[-,double,thick] (-1.5,4)--(2.5,0);

\draw[-,double,thick] (-1.5,4)--(-0.25,3.5);

\filldraw[fill=black,draw=black] (1,-1) circle (3.1pt) node[below=1.5pt]{\small $\ccc$};
\filldraw[fill=black,draw=black] (-0.25,-0.5) circle (3.1pt) node[below=1.5pt]{\small $\ccyb$};
\filldraw[fill=black,draw=black] (-1.5,0) circle (3.1pt) node[below=4pt]{\color{black}\small $\ccy$};
\filldraw[fill=black,draw=black] (-3,-1) circle (3.1pt) node[left=1.5pt]{\small $\ccb$};
\filldraw[fill=black,draw=black] (2.5,0) circle (3.1pt) node[right=1.5pt]{\color{black}\small $\ccd$};
\filldraw[fill=black,draw=black] (-3,3) circle (3.1pt) node[left=1.5pt]{\color{black}\small $\cca$};
\filldraw[fill=black,draw=black] (-1.5,4) circle (3.1pt) node[above=1.5pt]{\small $\ccf$};
\filldraw[fill=black,draw=black] (-2.25,1.5) circle (3.1pt) node[left=1.5pt]{\color{black}\small $\ccya$};
\filldraw[fill=black,draw=black] (0.5,2) circle (3.1pt) node[below=1.5pt]{\small $\ccyc$};
\filldraw[fill=black,draw=black] (2.5,4) circle (3.1pt) node[right=1.5pt]{\small $\cce$};

\filldraw[fill=white,draw=black] (1,3) circle (3.1pt) node[above=1.5pt]{\small $\ccz$};
\filldraw[fill=black,draw=black] (-0.25,3.5) circle (3.1pt) node[above=1.5pt]{\small $\cczb$};
\filldraw[fill=white,draw=black] (-1.0,1) circle (3.1pt) node[above=2.5pt]{\small $\cczc$};
\filldraw[fill=white,draw=black] (1.75,1.5) circle (3.1pt) node[right=2.5pt]{\small $\ccza$};

\fill[white] (-2,4.5) circle (0.01pt) node[left=1pt]{\color{black} Step 2.};

\end{scope}

\begin{scope}[xshift=250,yshift=-190]


\draw[-,gray,thick,dashed] (2.5,4)--(-1.5,4)--(-3,3)--(1,3)--(2.5,4);
\draw[-,gray,thick,dashed] (-3,-1)--(1,-1)--(2.5,0);\draw[-,gray,very thin,dashed] (2.5,0)--(-1.5,0)--(-3,-1);

\draw[-,gray,very thin,dashed] (-3,-1)--(2.5,0);\draw[-,gray,very thin,dashed] (-1.5,0)--(1,-1);
\draw[-,gray,thick,dashed] (-3,3)--(-3,-1);\draw[-,gray,very thin,dashed] (-3,-1)--(-1.5,4);\draw[-,gray,very thin,dashed] (-3,3)--(-1.5,0)--(-1.5,4);
\draw[-,gray,very thin,dashed] (-1.5,0)--(2.5,4);\draw[-,gray,very thin,dashed] (-1.5,4)--(2.5,0);

\draw[-,gray,thick,dashed] (-3,-1)--(1,3);\draw[-,gray,thick,dashed] (-3,3)--(1,-1);
\draw[-,gray,thick,dashed] (1,-1)--(2.5,4)--(2.5,0);\draw[-,gray,thick,dashed] (1,-1)--(1,3)--(2.5,0);
\draw[-,gray,thick,dashed] (-3,3)--(2.5,4);\draw[-,gray,thick,dashed] (-1.5,4)--(1,3);

\draw[-,double,thick] (-3,-1)--(2.5,0);\draw[-,double,thick] (1,-1)--(-1.5,0);
\draw[-] (-3,3)--(-3,-1)--(-1.5,4);\draw[-] (-3,3)--(-1.5,0)--(-1.5,4);
\draw[-,double,thick] (-1.5,0)--(2.5,4);\draw[-,double,thick] (-1.5,4)--(2.5,0);

\draw[-,double,thick] (-1.5,4)--(-0.25,3.5);

\draw[-,double,thick] (-1.5,4)--(-0.25,3.5);
\draw[-,double,thick] (-3,-1)--(-1,1)--(-3,3)--(2.5,4);\draw[-] (1.75,1.5)--(2.5,4)--(2.5,0)--(1.75,1.5);

\filldraw[fill=black,draw=black] (1,-1) circle (3.1pt) node[below=1.5pt]{\small $\ccc$};
\filldraw[fill=black,draw=black] (-0.25,-0.5) circle (3.1pt) node[below=1.5pt]{\small $\ccyb$};
\filldraw[fill=black,draw=black] (-1.5,0) circle (3.1pt) node[below=4pt]{\color{black}\small $\ccy$};
\filldraw[fill=black,draw=black] (-3,-1) circle (3.1pt) node[left=1.5pt]{\small $\ccb$};
\filldraw[fill=black,draw=black] (2.5,0) circle (3.1pt) node[right=1.5pt]{\color{black}\small $\ccd$};
\filldraw[fill=black,draw=black] (-3,3) circle (3.1pt) node[left=1.5pt]{\color{black}\small $\cca$};
\filldraw[fill=black,draw=black] (-1.5,4) circle (3.1pt) node[above=1.5pt]{\small $\ccf$};
\filldraw[fill=black,draw=black] (-2.25,1.5) circle (3.1pt) node[left=1.5pt]{\color{black}\small $\ccya$};
\filldraw[fill=black,draw=black] (0.5,2) circle (3.1pt) node[below=1.5pt]{\small $\ccyc$};
\filldraw[fill=black,draw=black] (2.5,4) circle (3.1pt) node[right=1.5pt]{\small $\cce$};

\filldraw[fill=white,draw=black] (1,3) circle (3.1pt) node[above=1.5pt]{\small $\ccz$};
\filldraw[fill=black,draw=black] (-0.25,3.5) circle (3.1pt) node[above=1.5pt]{\small $\cczb$};
\filldraw[fill=black,draw=black] (-1.0,1) circle (3.1pt) node[above=2.5pt]{\small $\cczc$};
\filldraw[fill=black,draw=black] (1.75,1.5) circle (3.1pt) node[right=2.5pt]{\small $\ccza$};

\fill[white] (-2,4.5) circle (0.01pt) node[left=1pt]{\color{black} Steps 3 \& 4.};

\end{scope}

\begin{scope}[xshift=125,yshift=-380]


\draw[-,gray,thick,dashed] (2.5,4)--(-1.5,4)--(-3,3)--(1,3)--(2.5,4);
\draw[-,gray,thick,dashed] (-3,-1)--(1,-1)--(2.5,0);\draw[-,gray,very thin,dashed] (2.5,0)--(-1.5,0)--(-3,-1);

\draw[-,gray,very thin,dashed] (-3,-1)--(2.5,0);\draw[-,gray,very thin,dashed] (-1.5,0)--(1,-1);
\draw[-,gray,thick,dashed] (-3,3)--(-3,-1);\draw[-,gray,very thin,dashed] (-3,-1)--(-1.5,4);\draw[-,gray,very thin,dashed] (-3,3)--(-1.5,0)--(-1.5,4);
\draw[-,gray,very thin,dashed] (-1.5,0)--(2.5,4);\draw[-,gray,very thin,dashed] (-1.5,4)--(2.5,0);

\draw[-,gray,thick,dashed] (-3,-1)--(1,3);\draw[-,gray,thick,dashed] (-3,3)--(1,-1);
\draw[-,gray,thick,dashed] (1,-1)--(2.5,4)--(2.5,0);\draw[-,gray,thick,dashed] (1,-1)--(1,3)--(2.5,0);
\draw[-,gray,thick,dashed] (-3,3)--(2.5,4);\draw[-,gray,thick,dashed] (-1.5,4)--(1,3);

\draw[-,double,thick] (-3,-1)--(2.5,0);\draw[-,double,thick] (1,-1)--(-1.5,0);
\draw[-] (-3,3)--(-3,-1)--(-1.5,4);\draw[-] (-3,3)--(-1.5,0)--(-1.5,4);
\draw[-,double,thick] (-1.5,0)--(2.5,4);\draw[-,double,thick] (-1.5,4)--(2.5,0);

\draw[-,double,thick] (-3,-1)--(1,3);\draw[-,double,thick] (-3,3)--(1,-1);
\draw[-] (1,-1)--(2.5,4)--(2.5,0);\draw[-] (1,-1)--(1,3)--(2.5,0);
\draw[-,double,thick] (-3,3)--(2.5,4);\draw[-,double,thick] (-1.5,4)--(1,3);

\filldraw[fill=black,draw=black] (1,-1) circle (3.1pt) node[below=1.5pt]{\small $\ccc$};
\filldraw[fill=black,draw=black] (-0.25,-0.5) circle (3.1pt) node[below=1.5pt]{\small $\ccyb$};
\filldraw[fill=black,draw=black] (-1.5,0) circle (3.1pt) node[below=4pt]{\color{black}\small $\ccy$};
\filldraw[fill=black,draw=black] (-3,-1) circle (3.1pt) node[left=1.5pt]{\small $\ccb$};
\filldraw[fill=black,draw=black] (2.5,0) circle (3.1pt) node[right=1.5pt]{\color{black}\small $\ccd$};
\filldraw[fill=black,draw=black] (-3,3) circle (3.1pt) node[left=1.5pt]{\color{black}\small $\cca$};
\filldraw[fill=black,draw=black] (-1.5,4) circle (3.1pt) node[above=1.5pt]{\small $\ccf$};
\filldraw[fill=black,draw=black] (-2.25,1.5) circle (3.1pt) node[left=1.5pt]{\color{black}\small $\ccya$};
\filldraw[fill=black,draw=black] (0.5,2) circle (3.1pt) node[below=1.5pt]{\small $\ccyc$};
\filldraw[fill=black,draw=black] (2.5,4) circle (3.1pt) node[right=1.5pt]{\small $\cce$};

\filldraw[fill=black,draw=black] (1,3) circle (3.1pt) node[above=1.5pt]{\small $\ccz$};
\filldraw[fill=black,draw=black] (-0.25,3.5) circle (3.1pt) node[above=1.5pt]{\small $\cczb$};
\filldraw[fill=black,draw=black] (-1.0,1) circle (3.1pt) node[above=2.5pt]{\small $\cczc$};
\filldraw[fill=black,draw=black] (1.75,1.5) circle (3.1pt) node[right=2.5pt]{\small $\ccza$};

\fill[white] (-2.5,4.0) circle (0.01pt) node[left=1pt]{\color{black} Steps 5 \& 6.};

\end{scope}

\end{tikzpicture}
\caption{The steps of the CAFCC algorithm, where unfilled vertices represent variables to be determined in subsequent steps.  Step 0 here represents the choice of initial variables.  In steps 1 and 2 five equations are used to uniquely determine five respective variables $\ccc,\ccf,\cca,\cce,\cczb$.  Steps 3 and 4 require consistency of two equations solving for $\ccza$, and consistency of two equations solving for $\cczc$, respectively.  Step 5 requires consistency of four equations solving for $\ccz$.  All fourteen variables have been determined in steps 1--5, and step 6 requires that the remaining equation centered at $\ccz$ is automatically satisfied.}
\label{CAFCCALL}
\end{figure}

For the above initial conditions the CAFCC property can be checked with the following six steps (also pictured graphically in Figure \ref{CAFCCALL}):

\begin{enumerate}

\item
The following two equations centered at $\ccy$ and $\ccyb$
\begin{align}
\begin{split}
\C{\ccy}{\ccya}{\ccf}{\ccyb}{\ccyc}{(\ccqa,\ccrb)}{(\ccpb,\ccra)}=0,\\
\B{\ccyb}{\ccy}{\ccd}{\ccb}{\ccc}{\ccpp}{\ccqq}=0,
\end{split}
\end{align}
may be solved respectively to uniquely determine the two variables $\ccf$, and $\ccc$.

\item
The following three equations centered at $\ccya$, $\ccyc$, and $\ccf$
\begin{align}
\begin{split}
\A{\ccya}{\ccf}{\ccy}{\cca}{\ccb}{\ccpp}{\ccrr}=0, \\
\B{\ccyc}{\ccf}{\cce}{\ccy}{\ccd}{\ccrr}{\ccqq}=0, \\
\C{\ccf}{\ccya}{\ccy}{\cczb}{\ccyc}{(\ccqa,\ccra)}{(\ccpb,\ccrb)}=0,
\end{split}
\end{align}
may be solved respectively to uniquely determine the three variables $\cca$, $\cce$, and $\cczb$.

\item 
For the first consistency check both of the following equations
\begin{align}
\begin{split}
\C{\ccd}{\ccza}{\cce}{\ccyb}{\ccyc}{(\ccqb,\ccrb)}{(\ccpb,\ccra)}=0, \\
\C{\cce}{\ccza}{\ccd}{\cczb}{\ccyc}{(\ccqb,\ccra)}{(\ccpb,\ccrb)}=0,
\end{split}
\end{align}
may be used to solve for the variable $\ccza$, and the two solutions must be in agreement.

\item 
For the second consistency check both of the following equations
\begin{align}
\begin{split}
\C{\cca}{\ccya}{\ccb}{\cczb}{\cczc}{(\ccqa,\ccra)}{(\ccpa,\ccrb)}=0,\\
\C{\ccb}{\ccya}{\cca}{\ccyb}{\cczc}{(\ccqa,\ccrb)}{(\ccpa,\ccra)}=0,
\end{split}
\end{align}
 may be used to solve for the variable $\cczc$, and the two solutions must be in agreement.

\item 
For the third consistency check each of the following four equations
\begin{align}
\begin{split}
\A{\ccza}{\cce}{\ccd}{\ccz}{\ccc}{\ccpp}{\ccrr}=0, \\
\B{\cczb}{\ccf}{\cce}{\cca}{\ccz}{\ccpp}{\ccqq}=0, \\
\B{\cczc}{\cca}{\ccz}{\ccb}{\ccc}{\ccrr}{\ccqq}=0, \\
\C{\ccc}{\ccza}{\ccz}{\ccyb}{\cczc}{(\ccqb,\ccrb)}{(\ccpa,\ccra)}=0.
\end{split}
\end{align}
may be used to solve for the final variable $\ccz$, and the four solutions must be agreement.

\item 
For the final consistency check the remaining equation centered at $\ccz$
\begin{align}
\begin{split}
\C{\ccz}{\ccza}{\ccc}{\cczb}{\cczc}{(\ccqb,\ccra)}{(\ccpa,\ccrb)}=0,
\end{split}
\end{align}
must be satisfied by the variables that have been determined in the previous steps.

\end{enumerate}



\subsection{Examples of equations that satisfy CAFCC}\label{sec:FCClist}

Examples of combinations of the type-A, -B, and -C face-centered quad equations that satisfy CAFCC are given in Table \ref{table-BC}.  Tables \ref{table-A} and \ref{table-BC2} give the expressions for $a(x;y;\alpha,\beta)$, $b(x;y;\alpha,\beta)$, and $c(x;y;\alpha,\beta)$, in the four-leg forms \eqref{4leg}, \eqref{4legb}, and \eqref{4legc}, corresponding toe type-A ,-B, and -C equations respectively. Each of the equations in Tables \ref{table-BC}--\ref{table-BC2} are also given in their affine-linear polynomial forms in Appendix \ref{app:afflin}.  The derivation of these equations from the IRF equations of Section \ref{sec:YBE} will be given in Section \ref{sec:CAFCCIRF}.

\begin{table}[htb!]
\centering
\begin{tabular}{c|c|c|c}

  Type-A & Type-B & Type-C & $P_1(x_a,x_b,x_c,x_d)$
 
 \\
 
 \hline
 
 \\[-0.4cm]

 $A3_{(\delta=1)}$ & $B3_{(\delta_1=\frac{1}{2};\,\delta_2=\frac{1}{2};\,\delta_3=0)}$ & $C3_{(\delta_1=\frac{1}{2};\,\delta_2=\frac{1}{2};\,\delta_3=0)}$
& $H3_{(\delta=1;\,\varepsilon=1)}$

\\[0.2cm]

  $A3_{(\delta=0)}$ & $B3_{(\delta_1=\frac{1}{2};\,\delta_2=0;\,\delta_3=\frac{1}{2})}$ & $C3_{(\delta_1=\frac{1}{2};\,\delta_2=0;\,\delta_3=\frac{1}{2})}$
& $H3_{(\delta=1;\,\varepsilon=1)}$

\\[0.2cm]

 $A3_{(\delta=0)}$ & $B3_{(\delta_1=1;\,\delta_2=0;\,\delta_3=0)}$ & $C3_{(\delta_1=1;\,\delta_2=0;\,\delta_3=0)}$
& $H3_{(\delta=0,1;\,\varepsilon=1-\delta)}$

\\[0.2cm]

 $A3_{(\delta=0)}$ & $B3_{(\delta_1=0;\,\delta_2=0;\,\delta_3=0)}$ ($D4$) & $C3_{(\delta_1=0;\,\delta_2=0;\,\delta_3=0)}$
& $H3_{(\delta=0;\,\varepsilon=0)}$

\\[0.2cm]

 $A2_{(\delta_1=1;\,\delta_2=1)}$ & $B2_{(\delta_1=1;\,\delta_2=1;\,\delta_3=0)}$ & $C2_{(\delta_1=1;\,\delta_2=1;\,\delta_3=0)}$
& $H2_{(\varepsilon=1)}$

\\[0.2cm]

 $A2_{(\delta_1=1;\,\delta_2=0)}$ & $B2_{(\delta_1=1;\,\delta_2=0;\,\delta_3=1)}$ & $C2_{(\delta_1=1;\,\delta_2=0;\,\delta_3=1)}$
& $H2_{(\varepsilon=1)}$

\\[0.2cm]

 $A2_{(\delta_1=1;\,\delta_2=0)}$ & $B2_{(\delta_1=1;\,\delta_2=0;\,\delta_3=0)}$ & $C2_{(\delta_1=1;\,\delta_2=0;\,\delta_3=0)}$
& $H2_{(\varepsilon=0)}$

\\[0.2cm]

 $A2_{(\delta_1=0;\,\delta_2=0)}$ & $B2_{(\delta_1=0;\,\delta_2=0;\,\delta_3=0)}$ ($D4$) & $C2_{(\delta_1=0;\,\delta_2=0;\,\delta_3=0)}$
& $H1_{(\varepsilon=1)}$

\\[0.2cm]

 $A2_{(\delta_1=0;\,\delta_2=0)}$ & $D1$ & $C1$
& $H1_{(\varepsilon=0)}$

\\[0.0cm]

\hline 
\end{tabular}
\caption{Examples of different combinations of type-A, -B, and -C equations which satisfy CAFCC, where type-A equations also satisfy CAFCC on their own.  These equations are given in Tables \ref{table-A} and \ref{table-BC2}.  $P_1$ is the CAC equation from the ABS list that arises as the coefficient of $x$ for the affine-linear form of the type-C equation.}
\label{table-BC}
\end{table}

\begin{table}[htb!]
\centering
\begin{tabular}{c|c|c}

 Type-A & $a(x;y;\alpha,\beta)$ & $P_1(x_a,x_b,x_c,x_d)$
 
 \\
 
 \hline
 
 \\[-0.4cm]

$A4$ & $\displaystyle
\frac{(G_{+}(x,\alpha,\beta)-F(x,y,\alpha,\beta))S_{-}(x,\alpha,\beta)}
     {(G_{-}(x,\alpha,\beta)-F(x,y,\alpha,\beta))S_{+}(x,\alpha,\beta)}$
& $Q4$

\\[0.4cm]

$A3_{(\delta=1)}$ & $\displaystyle
\frac{\alpha^2+\beta^2\overline{x}^2-2\alpha\beta \overline{x}y}
     {\beta^2+\alpha^2\overline{x}^2-2\alpha\beta \overline{x}y}$
& $Q3_{(\delta=1)}$

\\[0.4cm]

$A3_{(\delta=0)}$ & $\displaystyle
\frac{\beta x-\alpha y}{\alpha x-\beta y}$
& $Q3_{(\delta=0)}$

\\[0.4cm]

$A2_{(\delta_1=1;\,\delta_2=1)}$ & $\displaystyle
\frac{(\sqrt{x}+\alpha-\beta)^2-y}{(\sqrt{x}-\alpha+\beta)^2-y}$
& $Q2$

\\[0.4cm]

$A2_{(\delta_1=1;\,\delta_2=0)}$ & $\displaystyle
\frac{-x+y+\alpha-\beta}{x-y+\alpha-\beta}$
& $Q1_{(\delta=1)}$

\\[0.4cm]

$A2_{(\delta_1=0;\,\delta_2=0)}$ & $\displaystyle
\phantom{(add.) }\quad  \frac{\alpha-\beta}{x-y} \quad (add.)$
& $Q1_{(\delta=0)}$

\\[0.0cm]

\hline 
\end{tabular}
\caption{A list of functions $a(x;y;\alpha,\beta)$ for the type-A equations \eqref{4leg} in Table \ref{table-BC}, as well as for $A4$ which satisfies CAFCC on its own. $P_1$ is the CAC equation from the ABS list that arises as the coefficient of $x$ for the affine-linear form of the type-A equation \eqref{quadseries}.  The abbreviation ``add.'' indicates an additive form of the equation (see \eqref{fccexaadd} or \eqref{fceA2}).}
\label{table-A}
\end{table}

\begin{table}[htb!]
\centering
\begin{tabular}{c|c}

Type-B  & $b(x;y;\alpha,\beta)$ 
 
 \\
 
 \hline
 
 \\[-0.4cm]

$B3_{(\delta_1=\frac{1}{2};\,\delta_2=\frac{1}{2};\,\delta_3=0)}$ &  $\displaystyle
\beta^2+\alpha^2x^2-2\alpha\beta xy$

\\[0.2cm]

$B3_{(\delta_1=\frac{1}{2};\,\delta_2=0;\,\delta_3=\frac{1}{2})}$  & $\displaystyle
\frac{\alpha y-\beta\overline{x}}{\alpha\overline{x}y-\beta}$

\\[0.2cm]

$B3_{(\delta_1=1;\,\delta_2=0;\,\delta_3=0)}$ &  $\displaystyle
\beta-\alpha xy$

\\[0.2cm]

$B3_{(\delta_1=0;\,\delta_2=0;\,\delta_3=0)}$ ($D4$)  & $\displaystyle
y$

\\[0.2cm]

$B2_{(\delta_1=1;\,\delta_2=1;\,\delta_3=0)}$ & $\displaystyle
(x+\alpha-\beta)^2-y$

\\[0.2cm]

$B2_{(\delta_1=1;\,\delta_2=0;\,\delta_3=1)}$  & $\displaystyle
\frac{\sqrt{x}+y+\alpha-\beta}{-\sqrt{x}+y+\alpha-\beta}$

\\[0.2cm]

$B2_{(\delta_1=1;\,\delta_2=0;\,\delta_3=0)}$ & $\displaystyle
x+y+\alpha-\beta$

\\[0.2cm]

$B2_{(\delta_1=0;\,\delta_2=0;\,\delta_3=0)}$ ($D4$) & $\displaystyle
y$

\\[0.2cm]

$D1$  & $\displaystyle
\phantom{(add.) }\quad  y \quad (add.)$

\\[0.0cm]

\hline 
\end{tabular}
\hspace{0.0cm}
\begin{tabular}{c|c}

 Type-C & $c(x;y;\alpha,\beta)$ 
 
 \\
 
 \hline
 
 \\[-0.4cm]

 $C3_{(\delta_1=\frac{1}{2};\,\delta_2=\frac{1}{2};\,\delta_3=0)}$  
&
$\displaystyle
\frac{\alpha-\beta\overline{x}y}{\alpha\overline{x}-\beta y}$

\\[0.2cm]

$C3_{(\delta_1=\frac{1}{2};\,\delta_2=0;\,\delta_3=\frac{1}{2})}$
&
$\bigl(\tfrac{\alpha^2}{\beta}+\beta x^2-2\alpha x y\bigr)$

\\[0.2cm]

$C3_{(\delta_1=1;\,\delta_2=0;\,\delta_3=0)}$      
&
$xy-\tfrac{\alpha}{\beta}$

\\[0.2cm]

$C3_{(\delta_1=0;\,\delta_2=0;\,\delta_3=0)}$ 
&
$y$

\\[0.2cm]

$C2_{(\delta_1=1;\,\delta_2=1;\,\delta_3=0)}$
&
$\displaystyle
\frac{-\sqrt{x}+y-\alpha+\beta}{\sqrt{x}+y-\alpha+\beta}$

\\[0.2cm]

$C2_{(\delta_1=1;\,\delta_2=0;\,\delta_3=1)}$ 
&
$(x-\alpha+\beta)^2-y$

\\[0.2cm]

$C2_{(\delta_1=1;\,\delta_2=0;\,\delta_3=0)}$ 
&
$\displaystyle
x+y-\alpha+\beta$

\\[0.2cm]

$C2_{(\delta_1=0;\,\delta_2=0;\,\delta_3=0)}$ 
&
$\phantom{(a) }$ \,  $-\frac{y+\beta}{2x}  \quad (add.)$

\\[0.2cm]

$C1$ 
& 
$\phantom{(add.) }\quad  y \quad (add.)$

\\[0.0cm]

\hline 
\end{tabular}
\caption{A list of functions $b(x;y;\alpha,\beta)$ and $c(x;y;\alpha,\beta)$ for the type-B and type-C equations \eqref{4legb} and \eqref{4legc} respectively in Table \ref{table-BC}.  The function $a(x;y;\alpha,\beta)$ for the type-C equation \eqref{4legc} should be taken from the corresponding type-A entry in Tables \ref{table-BC} and \ref{table-A}.  The abbreviation ``add.'' indicates an additive form of the equation (see \eqref{fceB2} and \eqref{fceC2}).  }
\label{table-BC2}
\end{table}

In both Tables \ref{table-A} and \ref{table-BC2}, the notation $\overline{x}$ is used to denote
\begin{align}
\overline{x}=x+\sqrt{x^2-1},
\end{align}
for a variable $x$.  In Table \ref{table-A}, the functions appearing for $A4$ are defined by
\begin{align}\label{fgdef}
\begin{split}
F(\mx_a,\mx_b,\mpc,\mqc)=&\,\bigl(4(\mx_a+\mx_b)(\mpc-\mqc)^2+Q(\mpc,\mqc)\bigr)\bigl(4\mx_a(\mpc-\mqc)^2-Q(\mpc,\mqc)\bigr)^2, \\
G_{\pm}(\mx,\mpc,\mqc)=&\,\bigl(4\dmx(\mpc-\mqc)^3\pm R(\mpc,\mqc)\bigr)^2,
\end{split}
\end{align}
where
\begin{align}\label{sqrdef}
\begin{split}
 S_{\pm}(\mx,\mpc,\mqc)=&\,\dmx(\mqc-\mpc)\pm\mx(\dmp+\dmq)\mp(\dmq\mpc+\dmp\mqc), \\
Q(\mpc,\mqc)=&\,(\dmp+\dmq)^2-4(\mpc+\mqc)(\mpc-\mqc)^2, \\
R(\mpc,\mqc)=&\,4(\dmp\mqc+\dmq\mpc)(\mpc-\mqc)^2-(\dmp+\dmq)Q(\mpc,\mqc).
\end{split}
\end{align}
In \eqref{fgdef} and \eqref{sqrdef}, the notation $\dmx$ is used to denote
\begin{align}\label{xdotdef}
\dmx=4\mx^3-g_2\mx-g_3,
\end{align}
for a variable $\mx$, where $g_2$ and $g_3$ are Weierstrass elliptic invariants \cite{WW}.

Table \ref{table-BC} also lists the ABS quad equation denoted as $P_1(x_a,x_b,x_c,x_d)$, which is found to arise as the coefficient of $x$ in the affine-linear form of the type-C equations (see Appendix \ref{app:afflin}). Table \ref{table-A} similarly lists the ABS quad equation for the affine-linear from of the type-A equations \eqref{quadseries}.  The type-B equations do not seem to have an interesting quad equation appear for this same coefficient.

In Table \ref{table-A}, the abbreviation ``add.'' is used to indicate an edge function for an additive form of a type-A equation \eqref{4leg}, which is given by
\begin{align}\label{fccexaadd}
a(x;x_a;\alpha_2,\beta_1)+a(x;x_d;\alpha_1,\beta_2)-a(x;x_b;\alpha_2,\beta_2)-a(x;x_c;\alpha_1,\beta_1)=0.
\end{align}
Similarly, the appearance of ``add.'' in Table \eqref{table-BC2} indicates an additive form for the type-B and type-C equations (see \eqref{fceB2} and \eqref{fceC2}).

\subsubsection{Remarks}

\begin{enumerate}

\item 
        
In addition to the combinations of equations given in Table \ref{table-BC}, the type-A equations of Table \ref{table-A}  also satisfy CAFCC on their own.  That is, CAFCC will also be satisfied with the use of a type-A equation \eqref{4leg} of Table \ref{table-A}, replacing
\begin{align}
\begin{gathered}
B\to A,\qquad C\to A,
\end{gathered}
\end{align}
for each type-B and -C equations appearing in \eqref{6face} and \eqref{8corner} (which correspond to the equations in the six steps of CAFCC).  Furthermore, the equation $A4$ in Table \ref{table-A} is only known to satisfy CAFCC on its own.

\item

For the equations in Tables \ref{table-BC}--\ref{table-BC2} there are up to three parameters $\delta_1,\delta_2,\delta_3$ in the subscripts, which appear for their affine-linear expressions given in Appendix \ref{app:afflin}.  For each case the indicated values should be chosen for the equations to satisfy CAFCC.

\item

Although there appear square roots for the face variable $x$ in the expressions given in Tables \ref{table-A} and \ref{table-BC2}, any square roots cancel out of the respective affine-linear polynomial forms (see Appendix \ref{app:afflin}). 

\item

The affine-linear form \eqref{quadseries} for $A4$, is the only case where the degree $n$ of the face variable $x$ appears to be $n>2$.  Since all other cases of equations have degree $n\leq2$, it is expected that a simpler expression for $A4$ may be found for which $n=2$, but at the moment it is not known.



\item

Some of the equations in Tables \ref{table-A} and \ref{table-BC2} have appeared previously outside of the context of CAFCC.  The equations $D1$ and $D4$ are independent of the face variable $x$, and may be identified with two of the $H^6$-type quad equations classified by Boll \cite{BollThesis}.  The expressions for the four-leg forms \eqref{4leg} for A-type equations, may also be identified with expressions for discrete Toda-type equations associated to type-Q ABS equations \cite{AdlerPlanarGraphs,BobSurQuadGraphs,AdlerSurisQ4}.  The equation $A2_{(\delta_1=0;\,\delta_2=0)}$ also has a different interpretation as the simplest $n=1$ case of a set of $n$-component CAC quad equations \cite{Kels:2018qzx}.   In fact one of the initial motivations for formulating multidimensional consistency as CAFCC was because other equations given in Tables \ref{table-A} and \ref{table-BC2} do not satisfy CAC, although they are obtained in the same way as $A2_{(\delta_1=0;\,\delta_2=0)}$.  





\end{enumerate}

\section{CAFCC from IRF}
\label{sec:CAFCCIRF}



 \subsection{IRF Lagrangian functions to face-centered quad equations}\label{sec:method}

In this section it will be shown how the face-centered quad equations of Section \ref{sec:FCC}, are derived from the IRF equations of Section \ref{sec:YBE}.   In Section \ref{sec:YBE} there was given the expressions for the classical IRF Yang-Baxter equations \eqref{CAYBE1} and \eqref{CAYBE2}, which play a central role.  The derivatives of \eqref{CAYBE1} and \eqref{CAYBE2} with respect to their fourteen variables, respectively lead to fourteen expressions for the discrete Laplace-type equations.\footnote{The discrete Laplace-type equations usually refer to equations given in terms of a sum of four Lagrangians, where the Lagrangian functions are all the same.  Recall that in this paper even if the four Lagrangian functions are different, the sum of four Lagrangian functions obtained as the derivatives of the IRF YBE will also be referred to as discrete Laplace-type equations.}   
The examples of the fourteen discrete Laplace-type equations for \eqref{CAYBE1} are listed in Appendix \ref{app:FCCequations}. These equations can be grouped into three different types that correspond to one of the face-centered quad equations of types-A, -B, or -C, respectively.  Type-A equations involve only derivatives of $\lag$ and $\ol$, as given in \eqref{ey1} and \eqref{ez1}.  Type-B equations involve only derivatives of $\lam$ and $\olam$, as given in \eqref{ey2}, \eqref{ey3}, \eqref{ez2}, and \eqref{ez3}.  Type-C equations involve derivatives of two of $\lag$ and $\ol$, as well as two of $\lam$ and $\olam$, as given in the eight equations \eqref{ey}, \eqref{ez}, and \eqref{ea}--\eqref{ef}.

There are fourteen variables involved in the discrete Laplace-type equations of Appendix \ref{app:FCCequations}, which are associated to the fourteen vertices of the Yang-Baxter equation in Figure \ref{SSRYBE}.  For convenience, these fourteen variables are divided into the following two sets
\begin{align}
X_A=\{x_a,x_b,x_c,x_d,x_e,x_f,x_h,x_h',x_i,x_i'\},\qquad X_B=\{x_j,x_j',x_k,x_k'\},
\end{align}
according to whether or not they appear as one of the face variables of a discrete Laplace-type equation of type-B.  
%
%
Let also $Z$ denote the set of components of the parameters $\bu,\bv,\bw$, as
\begin{align}
Z=\{u_1,u_2,v_1,v_2,w_1,w_2\}.
\end{align}
There are three different types of changes of variables, denoted by $f(x)$, $g(x)$, $h(x)$, for the variables and parameters in $X_A$, $X_B$, $Z$, respectively, indicated as follows
\begin{align}\label{cov}
x\mapsto\left\{\begin{array}{ll} y=f(x), & x\in X_A, \\ y=g(x), & x\in X_B, \\ y=h(x), & x\in Z. \end{array} \right.
\end{align}
The transformed variables and parameters denoted by $y$ in \eqref{cov}, then become the variables and parameters of the face-centered quad equations.  The specific forms of $f(x)$, $g(x)$, $h(x)$, depend on the types of functions involved in the explicit expressions for the Lagrangian functions that are used in Section \ref{sec:FCElist}.  These functions are always one of elliptic-, hyperbolic-, rational-, or algebraic-types, and the typical associated changes of variables for these four types is indicated in Table \ref{tab-covs}.
\begin{table}[htb!]
\centering
\begin{tabular}{c c c}
Case & $f(x)$, $g(x)$ & $h(x)$ \\[0.0cm]
\hline \\[-0.3cm]
Elliptic & $\wp(x)$ & $\wp(x)$ \\[0.1cm]
Hyperbolic & $\EXP^{x}$ or $\cosh(x)$ & $\EXP^{x}$  \\[0.1cm]
Rational & $x$ or $x^2$ & $x$ \\[0.1cm]
Algebraic & $x$ & $x$ \\[0.1cm]
\hline 
\end{tabular}
 \caption{The different cases for the changes of variables \eqref{cov} for the different cases of Lagrangian functions used in Section \ref{sec:FCElist}.  The function $\wp(z)$ is the Weierstrass elliptic function \cite{WW}.}
\label{tab-covs}
\end{table}

There is another symmetry that will be used to simplify the expressions of the fourteen Laplace-type equations in Appendix \ref{app:FCCequations}.  
Namely, the derivatives of the Lagrangian functions will be found to satisfy (except for the elliptic case)
\begin{align}\label{covshift}
\frac{\partial \ol_{u-v}(x,y)}{\partial x}=-\frac{\partial\lag_{\pm\eta_0+u-v}(x,y)}{\partial x}+k_1\pi\ii,\qquad \frac{\partial\olam_{u-v}(x,y)}{\partial x}=-\frac{\partial \lam_{\pm\eta_0+u-v}(x,y)}{\partial x}+k_2\pi\ii,
\end{align}
for some constant $\eta_0$, where $k_1+k_2$ is an even integer.  For the rational and algebraic cases, $\eta_0=0$, while for the hyperbolic cases, $\eta_0=\pi$.  For the elliptic case, $2\eta_0$ is a quasi-period of the Weierstrass sigma function, and there will appear an additional $x$ dependent term for the difference of derivatives of Lagrangian functions taken in \eqref{covshift}.  However, these additional terms always cancel out of the combinations of Lagrangian functions in the expressions for the Laplace-type equations of Appendix \ref{app:FCCequations}, and particularly have no overall contribution to the derivation of the face-centered quad equations.


Motivated by \eqref{covshift}, the first components of the parameters of the fourteen Laplace-type equations of Appendix \ref{app:FCCequations} will be shifted by
\begin{align}\label{parshift}
    \bu\to\bu'=(u_1+\eta_0,u_2),\qquad \bv\to\bv'=(v_1+\eta_0,v_2),\qquad \bw\to\bw'=(w_1+\eta_0,w_2).
\end{align}
The shift of the form \eqref{parshift} may simply be regarded as a part of the change of variables given in \eqref{cov}, but it is also useful to consider its individual effect on the equations.  For the equations in Appendix \ref{app:FCCequations} which involve the Lagrangian functions $\lag$ and $\lam$, the components $u_1$, $v_1$, and $w_1$, only appear as a difference with one of $u_2$, $v_2$, or $w_2$.  Similarly, for the equations in Appendix \ref{app:FCCequations} which involve $\ol$ and $\olam$, the components $u_1$, $v_1$, and $w_1$, only appear as a difference with one of $u_1$, $v_1$, or $w_1$.  Then by the symmetries \eqref{covshift}, the shifts \eqref{parshift} effectively take $\lag\to\ol$ and $\lam\to\olam$ for each equation in Appendix \ref{app:FCCequations}, and leave $\ol$ and $\olam$ unchanged.  Thus applying this shift in combination with an appropriate change of variables \eqref{cov}, gives a slightly more convenient form of the equations which involve fewer Lagrangian functions. 

The expressions for the IRF Lagrangian functions $\lag_{\bu'\bv'}$ in \eqref{IRFlag1}, and $\lam_{\bu'\bv'}$ in \eqref{IRFlam1}, will be used with the change of variables \eqref{cov} and \eqref{parshift} to derive the face-centered quad equations of types A and B, respectively.  For the face-centered quad equation of type-C, the following IRF Lagrangian function  will be used
\begin{align}\label{ea2}
    \Xi_{\bu\bv}\!\left(\!x\Big|\,\arraycolsep=0.7pt\begin{array}{ll}x_a & x_b \\  x_c & x_d\end{array}\right)
    =\ol_{v_1-u_2}(x,x_a)+\lag_{v_2-u_2}(x,x_b)-\olam_{v_1-u_1}(x,x_c)-\lam_{v_2-u_1}(x,x_d).
\end{align}
The equation
\begin{align}
    \frac{\partial}{\partial x}\Xi_{\bu\bv}\!\left(\!x\Big|\,\arraycolsep=0.7pt\begin{array}{ll}x_a & x_b \\  x_c & x_d\end{array}\right)=0,
\end{align}
is equivalent to the type-C discrete Laplace-type equation \eqref{ea}, with the change of variables in the latter given by
\begin{align}
\begin{gathered}
x_a\to x,\quad x_i\to x_a,\quad x_j'\to x_c,\quad x_k'\to x_d, \\
w_1\to u_1,\quad v_1\to u_2,\quad u_1\to v_1.
\end{gathered}
\end{align}
Thus the expression \eqref{ea2} may be used to derive the face-centered quad equations of type-C.  For the second form of the classical Yang-Baxter equation \eqref{CASTR2}, \eqref{ea2} should be replaced by 
\begin{align}\label{ea3}
    \hat{\Xi}_{\bu\bv}\!\left(\!x\Big|\,\arraycolsep=0.7pt\begin{array}{ll}x_a & x_b \\  x_c & x_d\end{array}\right)
    =\ol_{v_1-u_2}(x_a,x)+\lag_{v_2-u_2}(x_b,x)-\olam_{v_1-u_1}(x_c,x)-\lam_{v_2-u_1}(x_d,x).
\end{align}
Note that with the appropriate change of variables any of the eight equations \eqref{ey}, \eqref{ez}, \eqref{ea}-\eqref{ef}, would lead to the same expression for a type-C face-centered quad equation.  

To summarise for the case of the classical Yang-Baxter equation \eqref{CASTR1},  using \eqref{cov} and \eqref{parshift} on  \eqref{IRFlag1} 
gives a face-centered quad equation of type-A as\footnote{For the degenerate additive type equations of Tables \ref{table-A} and \ref{table-BC2}, the exponential would not be used on the right hand sides of \eqref{teq1}--\eqref{teq3}.}
\begin{align}\label{teq1}
\A{y_e}{y_a}{y_b}{y_c}{y_d}{\ccpp}{\ccqq} = \exp\left\{\frac{\partial}{\partial x_e}\lag_{\bu'\bv'}\!\left(\!x_e\Big|\,\arraycolsep=0.7pt\begin{array}{ll}x_a & x_b \\  x_c & x_d\end{array}\right)\right\}\!,
\end{align}
where $ y_I=f(x_I)$, $(I=a,b,c,d,e)$.  Using \eqref{cov} and \eqref{parshift} on  \eqref{IRFlam1} gives a face-centered quad equation of type-B as
\begin{align}
\B{y_e}{y_a}{y_b}{y_c}{y_d}{\ccpp}{\ccqq} = \exp\left\{\frac{\partial}{\partial x_e}\lam_{\bu'\bv'}\!\left(\!x_e\Big|\,\arraycolsep=0.7pt\begin{array}{ll}x_a & x_b \\  x_c & x_d\end{array}\right)\right\}\!,
\end{align}
where $ y_I=f(x_I)$, $y_e=g(x_e)$, $(I=a,b,c,d)$.  Using \eqref{cov} and \eqref{parshift} on  \eqref{ea2} gives a face-centered quad equation of type-C as
\begin{align}\label{teq3}
\C{y_e}{y_a}{y_b}{y_c}{y_d}{\ccpp}{\ccqq} = \exp\left\{\frac{\partial}{\partial x_e}\Xi_{\bu'\bv'}\!\left(\!x_e\Big|\,\arraycolsep=0.7pt\begin{array}{ll}x_a & x_b \\  x_c & x_d\end{array}\right)\right\}\!,
\end{align}
where $ y_I=f(x_I)$, $(I=a,b,e)$, $y_J=g(x_J)$, $(J=c,d)$.  For each of \eqref{teq1}--\eqref{teq3}, the parameters are
\begin{align}
\alpha_1=h(u_1),\quad \alpha_2=h(u_2),\quad \beta_1=h(v_1),\quad \beta_2=h(v_2).
\end{align}


\subsection{CAFCC and the IRF YBE}\label{sec:MDC}

The face-centered quad equations which are derived using the above approach are found to satisfy the property of CAFCC given in Section \ref{sec:CAFCC}.  
The underlying reason for this is that the resulting face-centered quad equations are effectively a reinterpretation of the fourteen discrete Laplace-type equations of Appendix \ref{app:FCCequations}, which are implied by the classical IRF YBE \eqref{CAYBE1} that is satisfied by the Lagrangian functions.  Consequently, it is straightforward to see that CAFCC will be satisfied if the IRF YBE is satisfied, by comparing the six steps of CAFCC in Section \ref{sec:CAFCC} with the discrete Laplace-type equations of Appendix \ref{app:FCCequations}.

First consider the CAFCC algorithm of Section \ref{sec:CAFCC}, and note that the initial conditions along with the four equations centered at $\ccy$, $\ccya$, $\ccyb$, $\ccyc$, uniquely determine all of the variables that appear on the left hand side of the face-centered cube of Figure \ref{CAFCC5}. These variables are determined in steps 1 and 2 of CAFCC, and correspond to the variables that appear on the left hand side of the classical IRF YBE \eqref{CAYBE1}, up to the linearising change of variables \eqref{cov}.  Since the four equations centered at $\ccy$, $\ccya$, $\ccyb$, $\ccyc$, correspond to the four discrete Laplace-type equations in \eqref{ey1}-\eqref{ey}, the results of Section \ref{sec:CIRFYBEs} implies that there is a classical IRF YBE \eqref{CAYBE1} that is satisfied for the same (transformed) variables $\cca,\ccb,\ccc,\ccd,\cce,\ccf$, which have been determined through CAFCC.  

For the CAFCC algorithm of Section \ref{sec:CAFCC}, it remains to show that the face-centered quad equations are consistent in steps 3--6.
Consider first the two equations centered at $\ccd$ and $\cce$, either of which may be used to determine the variable $\ccza$ (step 3).  Let the different expressions for $\ccza$ resulting from these two equations be denoted by $\ccza^A$ and $\ccza^B$ respectively, and assume that $\ccza^A\neq\ccza^B$.  This implies that there are two solutions for the right hand side of the classical IRF YBE \eqref{CAYBE1}; one solution with $\ccza=\ccza^A$ and the other solution with $\ccza=\ccza^B$.  Now the classical IRF YBE with $\ccza=\ccza^A$ implies a set of fourteen discrete Laplace-type equations, for which the equation centered at $\cce$ is satisfied by $\ccza=\ccza^A$ (with the variable transformation \eqref{cov}).  However this same equation was determined through the CAFCC method to have the solution corresponding to $\ccza=\ccza^B$, where $\ccza^B\neq\ccza^A$, which is a contradiction.  Then it must be the case that $\ccza^A=\ccza^B$, and thus the two equations centered at $\ccd$ and $\cce$ are consistent.  Similar arguments may be used to show that the two equations centered at $\cca$ and $\ccb$ that are used to determine the variable $\cczc$ (step 4) must be consistent.

For step 5 of the CAFCC algorithm, any of the four face-centered quad equations centered at $\ccza,\cczb,\cczc,\ccc$, may next be used to determine the final variable $\ccz$. The consistency of these four equations can be shown using similar arguments to the above.  
Let any two distinct solutions of the above four equations be denoted by $\ccz^A$ and $\ccz^B$, and assume that $\ccz^A\neq\ccz^B$.  This means that there are at least two solutions for the right hand side of the classical IRF YBE \eqref{CAYBE1}; one solution with $\ccz=\ccz^A$, and the other solution with $\ccz=\ccz^B$.  This again leads to a contradiction where the discrete Laplace-type equation for the YBE which is solved with $\ccz^A$, has been determined through CAFCC to have a solution corresponding to $\ccz^B$, where $\ccz^B\neq\ccz^A$.  Thus it must be the case that $\ccz^A=\ccz^B$, and the two equations that were used to determine the two expressions for $\ccz$ must be consistent.  Consequently the four face-centered quad equations centered at $\ccza,\cczb,\cczc,\ccc$ are consistent.

The final step of the CAFCC algorithm (step 6) requires that the equation centered at $\ccz$ is automatically satisfied by the variables that were obtained in previous steps.  With the change of variables \eqref{cov}, this equation corresponds to the derivative of the classical IRF YBE given in \eqref{ez}, which must be satisfied since the variables that have been determined through CAFCC were seen above to correspond to the variables that satisfy the YBE.

\subsection{Explicit cases}\label{sec:FCElist}

The approach of Section \ref{sec:method} will be used here to derive the face-centered quad equations which satisfy CAFCC that are listed in Tables \ref{table-BC}--\ref{table-BC2}.  The starting point for each case is the expressions for the Lagrangian functions which satisfy the IRF YBEs of Section \ref{sec:CIRFYBEs}.  All face-centered quad equations can be derived by considering just the mixed cases of the Yang-Baxter equations \eqref{CAYBE1} and \eqref{CAYBE2}, together with an elliptic symmetric case for \eqref{CYBE1} that corresponds to $A4$ in Table \ref{table-A} (there is no known elliptic mixed case).

For a function $f(z)$, the notation $f(z_1\pm z_2)$ is used in this section to denote
\begin{align}
f(z_1\pm z_2)=f(z_1+z_2)+f(z_1-z_2).
\end{align}
Furthermore, the $\simeq$ sign will be used to indicate that the expressions for the IRF Lagrangian functions are given up to terms that are independent of the face variable $x$.  These terms can be disregarded since the face-centered quad equations are obtained from derivatives with respect to $x$.  Note that each equation is also labelled by a hypergeometric integral which has been previously identified with the star-triangle relations in \cite{Kels:2018xge}.

\subsubsection{Elliptic case}

Here $\wp(z)$ and $\sigma(z)$ denote the Weierstrass elliptic and sigma functions respectively, with associated elliptic invariants $g_2,g_3$, or half-periods $\omega_1,\omega_2$.  For \eqref{parshift}, $\eta_0=\frac{-\ii\pi\omega_2}{2\omega_1}$.  Also recall from \eqref{xdotdef} that the notation $\dmx$ denotes
\begin{align}
\dmx=4\mx^3-g_2\mx-g_3.
\end{align}

For $\im\bigl(\frac{\omega_2}{\omega_1}\bigr)>0$, the Lagrangian functions for the elliptic case can be written in terms of
\begin{align}
\phi(z)=\sum_{n=0}^\infty\Bigl(\lie\bigl(\EXP^{2\ii z}\EXP^{\pi\ii(2n+1)\omega_2/\omega_1}\bigr)-\lie\bigl(\EXP^{-2\ii z}\EXP^{\pi\ii(2n+1)\omega_2/\omega_1}\bigr)\!\Bigr),
\end{align}
where $\lie(z)$ is the dilogarithm function defined by
\begin{align}\label{dilogdef}
\lie(z)=-\int^z_0\frac{\Log(1-t)}{t}dt,\qquad z\in\mathbb{C}\setminus [1,\infty).
\end{align}


\paragraph{$A4$ (Elliptic beta integral) [Master solution model]}

The Lagrangian functions for the classical star-triangle relation \eqref{CSTR1} are given by \cite{Bazhanov:2010kz}
\begin{align}
\label{lagq4}
\begin{split}
\ds\mathcal{L}_{\alpha}(x_i,x_j)=&\frac{-\ii\omega_1\alpha\bigl(C(x_i)+C(x_j)\bigr)}{\pi\omega_2}  + \phi(\pm x_i\pm x_j+\ii\alpha), \\[0.1cm]
\ol_{\alpha}(x_i,x_j)=&\lag_{\eta_0-\alpha}(x_i,x_j)-\phi\bigl(\ii(\eta_0-2\alpha)\bigr),  \qquad C(x_i)=\frac{(\pi-4x_i)^2}{4}.
\end{split}
\end{align}
The saddle point equation \eqref{sym3leg1} for these Lagrangians is equivalent to $Q4$ from the ABS list.  In terms of \eqref{lagq4} the derivative with respect to the face variable of the IRF Lagrangian function \eqref{IRFlag1} is given by
\begin{align}\label{derq4}
\begin{split}
\frac{\partial}{\partial x}\lag_{\bu'\bv'}\!\left(\!x\Big|\,\arraycolsep=0.7pt\begin{array}{ll}x_a & x_b \\  x_c & x_d\end{array}\right)=
\varphi(x,x_a,u_2-v_1)+\varphi(x,x_d,u_1-v_2)\phantom{,} \\[-0.3cm] +\varphi(x,x_b,v_2-u_2)+\varphi(x,x_c,v_1-u_1),
\end{split}
\end{align}
where 
\begin{align}
\varphi(x_i,x_j,t)=\Log\frac{\sigma\bigl(\frac{2\omega_1}{\pi}(x_i-x_j+t)\bigr)\sigma\bigl(\frac{2\omega_1}{\pi}(x_i+x_j+t)\bigr)}
                            {\sigma\bigl(\frac{2\omega_1}{\pi}(x_i-x_j-t)\bigr)\sigma\bigl(\frac{2\omega_1}{\pi}(x_i+x_j-t)\bigr)}.
\end{align}

A linearising change of variables of the form \eqref{cov} is given by
\begin{align}\label{q4cov}
f(x)=\wp(2\omega_1 x\pi^{-1}),\qquad g(x)=\wp(2\omega_1 x\pi^{-1}),\qquad h(x)=\wp(2\omega_1 x\pi^{-1}).
\end{align}

Applying \eqref{q4cov} on \eqref{derq4} results in the type-A equation \eqref{4leg} where 
\begin{align}\label{dlq4}
a(x;y;\alpha,\beta)=
\frac{(G_{+}(x,\alpha,\beta)-F(x,y,\alpha,\beta))\bigl(\dmx(\mqc-\mpc)-\mx(\dmp+\dmq)+(\dmq\mpc+\dmp\mqc)\bigr)}
     {(G_{-}(x,\alpha,\beta)-F(x,y,\alpha,\beta))\bigl(\dmx(\mqc-\mpc)+\mx(\dmp+\dmq)-(\dmq\mpc+\dmp\mqc)\bigr)}.
\end{align}
The functions $F$ and $G_\pm$ were defined in \eqref{fgdef}.

There can be found other rather different looking (but equivalent) forms of \eqref{dlq4}, simply due to the wide variety of identities that are known for the Weierstrass (or Jacobi) elliptic functions.   It is likely that there could be found a simpler form of \eqref{dlq4}, particularly one that has at most quadratic dependence on the face variable $x$ similarly to the other equations, but at present a simpler expression is not known.

As was noted in Section \ref{sec:FCClist}, the CAC polynomials in the ABS list arise as the coefficient $P_1(x_a,x_b,x_c,x_d)$ of the linear term in $x$ in \eqref{quadseries}.  For the equation \eqref{4leg} with \eqref{dlq4}, $P_1(x_a,x_b,x_c,x_d)$ will correspond to $Q4$, although it appears difficult to see this explicitly due to the complicated form of the equations.  One way is to observe that the discriminants
\begin{align}\label{discdef}
r_k(x_k)=\left(\frac{\partial P^{ij}}{\partial x_l}\right)^2-2P^{ij}\frac{\partial^2 P^{ij}}{\partial x_l^2},
\end{align}
of $P_1(x_a,x_b,x_c,x_d)$, where $i,j,k,l$, are distinct elements of $\{a,b,c,d\}$, and $P^{ij}$ is the biquadratic polynomial
\begin{align}
P^{ij}=\frac{\partial P_1}{\partial x_i}\frac{\partial P_1}{\partial x_j}-P_1\frac{\partial^2 P_1}{\partial x_i\partial x_j},\qquad i,j=a,b,c,d,\quad i\neq j,
\end{align}
are quartic polynomials in $x_k$, for $k=a,b,c,d$. Then according to the classification result of \cite{ABS,ABS2}, this implies that $P_1(x_a,x_b,x_c,x_d)$ is M\"{o}bius equivalent to the polynomial for $Q4$ from the ABS list. 

\subsubsection{Hyperbolic cases}

The Lagrangian functions for the hyperbolic cases are given in terms of the dilogarithm \eqref{dilogdef}.  For \eqref{parshift}, $\eta_0=\pi$.   The notation $\overline{x}$ is used to denote
\begin{align}
\overline{x}=x+\sqrt{x^2-1}.
\end{align}


\paragraph{$A3_{(\delta=1)}$, $B3_{(\delta_1=\frac{1}{2};\,\delta_2=\frac{1}{2};\,\delta_3=0)}$, $C3_{(\delta_1=\frac{1}{2};\,\delta_2=\frac{1}{2};\,\delta_3=0)}$ (Hyperbolic Askey-Wilson integral and Hyperbolic Saalsch\"utz integral) case 1}

The Lagrangian functions for the classical star-triangle relations \eqref{CASTR1} and \eqref{CASTR2} are 
\begin{align}
\label{lagh3d1f}
\begin{split}
&\Lambda_{\alpha}(x_i,x_j)=\ds\lie(-\EXP^{x_i\pm x_j+\ii\alpha})+\tfrac{(x_i+\ii\alpha)^2+x_j^2}{2}, \\
&\ds\lag_{\alpha}(x_i,x_j)=\lie(-\EXP^{\pm (x_i-x_j)+\ii\alpha})-2\,\lie(-\EXP^{\ii\alpha})+\tfrac{(x_i-x_j)^2}{2}, \\
&\ds\lagh_{\alpha}(x_i,x_j)=\ds\lie(-\EXP^{\pm x_i\pm x_j+\ii\alpha})-2\,\lie(-\EXP^{\ii\alpha})+x_i^2+x_j^2-\tfrac{\alpha^2}{2}+\tfrac{\pi^2}{6}, \\
&\olam_{\alpha}(x_i,x_j)=\Lambda_{\pi-\alpha}(-x_i,x_j),\;\; \ol_\alpha(x_i,x_j)=\lag_{\pi-\alpha}(x_i,x_j),\;\; \olh_\alpha(x_i,x_j)=\lagh_{\pi-\alpha}(x_i,x_j).
\end{split}
\end{align}
The saddle point equations \eqref{asym3leg1} and \eqref{asym3leg2} are both equivalent to $H3_{(\delta=1;\,\varepsilon=1)}$ from the ABS list.  In terms of \eqref{lagh3d1f} the IRF Lagrangian functions \eqref{IRFlag2}, \eqref{IRFlam2}, and \eqref{ea3}, are respectively given by
\begin{align}
\begin{split}\label{IRFq3d1}
\lagh_{\bu'\bv'}\!\left(\!x\Big|\,\arraycolsep=0.7pt\begin{array}{ll}x_a & x_b \\  x_c & x_d\end{array}\right)
\simeq\, 4x^2  +\lie(\EXP^{\ii(u_2-v_1)\pm x\pm x_a}) + \lie(\EXP^{\ii(v_2-u_2)\pm x\pm x_b})\phantom{,} \\[-0.25cm]
\phantom{4x^2,}  +\lie(\EXP^{\ii(v_1-u_1)\pm x\pm x_c}) + \lie(\EXP^{\ii(u_1-v_2)\pm x\pm x_d}),
\end{split} \\
\label{IRFh3d11b}
\begin{split}
\lamh_{\bu'\bv'}\!\left(\!x\Big|\,\arraycolsep=0.7pt\begin{array}{ll}x_a & x_b \\  x_c & x_d\end{array}\right)
\simeq\, 2\ii(u_1+u_2-v_1-v_2)x   
  +\lie(\EXP^{\ii(u_2-v_1) + x\pm x_a})  + \lie(\EXP^{\ii(v_2-u_2) - x\pm x_b})\phantom{,}  \\[-0.25cm]
 + 2x(x-\ii\pi) +\lie(\EXP^{\ii(v_1-u_1) - x\pm x_c}) + \lie(\EXP^{\ii(u_1-v_2) + x\pm x_d}),
\end{split} \\
\label{IRFh3d11c}
\begin{split}
\hat{\Xi}_{\bu'\bv'}\!\left(\!x\Big|\,\arraycolsep=0.7pt\begin{array}{ll}x_a & x_b \\  x_c & x_d\end{array}\right)
\simeq - x^2  - \lie(\EXP^{\ii (u_2-v_1) \pm x \pm x_a}) - \lie(\EXP^{\ii (u_2-v_2) \pm x \pm x_b})\phantom{,} \\[-0.25cm]
\phantom{-x^2,} - \lie(\EXP^{\ii (u_1-v_1) \pm x -x_c}) - \lie(\EXP^{\ii (v_2-u_1) \pm x + x_d}).
\end{split}
\end{align}

A linearising change of variables of the form \eqref{cov} is given by
\begin{align}\label{covh31}
f(x)=\cosh(x),\qquad g(x)=\EXP^{x},\qquad h(x)=\EXP^{\ii x}.
\end{align}

Applying \eqref{covh31} on the derivative with respect to $x$ of \eqref{IRFq3d1} results in the type-A equation \eqref{4leg} where 
\begin{align}\label{dlq31}
a(x;y;\alpha,\beta)=
\frac{\alpha^2+\beta^2\overline{x}^2-2\alpha\beta \overline{x}y}
     {\beta^2+\alpha^2\overline{x}^2-2\alpha\beta \overline{x}y}.
\end{align}

Applying \eqref{covh31} on the derivatives with respect to $x$ of \eqref{IRFh3d11b} and \eqref{IRFh3d11c} results in the type-B and -C equations \eqref{4legb} and \eqref{4legc} respectively, where 
\begin{align}\label{dlh311}
b(x;y;\alpha,\beta)=\beta^2+\alpha^2x^2-2\alpha\beta xy, \qquad
c(x;y;\alpha,\beta)=\frac{\alpha-\beta\overline{x}y}{\alpha\overline{x}-\beta y}.
\end{align}

The coefficient of $x$ in the affine-linear form of the type-A equation \eqref{4leg} with \eqref{dlq31}, is given by \eqref{q34} with $\delta=1$.  Setting $x_b\to-x_b$, $\alpha_1\to\beta$, $\beta_1\to\alpha$, $\alpha_2\to1$, and $\beta_2\to1$, in \eqref{q34} with $\delta=1$ gives
\begin{align}
\begin{split}
4\beta(\alpha^2-1)(x_ax_b + x_cx_d) - 4\alpha(\beta^2-1)(x_ax_c + x_bx_d) - 4(\alpha^2-\beta^2)(x_ax_d + x_bx_c)\phantom{.} \\ -(\alpha^2-1)(\beta^2-1)(\alpha\beta^{-1}-\beta\alpha^{-1}). 
\end{split}
\end{align}
This is a polynomial for $Q3_{(\delta=1)}$ from the ABS list.

The coefficient of $x$ in the affine-linear form of the type-C equation \eqref{4legc} with $a(x;y;\alpha,\beta)$ from \eqref{dlq31} and $c(x;y;\alpha,\beta)$ from \eqref{dlh311}, is given by \eqref{h34} with $(\delta_1,\delta_2,\delta_3)=(\frac{1}{2},\frac{1}{2},0)$.  Setting $x_a\leftrightarrow x_c$, $x_b\to-x_b$, $\alpha_1\to1$, $\beta_1\to\beta$, $\alpha_2\to\alpha$, and $\beta_2\to\beta$, in \eqref{h34} with $(\delta_1,\delta_2,\delta_3)=(\frac{1}{2},\frac{1}{2},0)$ gives
\begin{align}
\begin{split}
 2\alpha(x_ax_c + x_bx_d)-2\beta(x_ax_b + x_cx_d) - (\alpha\beta^{-1} - \beta\alpha^{-1})(1 + \alpha\beta x_ax_d).
\end{split}
\end{align}
This is a polynomial for $H3_{(\delta=1;\,\varepsilon=1)}$ from the ABS list.

\paragraph{$A3_{(\delta=0)}$, $B3_{(\delta_1=\frac{1}{2};\,\delta_2=0;\,\delta_3=\frac{1}{2})}$, $C3_{(\delta_1=\frac{1}{2};\,\delta_2=0;\,\delta_3=\frac{1}{2})}$ (Hyperbolic Askey-Wilson integral and Hyperbolic Saalsch\"utz integral) case 2}

For the Lagrangian functions \eqref{lagh3d1f} there is the second set of IRF Lagrangian functions  \eqref{IRFlag1}, \eqref{IRFlam1}, and \eqref{ea2}, that are respectively given by
\begin{align}\label{irflagq30}
\begin{split}
\lag_{\bu'\bv'}\!\left(\!x\Big|\,\arraycolsep=0.7pt\begin{array}{ll}x_a & x_b \\  x_c & x_d\end{array}\right)
\simeq\, \frac{(x - x_a)^2 + (x - x_b)^2 + (x - x_c)^2 + (x - x_d)^2}{2}\phantom{,}  \\[-0.25cm]
 + \lie(\EXP^{\ii(u_2-v_1)\pm(x-x_a)})  + \lie(\EXP^{\ii(v_2-u_2)\pm(x-x_b)})\phantom{,} \\
  + \lie(\EXP^{\ii(v_1-u_1)\pm(x-x_c)}) + \lie(\EXP^{\ii(u_1-v_2)\pm(x-x_d)}), 
\end{split} \\
\label{IRFh3d12b}
\begin{split}
\lam_{\bu'\bv'}\!\left(\!x\Big|\,\arraycolsep=0.7pt\begin{array}{ll}x_a & x_b \\  x_c & x_d\end{array}\right)
\simeq\, 2x^2  + \lie(\EXP^{\ii(u_2-v_1)\pm x+x_a}) + \lie(\EXP^{\ii(v_2-u_2)\pm x-x_b})\phantom{,} \\[-0.25cm]
\phantom{2x^2,}  + \lie(\EXP^{\ii(v_1-u_1)\pm x-x_c}) + \lie(\EXP^{\ii(u_1-v_2)\pm x+x_d}),
\end{split} \\
\label{IRFh3d12c}
\begin{split}
\Xi_{\bu'\bv'}\!\left(\!x\Big|\,\arraycolsep=0.7pt\begin{array}{ll}x_a & x_b \\  x_c & x_d\end{array}\right)
\simeq \, \ii(2(\pi-u_1)- v_1-v_2)x 
 + \lie(\EXP^{\ii (u_2-v_1) \pm (x-x_a)}) -\lie(\EXP^{\ii (u_2-v_2) \pm (x-x_b)})\phantom{,} \\[-0.25cm]
- ( x + x_a - x_b)x  -\lie(\EXP^{\ii (u_1-v_1) \pm x_c -x}) - \lie(\EXP^{\ii (v_2-u_1) \pm x_d + x}).
\end{split}
\end{align}

A linearising change of variables of the form \eqref{cov} is given by
\begin{align}\label{covh12}
f(x)=\EXP^{x},\qquad g(x)=\cosh(x),\qquad h(x)=\EXP^{\ii x}.
\end{align}

Applying \eqref{covh12} on the derivative with respect to $x$ of \eqref{irflagq30} results in the type-A equation \eqref{4leg} where 
\begin{align}\label{dlq30}
a(x;y;\alpha,\beta)=\frac{\beta x-\alpha y}{\alpha x-\beta y}.
\end{align}

Applying \eqref{covh12} on the derivatives with respect to $x$ of \eqref{IRFh3d12b} and \eqref{IRFh3d12c} results in the type-B and -C equations \eqref{4legb} and \eqref{4legc} respectively, where  
\begin{align}\label{dlh3112}
b(x;y;\alpha,\beta)=\frac{\alpha y-\beta\overline{x}}{\alpha\overline{x}y-\beta}, \qquad
c(x;y;\alpha,\beta)=\beta^{-1}\bigl(\alpha^2+\beta^2x^2-2\alpha\beta x y\bigr).
\end{align}

The coefficient of $x$ in the affine-linear form of the type-A equation \eqref{4leg} with \eqref{dlq30}, is given by \eqref{q34} with $\delta=0$.  Setting $x_b\to-x_b$, $\alpha_1\to\beta$, $\beta_1\to\alpha$, $\alpha_2\to1$,  and $\beta_2\to1$, in \eqref{q34} with $\delta=0$ gives
\begin{align}
\begin{split}
\beta(\alpha^2-1)(x_ax_b + x_cx_d) - \alpha(\beta^2-1)(x_ax_c + x_bx_d) - (\alpha^2-\beta^2)(x_ax_d + x_bx_c).
\end{split}
\end{align}
This is a polynomial for $Q3_{(\delta=0)}$ from the ABS list.

The coefficient of $x$ in the affine-linear form of the type-C equation \eqref{4legc} with $a(x;y;\alpha,\beta)$ from \eqref{dlq30} and $c(x;y;\alpha,\beta)$ from \eqref{dlh3112}, is given by \eqref{h34} with $(\delta_1,\delta_2,\delta_3)=(\frac{1}{2},0,\frac{1}{2})$.  Setting $x_a\leftrightarrow x_c$, $x_b\to-x_b$, $\alpha_1\to1$, $\beta_1\to\beta$, $\alpha_2\to\alpha$, and $\beta_2\to\beta$, in \eqref{h34} with $(\delta_1,\delta_2,\delta_3)=(\frac{1}{2},0,\frac{1}{2})$ gives
\begin{align}
\begin{split}
2\alpha(x_ax_c + x_bx_d)-2\beta(x_ax_b + x_cx_d) - (\alpha\beta^{-1} - \beta\alpha^{-1})(1 + \alpha\beta x_bx_c).
\end{split}
\end{align}
This is a polynomial for $H3_{(\delta=1;\,\varepsilon=1)}$ from the ABS list.

\paragraph{$A3_{(\delta=0)}$, $B3_{(\delta_1=1;\,\delta_2=0;\,\delta_3=0)}$, $C3_{(\delta_1=1;\,\delta_2=0;\,\delta_3=0)}$ (Hyperbolic Barnes's 1st lemma)}

The Lagrangian functions for the classical star-triangle relation \eqref{CASTR} are given by
\begin{align}
\label{lagh3d0}
\begin{split}
\ds\lag_{\alpha}(x_i,x_j)&=\ds\lie(-\EXP^{x_i-x_j+\ii\alpha})+\lie(-\EXP^{x_j-x_i+\ii\alpha})-2\,\lie(-\EXP^{\ii\alpha})+\tfrac{(x_i-x_j)^2}{2}, \\[0.1cm]
\Lambda_{\alpha}(x_i,x_j)&=\ds \lie(-\EXP^{x_i+x_j+\ii\alpha})+\ii(x_i+x_j)\alpha+x_i^2+x_j^2-\tfrac{\alpha^2}{2}+\tfrac{\pi^2}{6},  \\[0.1cm]
\ol_{\alpha}(x_i,x_j)&=\lag_{\pi-\alpha}(x_i,x_j),\qquad \olam_{\alpha}(x_i,x_j)=-\Lambda_{-\pi+\alpha}(x_i,x_j).
\end{split}
\end{align}
The saddle point equation \eqref{mix3leg} for these Lagrangians are equivalent to $H3_{(\delta=0,1;\,\varepsilon=1-\delta)}$ from the ABS list.  In terms of \eqref{lagh3d0} the IRF Lagrangian functions \eqref{IRFlag1}, \eqref{IRFlam1}, and \eqref{ea2}, are respectively given by \eqref{irflagq30} and
\begin{align}\label{irfh3d0b}
\begin{split}
\lam_{\bu'\bv'}\!\left(\!x\Big|\,\arraycolsep=0.7pt\begin{array}{ll}x_a & x_b \\  x_c & x_d\end{array}\right)
\simeq\, 2\pi\ii x + \lie(\EXP^{\ii(u_2-v_1)+x+x_a}) - \lie(\EXP^{\ii(u_2-v_2)+x+x_b})\phantom{,} \\[-0.25cm]
 \phantom{ 2\pi\ii x,} - \lie(\EXP^{\ii(u_1-v_1)+x+x_c}) + \lie(\EXP^{\ii(u_1-v_2)+x+x_d}),
\end{split} \\
\label{irfh3d0c}
\begin{split}
\Xi_{\bu'\bv'}\!\left(\!x\Big|\,\arraycolsep=0.7pt\begin{array}{ll}x_a & x_b \\  x_c & x_d\end{array}\right)
\simeq\, \frac{(x_a - x)^2- (x_b - x)^2}{2} - \ii (v_2(x_d+x) -v_1(x_c+x))\phantom{,} \\[-0.15cm]
   + \lie(\EXP^{\ii (u_2 - v_1) \pm ( x - x_a)}) - \lie(\EXP^{\ii (u_2 - v_2) \pm (x - x_b)})\phantom{,}  \\
 +\lie(\EXP^{\ii (v_1 - u_1) + x + x_c})  - \lie(\EXP^{\ii (v_2 - u_1) + x + x_d}).
\end{split}
\end{align}

A linearising change of variables of the form \eqref{cov} is given by
\begin{align}\label{covh10}
f(x)=\EXP^{x},\qquad g(x)=\EXP^{x},\qquad h(x)=\EXP^{\ii x}.
\end{align}

Applying \eqref{covh10} on the derivatives with respect to $x$ of \eqref{irfh3d0b} and \eqref{irfh3d0c} results in the type-B and -C equations \eqref{4legb} and \eqref{4legc} respectively, where $a(x;y;\alpha,\beta)$ is given by \eqref{dlq30}, and
\begin{align}\label{dlh310}
b(x;y;\alpha,\beta)=\beta-\alpha xy, \qquad
c(x;y;\alpha,\beta)=xy-\tfrac{\alpha}{\beta}.
\end{align}

The coefficient of $x$ in the affine-linear form of the type-C equation \eqref{4legc} with $a(x;y;\alpha,\beta)$ from \eqref{dlq30} and $c(x;y;\alpha,\beta)$ from \eqref{dlh310}, is given by \eqref{h34} with $(\delta_1,\delta_2,\delta_3)=(1,0,0)$.  Setting $x_a\leftrightarrow x_c$, $x_b\to-x_b$, $\alpha_1\to1$, $\beta_1\to\beta$, $\alpha_2\to\alpha$, and $\beta_2\to\beta$, in \eqref{h34} with $(\delta_1,\delta_2,\delta_3)=(1,0,0)$ gives
\begin{align}
\begin{split}
\alpha^2\beta(x_ax_c + x_bx_d) - \alpha\beta^2(x_ax_b + x_cx_d) + \beta^2-\alpha^2.
\end{split}
\end{align}
This is a polynomial for $H3_{(\delta=0,1;\,\varepsilon=1-\delta)}$ from the ABS list.

\paragraph{$A3_{(\delta=0)}$, $B3_{(\delta_1=0;\,\delta_2=0;\,\delta_3=0)}$, $C3_{(\delta_1=0;\,\delta_2=0;\,\delta_3=0)}$ (Hyperbolic Barnes's $_2F_1$ integral)}

The Lagrangian functions for the classical star-triangle relation \eqref{CASTR} are given by
\begin{align}
\label{lagh3d02}
\begin{split}
\ds\lag_{\alpha}(x_i,x_j)&=\ds \lie(-\EXP^{x_i-x_j+\ii\alpha})+\lie(-\EXP^{x_j-x_i+\ii\alpha})-2\lie(-\EXP^{\ii\alpha})+\tfrac{(x_i-x_j)^2}{2}, \\[0.1cm]
\ol_{\alpha}(x_i,x_j)&=\lag_{\pi-\alpha}(x_i,x_j), \qquad
\Lambda(x_i,x_j)=\ds-x_ix_j, \qquad
\olam(x_i,x_j)=-\Lambda(x_i,x_j). 
\end{split}
\end{align}
The saddle point equation \eqref{mix3leg} for these Lagrangians are equivalent to $H3_{(\delta=0;\,\varepsilon=0)}$ from the ABS list.  In terms of \eqref{lagh3d02} the IRF Lagrangian functions \eqref{IRFlag1}, \eqref{IRFlam1}, and \eqref{ea2}, are respectively given by \eqref{irflagq30} and
\begin{align}\label{irfh3d02b}
\begin{split}
 \text{\scalebox{0.99}{$\lam_{\bu'\bv'}\!\left(\!x\Big|\,\arraycolsep=0.7pt\begin{array}{ll}x_a & x_b \\  x_c & x_d\end{array}\right)$}}
& \text{\scalebox{0.99}{$= x(x_b + x_c - x_a - x_d),$}}
\end{split}
\\
\label{irfh3d02c}
\begin{split}
 \text{\scalebox{0.99}{$\Xi_{\bu'\bv'}\!\left(\!x\Big|\,\arraycolsep=0.7pt\begin{array}{ll}x_a & x_b \\  x_c & x_d\end{array}\right)$}}
& \text{\scalebox{0.99}{$\simeq x(x_b+x_c-x_a-x_d)
+ \lie(\EXP^{\ii (u_2 -v_1) \pm (x - x_a)}) - \lie(\EXP^{\ii (u_2 - v_2) \pm (x - x_b)}) .$}}
\end{split}
\end{align}

A linearising change of variables of the form \eqref{cov} is given by
\begin{align}\label{covh02}
f(x)=\EXP^{x},\qquad g(x)=\EXP^{x},\qquad h(x)=\EXP^{\ii x}.
\end{align}

Applying \eqref{covh02} on the derivatives with respect to $x$ of \eqref{irfh3d02b} and \eqref{irfh3d02c} results in the type-B and -C equations \eqref{4legb}, and \eqref{4legc} respectively, where $a(x;y;\alpha,\beta)$ is given by  \eqref{dlq30}, and
\begin{align}\label{dlh30}
b(x;y;\alpha,\beta)=y, \qquad
c(x;y;\alpha,\beta)=y.
\end{align}

The coefficient of $x$ in the affine-linear form of the type-C equation \eqref{4legc} with $a(x;y;\alpha,\beta)$ from \eqref{dlq30} and $c(x;y;\alpha,\beta)$ from \eqref{dlh30}, is given by \eqref{h34} with $(\delta_1,\delta_2,\delta_3)=(0,0,0)$.  Setting $x_a\leftrightarrow x_c$, $x_b\to-x_b$, $\alpha_1\to1$, $\beta_1\to\beta$, $\alpha_2\to\alpha$, and $\beta_2\to\beta$, in \eqref{h34} with $(\delta_1,\delta_2,\delta_3)=(0,0,0)$ gives
\begin{align}
\begin{split}
\alpha(x_ax_c + x_bx_d) - \beta(x_ax_b + x_cx_d).
\end{split}
\end{align}
This is a polynomial for $H3_{(\delta=0;\,\varepsilon=0)}$ from the ABS list.

\subsubsection{Rational cases}

The edge Lagrangian functions for the rational cases have appeared in the examples that were given in Section \ref{sec:YBE}.  Recall that these Lagrangian functions are defined in terms of the complex logarithm through
\begin{align}\label{gamdef}
\gamma(z)=\ii z\Log(\ii z),\qquad \ii z\in \mathbb{C}\setminus (-\infty,0].
\end{align}
Since $\eta=0$ for the rational cases, there is no shift of the form \eqref{parshift} that is applied to the parameters $\bu,\bv,\bw$.

\paragraph{$A2_{(\delta_1=1;\,\delta_2=1)}$, $B2_{(\delta_1=1;\,\delta_2=1;\,\delta_3=0)}$, $C2_{(\delta_1=1;\,\delta_2=1;\,\delta_3=0)}$ (De Branges-Wilson Integral and Barnes's 2nd lemma) case 1}


The Lagrangian functions for the classical star-triangle relations \eqref{CASTR1} and \eqref{CASTR2} were given in \eqref{lagex1af}, \eqref{lagex1bf}, and \eqref{lagex3a}, as
\begin{align}
\label{lagh2d1}
\begin{array}{rl}
\Lambda_{\alpha}(x_i,x_j)=\,\ds \gamma(x_i+x_j+\ii\alpha)+\gamma(x_i-x_j+\ii\alpha),& \quad \ds
\olam_{\alpha}(x_i,x_j)=\Lambda_{-\alpha}(-x_i,x_j), \\[0.1cm]
\lag_{\alpha}(x_i,x_j)=\,\ds \gamma(x_i-x_j+\ii\alpha)-\gamma(x_i-x_j-\ii\alpha),& \quad \ds \ol_{\alpha}(x_i,x_j)=\lag_{-\alpha}(x_i,x_j)-\gamma(-2\ii\alpha), \\[0.1cm]
\lagh_{\alpha}(x_i,x_j)=\,\ds \gamma(x_i\pm x_j+\ii\alpha)-\gamma(x_i\pm x_j-\ii\alpha),& \quad \ds 
\olh_{\alpha}(x_i,x_j)=\ds \lagh_{-\alpha}(x_i,x_j)-\gamma(-2\ii\alpha).
\end{array}
\end{align}
The saddle point equations \eqref{asym3leg1} and \eqref{asym3leg2}  for these Lagrangians are both equivalent to $H2_{(\varepsilon=1)}$ from the ABS list.  In terms of \eqref{lagh2d1} the IRF Lagrangian functions \eqref{IRFlag2}, \eqref{IRFlam2}, and \eqref{ea3}, are respectively given by
\begin{align}
\begin{split}\label{irflagq2}
\lagh_{\bu\bv}\!\left(\!x\Big|\,\arraycolsep=0.7pt\begin{array}{ll}x_a & x_b \\  x_c & x_d\end{array}\right)
=\gamma(\ii(u_2-v_1)\pm x+x_a)-\gamma(\ii(v_1-u_2)\pm x+x_a)\phantom{,} \\[-0.25cm]
+\gamma (\ii(v_2-u_2)+x\pm x_b)-\gamma(\ii(u_2-v_2)+x\pm x_b)\phantom{,} \\ +\gamma(\ii(v_1-u_1)\pm x+x_c) 
-\gamma (\ii(u_1-v_1)\pm x+x_c)\phantom{,} \\ +\gamma(\ii(u_1-v_2)+x\pm x_d)-\gamma(\ii(v_2-u_1)+x\pm x_d), 
\end{split} \\
\label{IRFh2d11b}
\begin{split}
\lamh_{\bu\bv}\!\left(\!x\Big|\,\arraycolsep=0.7pt\begin{array}{ll}x_a & x_b \\  x_c & x_d\end{array}\right)
= \gamma(\ii(u_2-v_1) +x\pm x_a)+\gamma(\ii(v_2-u_2) -x\pm x_b)\phantom{,} \\[-0.25cm]
 +\gamma(\ii(v_1-u_1) -x\pm x_c)+\gamma(\ii(u_1-v_2) +x\pm x_d),
\end{split} 
\\
\label{IRFh2d11c}
\begin{split}
\hat{\Xi}_{\bu\bv}\!\left(\!x\Big|\,\arraycolsep=0.7pt\begin{array}{ll}x_a & x_b \\  x_c & x_d\end{array}\right)
=\gamma(\ii(u_2-v_1)+x\pm x_a)  -\gamma(\ii(v_1-u_2)+x\pm x_a)\phantom{,}  \\[-0.25cm]
+ \gamma(\ii(v_2-u_2)+x\pm x_b) - \gamma(\ii(u_2-v_2)+x\pm x_b)\phantom{,} \\
- \gamma(\ii(u_1-v_1)-x_c\pm x) - \gamma(\ii(v_2-u_1)+x_d\pm x).
\end{split}
\end{align}

A linearising change of variables of the form \eqref{cov} is given by
\begin{align}\label{covq2}
f(x)=x^2,\qquad g(x)=x,\qquad h(x)=\ii x.
\end{align}

Applying \eqref{covq2} on the derivative with respect to $x$ of \eqref{irflagq2}  results in the type-A equation \eqref{4leg} where 
\begin{align}\label{dlq2}
a(x;y;\alpha,\beta)=\frac{(\sqrt{x}+\alpha-\beta)^2-y}{(\sqrt{x}-\alpha+\beta)^2-y}.
\end{align}

Applying \eqref{covq2} on the derivatives with respect to $x$ of \eqref{IRFh2d11b} and \eqref{IRFh2d11c} results in the type-B and -C equations \eqref{4legb} and \eqref{4legc} respectively, where
\begin{align}\label{dlh21}
b(x;y;\alpha,\beta)=(x+\alpha-\beta)^2-y, \qquad
c(x;y;\alpha,\beta)=\frac{y-\sqrt{x}-\alpha+\beta}{y+\sqrt{x}-\alpha+\beta}.
\end{align}

The coefficient of $x$ in the affine-linear form of the type-A equation \eqref{4leg} with $a(x;y;\alpha,\beta)$ from \eqref{dlq2}, is given by \eqref{q24} with $(\delta_1,\delta_2)=(1,1)$.  Setting $x_b\to-x_b$, $\alpha_1\to\beta$, $\beta_1\to\alpha$, $\alpha_2\to0$, and $\beta_2\to0$, in \eqref{q24} with $(\delta_1,\delta_2)=(1,1)$ gives
\begin{align}
\begin{split}
  \alpha(x_a - x_c)(x_b - x_d) - \beta(x_a - x_b)(x_c - x_d)  -  \alpha\beta(\alpha-\beta)(\!\!\sum_{I\in\{a,b,c,d\}}\!\!x_I - \alpha^2 + \alpha\beta - \beta^2). 
\end{split}
\end{align}
This is a polynomial for $Q2$ from the ABS list.

The coefficient of $x$ in the affine-linear form of the type-C equation \eqref{4legc} with $a(x;y;\alpha,\beta)$ from \eqref{dlq2} and $c(x;y;\alpha,\beta)$ from \eqref{dlh21}, is given by \eqref{h24} with $(\delta_1,\delta_2,\delta_3)=(1,1,0)$.  Setting $x_a\leftrightarrow x_c$, $x_b\to-x_b$, $\alpha_1\to0$, $\beta_1\to\beta$, $\alpha_2\to\alpha$, and $\beta_2\to\beta$, in \eqref{h24} with $(\delta_1,\delta_2,\delta_3)=(1,1,0)$ gives
\begin{align}
\begin{split}
(x_a - x_d)(x_b - x_c) + (\alpha^2-\beta^2)(x_a + x_d) - (\alpha-\beta)(x_b + x_c - 2x_ax_d -\alpha^2-\beta^2) .
\end{split}
\end{align}
This is a polynomial for $H2_{(\varepsilon=1)}$ from the ABS list.

\paragraph{$A2_{(\delta_1=1;\,\delta_2=0)}$, $B2_{(\delta_1=1;\,\delta_2=0;\,\delta_3=1)}$, $C2_{(\delta_1=1;\,\delta_2=0;\,\delta_3=1)}$ (De Branges-Wilson Integral and Barnes's 2nd lemma) case 2}

For the Lagrangian functions \eqref{lagh2d1} there is the second set of IRF Lagrangian functions  \eqref{IRFlag1}, \eqref{IRFlam1}, and \eqref{ea2}, that are respectively given by
\begin{align}\label{irflagq11}
\begin{split}
\lag_{\bu\bv}\!\left(\!x\Big|\,\arraycolsep=0.7pt\begin{array}{ll}x_a & x_b \\  x_c & x_d\end{array}\right)
= \gamma(\ii(u_2-v_1)-x+x_a)-\gamma(\ii(v_1-u_2)-x+x_a)\phantom{,} \\[-0.25cm]
 -\gamma(\ii(u_2-v_2)-x+x_b)+\gamma(\ii(v_2-u_2)-x+x_b)\phantom{,} \\ -\gamma(\ii(u_1-v_1)-x+x_c) 
 +\gamma(\ii(v_1-u_1)-x+x_c)\phantom{,} \\ +\gamma(\ii(u_1-v_2)-x+x_d)-\gamma(\ii(v_2-u_1)-x+x_d),
\end{split}
\\
\label{IRFh2d12b}
\begin{split}
\lam_{\bu\bv}\!\left(\!x\Big|\,\arraycolsep=0.7pt\begin{array}{ll}x_a & x_b \\  x_c & x_d\end{array}\right)
= \gamma(\ii(u_2-v_1)\pm x+x_a)+\gamma(\ii(v_2-u_2)\pm x-x_b)\phantom{,} \\[-0.25cm]
 +\gamma(\ii(v_1-u_1)\pm x-x_c)+\gamma(\ii(u_1-v_2)\pm x+x_d),
\end{split}
\\
\label{IRFh2d12c}
\begin{split}
\Xi_{\bu\bv}\!\left(\!x\Big|\,\arraycolsep=0.7pt\begin{array}{ll}x_a & x_b \\  x_c & x_d\end{array}\right)
 = \gamma(\ii(u_2-v_1)+x-x_a)  - \gamma(\ii(v_1-u_2)+x-x_a)\phantom{,} \\[-0.25cm]
+ \gamma(\ii(v_2-u_2)+x-x_b) - \gamma(\ii(u_2-v_2)+x-x_b)\phantom{,}  \\
- \gamma(\ii(u_1-v_1)-x\pm x_c) -\gamma(\ii(v_2-u_1)+x\pm x_d).
\end{split}
\end{align}

A linearising change of variables of the form \eqref{cov} is given by
\begin{align}\label{covq11}
f(x)=x,\qquad g(x)=x^2,\qquad h(x)=\ii x.
\end{align}

Applying \eqref{covq11} on the derivative with respect to $x$ of \eqref{irflagq11}  results in the type-A equation \eqref{4leg} where 
\begin{align}\label{dlq11}
a(x;y;\alpha,\beta)=\frac{y-x+\alpha-\beta}{x-y+\alpha-\beta}.
\end{align}

Applying \eqref{covq11} on the derivatives with respect to $x$ of \eqref{IRFh2d12b} and \eqref{IRFh2d12c} results in the type-B and -C equations \eqref{4legb} and \eqref{4legc} respectively, where
\begin{align}\label{dlh212}
b(x;y;\alpha,\beta)=\frac{y+\sqrt{x}+\alpha-\beta}{y-\sqrt{x}+\alpha-\beta}, \qquad
c(x;y;\alpha,\beta)=(x-\alpha+\beta)^2-y.
\end{align}

The coefficient of $x$ in the affine-linear form of the type-A equation \eqref{4leg} with $a(x;y;\alpha,\beta)$ from \eqref{dlq11}, is given by \eqref{q24} with $(\delta_1,\delta_2)=(1,0)$.  Setting $x_b\to-x_b$, $\alpha_1\to\beta$, $\beta_1\to\alpha$, $\alpha_2\to0$, and $\beta_2\to0$, in \eqref{q24} with $(\delta_1,\delta_2)=(1,0)$ gives
\begin{align}
\begin{split}
\alpha(x_a - x_c)(x_b - x_d) - \beta(x_a - x_b)(x_c - x_d) - \alpha\beta(\alpha - \beta).
\end{split}
\end{align}
This is a polynomial for $Q1_{(\delta=1)}$ from the ABS list \cite{ABS}.

The coefficient of $x$ in the affine-linear form of the type-C equation \eqref{4legc} with $a(x;y;\alpha,\beta)$ from \eqref{dlq11} and $c(x;y;\alpha,\beta)$ from \eqref{dlh212}, is given by \eqref{h24} with $(\delta_1,\delta_2,\delta_3)=(1,0,1)$.  Setting $x_a\leftrightarrow x_c$, $x_b\to-x_b$, $\alpha_1\to0$, $\beta_1\to\beta$, $\alpha_2\to\alpha$, and $\beta_2\to\beta$, in \eqref{h24} with $(\delta_1,\delta_2,\delta_3)=(1,0,1)$ gives
\begin{align}
\begin{split}
(x_a - x_d)(x_b - x_c)  + (\alpha^2 - \beta^2)(x_b + x_c) - (\alpha - \beta)(x_a + x_d - 2 x_bx_c - \alpha^2-\beta^2) .
\end{split}
\end{align}
This is a polynomial for $H2_{(\varepsilon=1)}$ from the ABS list.

\paragraph{$A2_{(\delta_1=1;\,\delta_2=0)}$, $B2_{(\delta_1=1;\,\delta_2=0;\,\delta_3=0)}$, $C2_{(\delta_1=1;\,\delta_2=0;\,\delta_3=0)}$ (Barnes's 1st lemma)}


The Lagrangian functions for the classical star-triangle relation \eqref{CASTR} were given in \eqref{lagex1bf} and \eqref{lagex2} as
\begin{align}
\label{lagh2d0}
\begin{array}{rll}
&\lag_{\alpha}(x_i,x_j)=\gamma(x_i-x_j+\ii\alpha)-\gamma(x_i-x_j-\ii\alpha),  \; &\Lambda_{\alpha}(x_i,x_j)=\gamma(x_i+x_j+\ii\alpha), \\[0.1cm]
&\ol_{\alpha}(x_i,x_j)=\lag_{-\alpha}(x_i,x_j)-\gamma(-2\ii\alpha), \; &\olam_{\alpha}(x_i,x_j)=\Lambda_{-\alpha}(-x_i,-x_j). 
\end{array}
\end{align}
The saddle point equation \eqref{mix3leg} for these Lagrangians is equivalent to $H2_{(\varepsilon=0)}$ in the ABS list.  In terms of \eqref{lagh2d0} the IRF Lagrangian functions \eqref{IRFlag1}, \eqref{IRFlam1}, and \eqref{ea2}, are respectively given by \eqref{irflagq11} and
\begin{align}
\label{IRFh2d0b}
\begin{split}
\lam_{\bu\bv}\!\left(\!x\Big|\,\arraycolsep=0.7pt\begin{array}{ll}x_a & x_b \\  x_c & x_d\end{array}\right)
= \gamma(\ii(u_2-v_1)+x+x_a)+\gamma(\ii(v_2-u_2)-x-x_b)\phantom{,} \\[-0.25cm]
 +\gamma(\ii(v_1-u_1)-x-x_c)+\gamma(\ii(u_1-v_2)+x+x_d),
\end{split}
\\
\label{IRFh2d0c}
\begin{split}
\Xi_{\bu\bv}\!\left(\!x\Big|\,\arraycolsep=0.7pt\begin{array}{ll}x_a & x_b \\  x_c & x_d\end{array}\right)
 \simeq \gamma(\ii(u_2-v_1)\pm (x-x_a)) + \gamma(\ii(v_2-u_2) + x-x_b)\phantom{,} \\[-0.25cm]
  - \gamma(\ii(u_2-v_2)+x-x_b) - \gamma(\ii(u_1-v_1)-x-x_c)\phantom{,} \\ + \gamma(\ii(v_2-u_1)+x+x_d).
\end{split}
\end{align}

These Lagrangians will already be linear after taking a derivative, thus an appropriate change of variables \eqref{cov} is
\begin{align}\label{covh20}
f(x)=x,\qquad g(x)=x,\qquad h(x)=\ii x.
\end{align}

Applying \eqref{covh20} on the derivatives with respect to $x$ of  \eqref{IRFh2d0b} and \eqref{IRFh2d0c} results in the type-B and type-C equations \eqref{4legb} and \eqref{4legc} respectively, where  $a(x;y;\alpha,\beta)$ is given by  \eqref{dlq11}, and
\begin{align}\label{dlh20}
b(x;y;\alpha,\beta)=x+y+\alpha-\beta, \qquad
c(x;y;\alpha,\beta)=x+y-\alpha+\beta.
\end{align}

The coefficient of $x$ in the affine-linear form of the type-C equation \eqref{4legc} with $a(x;y;\alpha,\beta)$ from \eqref{dlq11} and $c(x;y;\alpha,\beta)$ from \eqref{dlh20}, is given by \eqref{h24} with $(\delta_1,\delta_2,\delta_3)=(1,0,0)$.  Setting $x_a\leftrightarrow x_c$, $x_b\to-x_b$, $\alpha_1\to0$, $\beta_1\to\beta$, $\alpha_2\to\alpha$, and $\beta_2\to\beta$, in \eqref{h24} with $(\delta_1,\delta_2,\delta_3)=(1,0,0)$  gives
\begin{align}
(x_a - x_d)(x_b - x_c) - (\alpha-\beta)(x_a + x_b + x_c + x_d) -\alpha^2 + \beta^2.
\end{align}
This is a polynomial for $H2_{(\varepsilon=0)}$ from the ABS list.

\subsubsection{Algebraic cases}

The face-centered quad equations at the algebraic level are usually written in one of the following additive forms\footnote{However, the first case below involves a combination of both multiplicative and additive equations.}
\begin{align}
\label{fceA2}
%
a(x,x_a,\alpha_2,\beta_1)+a(x,x_d,\alpha_1,\beta_2)-a(x,x_b,\alpha_2,\beta_2)-a(x,x_c,\alpha_1,\beta_1)=0, 
\\
\label{fceB2}
b(x,x_a,\alpha_2,\beta_1)+b(x,x_d,\alpha_1,\beta_2)-b(x,x_b,\alpha_2,\beta_2)-b(x,x_c,\alpha_1,\beta_1)=0, 
\\
 \label{fceC2}
a(x,x_a,\alpha_2,\beta_1)+c(x,x_d,\alpha_1,\beta_2)-a(x,x_b,\alpha_2,\beta_2)-c(x,x_c,\alpha_1,\beta_1)=0.
\end{align}
for type-A, type-B, and type-C, respectively.  
For these cases $\eta=0$ and there is no shift of the form \eqref{parshift} that will be applied to the parameters $\bu,\bv,\bw$.

\paragraph{$A2_{(\delta_1=0;\,\delta_2=0)}$, $B2_{(\delta_1=0;\,\delta_2=0;\,\delta_3=0)}$, $C2_{(\delta_1=0;\,\delta_2=0;\,\delta_3=0)}$ (Barnes's $_2F_1$ integral)}


The Lagrangian functions for the classical star-triangle relation \eqref{CASTR} are given by
\begin{align}
\label{lagh1e0b}
\begin{split}
\ds\Lambda_{\alpha}(x_i,x_j)&=(\ii x_j-\alpha)\Log x_i , \qquad \olam_{\alpha}(x_i,x_j)=-\Lambda_{\alpha}(x_i,x_j), \\
\ds{\lag}_{\alpha}(x_i,x_j)&=-2\alpha\Log(x_i+x_j), \\ 
\ds{\ol}_{\alpha}(x_i,x_j)&=\gamma(x_i-x_j-\ii\alpha)+\gamma(x_j-x_i-\ii\alpha)-\gamma(-2\ii\alpha).
\end{split}
\end{align}
The saddle point equation \eqref{mix3leg} for these Lagrangians is equivalent to $H1_{(\varepsilon=1)}$ from the ABS list.  This case is unusual because unlike other solutions of the star-triangle relation \eqref{CASTR} there is no simple symmetry relation between $\Lambda_\alpha(x_i,x_j)$ and $\Lambda_\alpha(x_j,x_i)$, which is needed to construct the Yang-Baxter equation of Section \ref{sec:CIRFYBEs}.  Nevertheless this case can still be used to derive face-centered quad equations which satisfy CAFCC.

In terms of \eqref{lagh1e0b} the IRF Lagrangian functions \eqref{IRFlag1}, \eqref{IRFlam1}, and \eqref{ea2}, are respectively given by
\begin{align}\label{irflagq10}
\begin{split}
\lag_{\bu\bv}\!\left(\!x\Big|\,\arraycolsep=0.7pt\begin{array}{ll}x_a & x_b \\  x_c & x_d\end{array}\right)
= (u_2 - v_1)\Log(x + x_a) + (v_2 - u_2)\Log(x + x_b)\phantom{,} \\[-0.25cm]
 + (v_1 - u_1)\Log(x + x_c) + (u_1 - v_2)\Log(x + x_d), 
\end{split}
\\
\label{IRFh1d1b}
\begin{split}
\lam_{\bu\bv}\!\left(\!x\Big|\,\arraycolsep=0.7pt\begin{array}{ll}x_a & x_b \\  x_c & x_d\end{array}\right)
\simeq\ii x(\Log x_b+\Log x_c-\Log x_a-\Log x_d) ,
\end{split}
\\
\label{IRFh1d1c}
\begin{split}
\Xi_{\bu\bv}\!\left(\!x\Big|\,\arraycolsep=0.7pt\begin{array}{ll}x_a & x_b \\  x_c & x_d\end{array}\right)
= (v_1 - v_2 - \ii( x_c-x_d ) ) \Log x  + 2 (u_2 - v_1) \Log(x + x_a)\phantom{,}  \\[-0.25cm]
- 2 (u_2 - v_2) \Log(x + x_b).
\end{split}
\end{align}

These Lagrangians will already be linear after taking a derivative, thus an appropriate change of variables \eqref{cov} is
\begin{align}\label{covh0}
f(x)=x,\qquad g(x)=x,\qquad h(x)=\ii x.
\end{align}

Applying \eqref{covh0} on the derivative with respect to $x$ of \eqref{irflagq10} results in the type-A equation \eqref{fceA2} where
\begin{align}\label{dlq10a}
a(x;y;\alpha,\beta)=\frac{\alpha-\beta}{x+y}.
\end{align}
Note that in this form the type-A equation \eqref{fceA2} does not satisfy CAFCC on its own, but requires first negating a variable in the denominator.  Doing so leads to the type-A equation \eqref{fceA2} where 
\begin{align}\label{dlq10}
a(x;y;\alpha,\beta)&=\frac{\alpha-\beta}{x-y}.
\end{align}

Applying \eqref{covh0} on the derivatives with respect to $x$ of \eqref{IRFh1d1b} and \eqref{IRFh1d1c} results in the type-B and -C equations \eqref{fceB2} and \eqref{fceC2} respectively, where $a(x;y;\alpha,\beta)$ is given by  \eqref{dlq10a}, and
\begin{align}
\label{dlq10b2}
b(x;y;\alpha,\beta)&=\Log y, \\
c(x;y;\alpha,\beta)&=\frac{y+\beta}{2x}. \label{dlq10c}
\end{align}

To have an expression for $b(x;y;\alpha,\beta)$ which is linear in $y$ requires taking an exponential of \eqref{fceB2} with \eqref{dlq10b2}.  This results in the type-B equation in the multiplicative form \eqref{4legb} with
\begin{align}\label{dlq10b}
   b(x;y;\alpha,\beta)=y.
\end{align}

The multiplicative type-B equation \eqref{4legb} with \eqref{dlq10b}, along with the additive type-A and -C equations \eqref{fceA2} with \eqref{dlq10} (or \eqref{dlq10a}), and  \eqref{fceC2} with \eqref{dlq10c}, satisfy the CAFCC property.  

The coefficient of $x$ in the affine-linear form of \eqref{fceA2} with $a(x;y;\alpha,\beta)$ from \eqref{dlq10}, is given by \eqref{q24} with $(\delta_1,\delta_2)=(0,0)$.  Setting $x_b\to-x_b$, $\alpha_1\to\beta$, $\beta_1\to\alpha$, $\alpha_2\to0$, and $\beta_2\to0$ in  \eqref{q24} with $(\delta_1,\delta_2)=(0,0)$ gives
\begin{align}
\alpha(x_a - x_c)(x_b - x_d) - \beta(x_a - x_b)(x_c - x_d).
\end{align}
This is a polynomial for $Q1_{(\delta=0)}$ from the ABS list.

The coefficient of $x$ in the affine-linear form of \eqref{fceC2} with $a(x;y;\alpha,\beta)$ from \eqref{dlq10} and $c(x;y;\alpha,\beta)$ from \eqref{dlq10c}, is given by \eqref{h24} with $(\delta_1,\delta_2,\delta_3)=(0,0,0)$.  Setting $x_a\leftrightarrow x_c$, $x_b\to-x_b$, $\alpha_1\to0$, $\beta_1\to\beta$, $\alpha_2\to\alpha$, and $\beta_2\to\beta$, in \eqref{h24} with $(\delta_1,\delta_2,\delta_3)=(0,0,0)$ gives
\begin{align}
(x_a - x_d)(x_b - x_c) - (\alpha - \beta)(x_b + x_c).
\end{align}
This is a polynomial for $H1_{(\varepsilon=1)}$ from the ABS list.

\paragraph{$A2_{(\delta_1=0;\,\delta_2=0)}$, $D1$, $C1$ (Euler beta function)}

The Lagrangian functions for the classical star-triangle relation \eqref{CASTR} are given by
\begin{align}
\label{lagh1e0b4}
\begin{gathered}
\ds\Lambda(x_i,x_j)=\ii x_ix_j, \qquad \ds\olam(x_i,x_j)=-\Lambda(x_i,x_j), \qquad \ol_{\alpha}(x_i,x_j)=2\ii\alpha\Log(x_i-x_j), \\
\ds{\lag}_{\alpha}(x_i,x_j)=2\ii\alpha\bigl(\Log(\alpha)-\Log(x_i-x_j)\bigr). 
\end{gathered}
\end{align}
The saddle point equation \eqref{mix3leg} for these Lagrangians is equivalent to $H1_{(\varepsilon=0)}$ from the ABS list.  In terms of \eqref{lagh1e0b4} the IRF Lagrangian functions \eqref{IRFlag1}, \eqref{IRFlam1}, and \eqref{ea2}, are respectively given by \eqref{irflagq10} and
\begin{align}
\label{IRFh1d0b}
\begin{split}
\lam_{\bu\bv}\!\left(\!x\Big|\,\arraycolsep=0.7pt\begin{array}{ll}x_a & x_b \\  x_c & x_d\end{array}\right)
&= x(x_a - x_b - x_c + x_d),
\end{split}
\\
\label{IRFh1d0c}
\begin{split}
\Xi_{\bu\bv}\!\left(\!x\Big|\,\arraycolsep=0.7pt\begin{array}{ll}x_a & x_b \\  x_c & x_d\end{array}\right)
& \simeq x(x_c-x_d) + 2(v_1 - u_2)\Log(x - x_a) + 2(u_2 - v_2) \Log(x - x_b).
\end{split}
\end{align}

These Lagrangians will be linear after taking derivatives and do not require a change of variables.  
The derivatives with respect to $x$ of \eqref{IRFh1d0b} and \eqref{IRFh1d0c} may be written in the form of the type--B and type-C equations \eqref{fceB2} and \eqref{fceC2} respectively, where $a(x;y;\alpha,\beta)$ is given by  \eqref{dlq10a}, and
\begin{align}\label{dlq10d}
b(x;y;\alpha,\beta)=y, \qquad
c(x;y;\alpha,\beta)=-\frac{y}{2}.
\end{align}

The coefficient of $x$ in the affine-linear form of \eqref{fceC2} with $a(x;y;\alpha,\beta)$ from \eqref{dlq10} and $c(x;y;\alpha,\beta)$ from \eqref{dlq10d}, is given by \eqref{h14}.  Setting $x_a\to x_c$, $x_b\to-x_b$, $x_c\to-x_a$, $\alpha_1\to0$, $\beta_1\to\beta$, $\alpha_2\to\alpha$, and $\beta_2\to\beta$, in \eqref{h14} gives
\begin{align}
(x_a - x_d)(x_b - x_c)-2(\alpha-\beta) .
\end{align}
This is a polynomial for $H1_{(\varepsilon=0)}$ from the ABS list.

 \section{Conclusion}

In Section \ref{sec:FCC} of this paper, a new formulation of the multidimensional consistency integrability condition was established, called consistency-around-a-face-centered-cube (CAFCC), which is applicable to five-point face-centered quad equations which are defined on a vertex and its four nearest neighbours in the square lattice.  The formulation of CAFCC was motivated by interaction-round-a-face (IRF) forms of the Yang-Baxter equation that arise for integrable lattice models of statistical mechanics, which were presented in Section \ref{sec:YBE}.  Through this connection fifteen sets of face-centered quad equations which satisfy CAFCC  have been derived in Section \ref{sec:CAFCCIRF}. These face-centered quad equations include expressions for discrete Laplace-type equations associated to type-Q ABS equations, as well as other equations which have not previously been considered in the context of discrete integrability.  

It will be important to investigate other characteristics associated to the multidimensional consistency of the face-centered quad equations.  In this direction, it has recently been established how Lax pairs of the equations can be derived from the property of CAFCC \cite{KelsLax}.  It would be interesting if this result can be utilised in the computation of the solutions of the equations.  It is also an open problem to classify the face-centered quad equations, and doing so may lead to new equations or possibly reveal different combinations of the type-A, -B, or -C equations that will satisfy CAFCC.

It will also be important to further investigate the connection between integrable lattice equations and integrable lattice models of statistical mechanics.  As an example, a possible extension of the results of this paper could be obtained by considering a quasi-classical expansion of multi-component solutions of the Yang-Baxter equation \cite{Bazhanov:2011mz}.  The resulting classical lattice equations should correspond to multi-component extensions of the equation $A4$.  This potential application to multi-component equations was one of the main motivations for determining the multidimensional consistency of the face-centered quad equations in this paper.  




\begin{appendices}
\numberwithin{equation}{section}

\section{Affine-linear expressions}\label{app:afflin}

Here it is useful to first define the following quad polynomials:
\begin{align}
\begin{gathered}
D1(\mxa,\mxb,\mxc,\mxd)=    \mxa - \mxb - \mxc + \mxd,\qquad D4(\mxa,\mxb,\mxc,\mxd)=\mxa\mxd-\mxb\mxc,
\\
L(\mxa,\mxb,\mxc,\mxd,\alpha_1,\alpha_2,\alpha_3,\alpha_4)=\alpha_1\mxa +\alpha_2\mxb +\alpha_3\mxc + \alpha_4\mxd.
\end{gathered}
\end{align}
The $D1$ and $D4$ are two CAC quad polynomials in the $H^6$ list given by Boll \cite{BollThesis}, while for a choice of parameters $L$ is equivalent to a linear CAC quad polynomial given by Atkinson \cite{Atkinson09}.

For the parameters $\al=(\mpa,\mpb)$ and $\bt=(\mqa,\mqb)$ it is also useful to define 
\begin{align}
\theta(\al,\bt)=(\mpa-\mpb)(\mqa-\mqb),\qquad \phi(\al,\bt)=\mpa+\mpb-\mqa-\mqb.
\end{align}
The following notations for the parameter $\al$ will also be used (and similarly for $\bt$)
\begin{align}
-\al=(-\mpa,-\mpb),\quad \al^2=(\mpa^2,\mpb^2),\quad \al^*=(\mpa,-\mpb),\quad \al_i=(\alpha_i,\alpha_i),\;(i=1,2).
\end{align}

\subsection{Symmetric cases (type-A equations)}

For affine-linear forms of the type-A equations given in Table \ref{table-A}, there will appear the following four-parameter versions of the type-Q quad polynomials in the ABS list \cite{ABS}:
\begin{align}
\label{q34}
\begin{split}
\text{\scalebox{0.98}{$  Q3_{(\dl)}(\mxa,\mxb,\mxc,\mxd;\al,\bt)= \tfrac{\dl}{4}\theta(\al^2,\bt^2)(\tfrac{\mqa\mqb}{\mpa\mpb}-\tfrac{\mpa\mpb}{\mqa\mqb})+
\mqa\mqb(\mpa^2-\mpb^2)D4(\mxa,\mxb,\mxd,\mxc)  \phantom{,}$}}  \\
\text{\scalebox{0.98}{$ +  \mpa\mpb(\mqa^2-\mqb^2)D4(\mxa,\mxd,\mxc,\mxb) - (\mpa^2\mpb^2-\mqa^2\mqb^2)D4(\mxa,\mxb,\mxc,\mxd),$}}
\end{split}
\end{align}
\begin{align}
\label{q24}
\begin{split}
\text{\scalebox{0.98}{$ Q2_{(\dl_1;\,\dl_2)}(\mxa,\mxb,\mxc,\mxd;\al,\bt)= 
(\mpa-\mpb)D4(\mxa,\mxb,\mxd,\mxc)  +  (\mqa-\mqb)D4(\mxa,\mxd,\mxc,\mxb)\phantom{,}$}}  \\ 
\text{\scalebox{0.98}{$- \phi(\al,\bt)D4(\mxa,\mxb,\mxc,\mxd)+\dl_2L(\mxa,\mxb,\mxc,\mxd,\rho_1,\rho_2,\rho_3,\rho_4)\phantom{,}$}} \\
\text{\scalebox{0.98}{$+\dl_1\theta(\al,\bt)\phi(\al,\bt)\bigl(\theta(\al^*,\bt^*) - \phi(\al^2,-\bt^2)\bigr)^{\dl_2},$}}
\end{split}
\end{align}
where 
\begin{align}
\begin{split}
\text{\scalebox{1.0}{$\rho_1=$}}&
\text{\scalebox{1.0}{$\phi(\al,\bt)(\mpb(\mqb-\mqa)+\mpa(\mqa+\mqb))-2(\mpa\mpb(\mpa-\mqa)+\mqa\mqb(\mpb-\mqb)),$}} \\
\text{\scalebox{1.0}{$\rho_2=$}}&
\text{\scalebox{1.0}{$\phi(\al,\bt)(\mpb(\mqb-\mqa)-\mpa(\mqa+\mqb))+2(\mpa\mpb(\mpa-\mqb)+\mqa\mqb(\mpb-\mqa)),$}} \\
\text{\scalebox{1.0}{$\rho_3=$}}&
\text{\scalebox{1.0}{$\phi(\al,\bt)(\mpa(\mqa-\mqb)-\mpb(\mqa+\mqb))+2(\mpa\mpb(\mpb-\mqa)+\mqa\mqb(\mpa-\mqb)),$}} \\
\text{\scalebox{1.0}{$\rho_4=$}}&
\text{\scalebox{1.0}{$\phi(\al,\bt)(\mpa(\mqa-\mqb)+\mpb(\mqa+\mqb))-2(\mpa\mpb(\mpb-\mqb)+\mqa\mqb(\mpa-\mqa)).$}}
\end{split}
\end{align}

As noted in Section \ref{sec:FCElist}, for $Q3_{(\dl)}$ in \eqref{q34}, the case $\dl=1$, is associated to the ABS polynomial $Q3_{(\delta=1)}$, and the case $\dl=0$, is associated to the ABS polynomial $Q3_{(\delta=0)}$.

For  $Q2_{(\dl_1;\,\dl_2)}$ in \eqref{q24}, the case $(\dl_1,\dl_2)=(1,1)$, is associated to the ABS polynomial $Q2$, the case $(\dl_1,\dl_2)=(1,0)$, is associated to the ABS polynomial $Q1_{(\delta=1)}$, and the case $(\dl_1,\dl_2)=(0,0)$, is associated to the ABS polynomial $Q1_{(\delta=0)}$.

The affine-linear expressions of type-A equations given in Table \ref{table-A}, are as follows:
\begin{empheq}[box=\fbox]{align}\nonumber
\begin{split}
& A4: \\
& \phantom{\times}\bigl(G_{+}(\mx,\mpb,\mqb)-F(\mx,\mxb,\mpb,\mqb)\bigr)\bigl(G_{+}(\mx,\mpa,\mqa)-F(\mx,\mxc,\mpa,\mqa)\bigr) \\
& \times\bigl(G_{-}(\mx,\mpb,\mqa)-F(\mx,\mxa,\mpb,\mqa)\bigr)\bigl(G_{-}(\mx,\mpa,\mqb)-F(\mx,\mxd,\mpa,\mqb)\bigr) \\
&  \times \bigl(S_{-}(\mx,\mpb,\mqb)S_{-}(\mx,\mpa,\mqa)S_{+}(\mx,\mpb,\mqa)S_{+}(\mx,\mpa,\mqb)\bigr)^2  \\
&  -\bigl(G_{-}(\mx,\mpb,\mqb)-F(\mx,\mxb,\mpb,\mqb)\bigr)\bigl(G_{-}(\mx,\mpa,\mqa)-F(\mx,\mxc,\mpa,\mqa)\bigr) \\
& \times\bigl(G_{+}(\mx,\mpb,\mqa)-F(\mx,\mxa,\mpb,\mqa)\bigr)\bigl(G_{+}(\mx,\mpa,\mqb)-F(\mx,\mxd,\mpa,\mqb)\bigr) \\
&  \times \bigl(S_{+}(\mx,\mpb,\mqb)S_{+}(\mx,\mpa,\mqa)S_{-}(\mx,\mpb,\mqa)S_{-}(\mx,\mpa,\mqb)\bigr)^2   =0.
\end{split}
\\[0.1cm]
\nonumber
\begin{split}
& A3_{(\dl)}: \\
& \phantom{-}L(\mxa,\mxb,\mxc,\mxd,\rho_1,\rho_2,\rho_3,\rho_4)\mx^2  +Q3_{(\dl)}(\mxa,\mxb,\mxc,\mxd,\al,\bt)\mx \\
& -\mxa\mxb\mxc\mxd L\bigl(\mxd^{-1},\mxc^{-1},\mxb^{-1},\mxa^{-1},\rho_4,\rho_3,\rho_2,\rho_1)   +\tfrac{\dl}{4}L(\mxa,\mxb,\mxc,\mxd, \gamma_1,\gamma_2,\gamma_3,\gamma_4) =0.
\end{split}
\\[0.1cm]
\nonumber
\begin{split}
& A2_{(\dl_1;\,\dl_2)}: \\
& \phantom{+}\bigl(L(\mxa,\mxb,\mxc,\mxd,\mpb-\mqa,\mqb-\mpb,\mqa-\mpa,\mpa-\mqb) + \dl_2\theta(\al,\bt)\phi(\al,\bt)\bigr)\mx^2 \\
& +Q2_{(\dl_1;\,\dl_2)}(\mxa,\mxb,\mxc,\mxd,\al,\bt)\mx  +\dl_1 L(\mxa,\mxb,\mxc,\mxd,\rho_1\gamma_1^{\dl_2},\rho_2\gamma_2^{\dl_2},\rho_3\gamma_2^{\dl_2},\rho_4\gamma_2^{\dl_2}) \\
& +\dl_2\Bigl((\mpa-\mpb)\bigl( (\mpa-\mqa) (\mpb-\mqa)\mxb\mxd - (\mpa-\mqb) (\mpb-\mqb)\mxa\mxc\bigr) \\
& \phantom{+\dl_2}+  (\mqa-\mqb)\bigl((\mpb-\mqa) (\mpb-\mqb) \mxc\mxd - (\mpa-\mqa) (\mpa-\mqb) \mxa\mxb\bigr)   \\
& \phantom{+\dl_2}+ \phi(\al,\bt)\bigl((\mpb-\mqa) (\mpa-\mqb) \mxb\mxc-(\mpa-\mqa) (\mpb-\mqb) \mxa\mxd\bigr)  \\
& \phantom{+\dl_2}+ (\mpa-\mqa) (\mpb-\mqa) (\mpa-\mqb) (\mpb-\mqb) \phi(\al,\bt)\theta(\al,\bt)\Bigr) \\
& +\mxa\mxb\mxc\mxd L(\mxd^{-1},\mxc^{-1},\mxb^{-1},\mxa^{-1},\mqb-\mpa,\mpa-\mqa,\mpb-\mqb,\mqa-\mpb)    =0.
\end{split}
\end{empheq}

For $A4$ the definitions \eqref{fgdef}--\eqref{xdotdef}, are used. 
For $A3_{(\dl)}$, the $\rho_i$, $\gamma_i$, ($i=1,\ldots,4$), are given by:
\begin{align}
\begin{split}
& \rho_1=\mpa\mqb(\mpb^2-\mqa^2), \qquad\gamma_1=(\mpa^2-\mqa^2)(\mpb^2-\mqb^2)(\mpa\mqb^{-1}-\mqb\mpa^{-1}) , \\
& \rho_2=\mpa\mqa(\mqb^2-\mpb^2), \qquad\gamma_2=(\mpa^2-\mqb^2)(\mpb^2-\mqa^2)(\mqa\mpa^{-1}-\mpa\mqa^{-1}), \\
& \rho_3=\mpb\mqb(\mqa^2-\mpa^2), \qquad\gamma_3=(\mpa^2-\mqb^2)(\mpb^2-\mqa^2)(\mqb\mpb^{-1}-\mpb\mqb^{-1}), \\
& \rho_4=\mpb\mqa(\mpa^2-\mqb^2), \qquad\gamma_4=(\mpa^2-\mqa^2)(\mpb^2-\mqb^2)(\mpb\mqa^{-1}-\mqa\mpb^{-1}). 
\end{split}
\end{align}

For $A2_{(\dl_1;\,\dl_2)}$, the $\rho_i$, $\gamma_i$, ($i=1,\ldots,4$), are given by:
\begin{align}
\begin{split}
& \rho_1=(\mqa-\mpa)(\mqb-\mpb)(\mqb-\mpa), \qquad\gamma_1=\bigl( \mpa(\mqa + \mqb) - \mpb(\mqa - \mqb) - \mpa^2 - \mqb^2 \bigr), \\
& \rho_2=(\mpa-\mqb)(\mpb-\mqa)(\mpa-\mqa), \qquad\gamma_2=\bigl( \mpa(\mqa + \mqb) - \mpb(\mqb - \mqa) - \mpa^2 - \mqa^2 \bigr), \\
& \rho_3=(\mpa-\mqb)(\mpb-\mqa)(\mpb-\mqb), \qquad\gamma_3=\bigl( \mpb(\mqa + \mqb) - \mpa(\mqa - \mqb) - \mpb^2 - \mqb^2 \bigr), \\
& \rho_4=(\mqa-\mpa)(\mqb-\mpb)(\mqa-\mpb), \qquad\gamma_4=\bigl( \mpb(\mqa + \mqb) - \mpa(\mqb - \mqa) - \mpb^2 - \mqa^2 \bigr).
\end{split}
\end{align}

\subsection{Mixed cases (type-B and -C equations)}




\subsubsection{Type-B equations}

The affine-linear expressions of type-B equations given in Table \ref{table-BC2}, are as follows:
\begin{empheq}[box=\fbox]{align}\nonumber
\begin{split}
& B3_{(\dl_1;\,\dl_2;\,\dl_3)}: \\
& \phantom{+}\dl_2L\bigl(\mxa,\mxb,\mxc,\mxd,\tfrac{\mpa}{\mqb},-\tfrac{\mpa}{\mqa},-\tfrac{\mpb}{\mqb},\tfrac{\mpb}{\mqa}\bigr)\mx  - \tfrac{\dl_2}{2}\theta(\al^2,\bt^2)(\mpa\mpb\mqa\mqb)^{-1} \\
& +\dl_3\bigl(\mxa\mxb\mpb(\tfrac{\mxd}{\mqb}-\tfrac{\mxc}{\mqa})+\mxc\mxd\mpa(\tfrac{\mxa}{\mqa}-\tfrac{\mxb}{\mqb})\bigr)\mx^{-1} \\
& +\dl_1L\bigl(\mxa,\mxb,\mxc,\mxd,\tfrac{\mqb}{\mpa},-\tfrac{\mqa}{\mpa},-\tfrac{\mqb}{\mpb},\tfrac{\mqa}{\mpb}\bigr)\mx^{-1}-D4(\mxa,\mxb,\mxc,\mxd)  =0.
\end{split}
\\[0.1cm]\nonumber
\begin{split}
& B2_{(\dl_1;\,\dl_2;\,\dl_3)}: \\
& \phantom{+}  \dl_1 D1(\mxa,\mxb,\mxc,\mxd)\mx(-\mx)^{\dl_2} + (\dl_3-2\dl_2)\phi(\al,\bt)\theta(\al,\bt)\mx^{\dl_2}  \\
& \phantom{+} \dl_1\Bigl( (1-\dl_3)L(\mxa,\mxb,\mxc,\mxd,\mpa-\mqb,\mqa-\mpa,\mqb-\mpb,\mpb-\mqa)(-2\mx)^{\dl_2} \\
& \phantom{+\dl_1} -\theta(\al,\bt)\bigl(2(\mx^2+(\mpa\mpb-\mqa\mqb))-\theta(\al^*,\bt^*)\bigr)^{\dl_2}D1(-\mxa,\mxb,\mxc,-\mxd)^{\dl_3}\Bigr)  \\
& -(\dl_2+\dl_3)L\bigl(\mxa,\mxb,\mxc,\mxd,(\mpa-\mpb)^2,-(\mpa-\mqa)^2,-(\mpb-\mqb)^2,(\mpa-\mqa)^2\bigr)  \\
& +\dl_3\Bigl(\mxb\mxc(\mxa+\mxd)-\mxa\mxd(\mxb+\mxc)-D4(\mxa,\mxb,\mxd,\mxc)(\mpa-\mpb) \\
& \phantom{+\dl_3} -D4(\mxa,\mxc,\mxd,\mxb)(\mqa-\mqb)\Bigr)  +D4(\mxa,\mxb,\mxc,\mxd)\phi(\bt,\al)^{\dl_3}  =0.
\end{split}
\\[0.1cm]\nonumber
\begin{split}
& D1: D1(\mxa,\mxb,\mxc,\mxd)=0.
\end{split}
\end{empheq}


\subsubsection{Type-C equations}

For the type-C face-centered quad equations in Table \ref{table-BC2}, there will appear the following four-parameter versions of the type-H quad polynomials in the ABS list \cite{ABS2}:
\begin{align}\label{h34}
\begin{split}
H3_{(\dl_1;\,\dl_2;\,\dl_3)}(\mxa,\mxb,\mxc,\mxd;\al,\bt)=\mqa\mqb D4(\mxa,\mxb,\mxd,\mxc) - \mpb^2D4(\mxa,\mxb,\mxc,\mxd)\phantom{,} \\
  +(\tfrac{\mqb}{\mqa}-\tfrac{\mqa}{\mqb})\bigl(\dl_1\mpa\mpb+\tfrac{\mpb\mqa\mqb}{\mpa}(\dl_2\mxc\mxd-\dl_3\mxa\mxb)\bigr),
\end{split}
\end{align}
\begin{align}
\label{h24}
\begin{split}
H2_{(\dl_1;\,\dl_2;\,\dl_3)}(\mxa,\mxb,\mxc,\mxd;\al,\bt)=(\mxb-\mxa)\phi(\al_2,\bt)\phi(\al_1,\bt)^{\dl_3}+ (\mxa+\mxb)(\mxc-\mxd)\phantom{,} \\
  -\dl_1(\mqa-\mqb)\bigl(\phi(\al_1,\bt)(-\mqa-\mqb)^{\dl_2+\dl_3} - (\mxc+\mxd)\phi(\al_1,\bt)^{\dl_2} \bigr)\phantom{,} \\
  +2\dl_2\bigl((\mqb-\mqa)(\mxc\mxd-\mqa\mqb+\mpa^2) + (\mpb-\mqa)(\mpb-\mqb)(\mxc-\mxd) \bigr)\phantom{,} \\
  +2\dl_3(\mqa-\mqb)\bigl(\mxa\mxb  -\mpa^2 - \mpb(\mpb - \mqa - \mqb)\bigr),
\end{split}
\\
\label{h14}
\begin{split}
H1(\mxa,\mxb,\mxc,\mxd;\al,\bt)=2\phi(\bt,\al_2)-(\mxa + \mxb) (\mxc + \mxd).
\end{split}
\end{align}

The affine-linear expressions of type-C equations given in Table \ref{table-BC2}, are as follows:
\begin{empheq}[box=\fbox]{align}\nonumber
\begin{split}
& C3_{(\dl_1;\,\dl_2;\,\dl_3)}: \\
& \phantom{+} \Bigl(\mpb (\mqa\mxd-\mqb\mxc) - \dl_3\mpa^{-1}\bigl(\mpb^2(\mqa\mxb-\mqb\mxa) + \mqa\mqb(\mqa\mxa - \mqb\mxb)\bigr)\Bigr)\mx^2 \\
& +H3_{(\dl_1;\,\dl_2;\,\dl_3)}(\mxa,\mxb,\mxc,\mxd;\al,\bt)x  + \mpb\mxa\mxb(\mqb\mxd-\mqa\mxc)  \\
& +\dl_1\bigl(\mpa(\mqa\mxb-\mqb\mxa) + \mpa\mpb^2(\mxa\mqb^{-1} - \mxb\mqa^{-1})\bigr) \\
& +\tfrac{\dl_2}{2}\Bigl(\mpb(\tfrac{\mpb}{\mqa} - \tfrac{\mqa}{\mpb})(\tfrac{\mpb}{\mqb} - \tfrac{\mqb}{\mpb})(\mqb\mxd-\mqa\mxc) \\
&  \phantom{+\dl_2}+\tfrac{2\mxc\mxd}{\mpa}\bigl(\mqa\mqb(\mqa\mxb-\mqb\mxa) + \mpb^2(\mqa\mxa - \mqb\mxb)\bigr)\Bigr)  =0.
\end{split}
\\[0.1cm]\nonumber
\begin{split}
& C2_{(\dl_1;\,\dl_2;\,\dl_3)}: \\
& \phantom{+} \Bigl((\mqa - \mqb)\phi(\al_1,\bt)^{\dl_3} - \mxc + \mxd + 2\dl_3\bigl((\mpb-\mqa)\mxa - (\mpb-\mqb)\mxb \bigr) \Bigr)\mx^2  \\
& +H2_{(\dl_1;\,\dl_2;\,\dl_3)}(\mxa,\mxb,\mxc,\mxd;\al,\bt)\mx + \mxa\mxb\bigl((\mqb-\mqa)\phi(\al_1,\bt)^{\dl_3} + \mxd - \mxc\bigr) \\
& +\dl_1\Bigl(
 (\mpb-\mqa)\mxb\mxd\bigl(2(\mpa-\mxc) - \mpb - \mqa)^{\dl_2} -(\mpb-\mqb)\mxa\mxc\bigl(2(\mpa-\mxd)-\mpb-\mqb\bigr)^{\dl_2} \\
& \phantom{+\dl_1} -(\mpb-\mqb)\mxa\mxd\bigl(2(\mpa-\mqa)+\mpb-\mqb\bigr)^{\dl_2} +(\mpb-\mqa)\mxb\mxc\bigl(2(\mpa-\mqb) + \mpb - \mqa\bigr)^{\dl_2} \\
& \phantom{+\dl_1}+\phi(\al_1,\bt)\bigl(\mqa\mxb - \mqb\mxa + \mpb(\mxa - \mxb)\bigr)(-\mqa - \mqb)^{\dl_2+\dl_3} \\
& \phantom{+\dl_1}+(\mpb-\mqa)(\mpb-\mqb)(\mxc-\mxd)\bigl((\mpb-\mqa) (\mqb-\mpb) - (\mqa-\mqb)^2\bigr)^{\dl_2} \\
& \phantom{+\dl_1}+(\mpb-\mqa)(\mpb-\mqb)(\mqa-\mqb)\bigl(\mqa\mqb-\mpa\mpb + (2\mpa-\mpb)\phi(\al,\bt)\bigr)^{\dl_2}\phi(\al_1,\bt)^{\dl_3}\Bigr) \\
& +\dl_2\Bigl(
2(\mpb - \mqa)(\mpb - \mqb)(\mqa - \mqb)\mxc\mxd +(\mqa-\mqb)\bigl(\mpa^2+\mpb^2-\mpb(\mqa+\mqb)\bigr)(\mxa+\mxb) \\
& \phantom{+\dl_2} + \phi(\al_2,\bt)(\mpa^2-\mqa\mqb)(\mxa-\mxb) - (\mpb - \mqa)(\mpb - \mqb)(\mqa-\mqb)\phi(\al_1,\bt)(\mxc+\mxd)\Bigr) \\
& +2\dl_3(\mpa^2 - \mqa\mqb)\bigl(\mqa\mxb - \mqb\mxa + \mpb(\mxa - \mxb)\bigr) \! =0.
\end{split}
\\[0.1cm]\nonumber
\begin{split}
& C1: \\
& (\mxc+\mxd)\mx^2 
 +H1(\mxa,\mxb,\mxc,\mxd;\al,\bt)\mx 
+\bigl(2(\mpb-\mqb)+\mxb\mxc\bigr)\mxa+\bigl(2(\mpb-\mqa)+\mxa\mxd\bigr)\mxb \\
& =0.
\end{split}
\end{empheq}

\section{Fourteen discrete Laplace-type equations on the face-centered cube}
\label{app:FCCequations}


In Section \ref{sec:YBE}, it was seen how the interaction-round-a-face form of the Yang-Baxter equation implies fourteen discrete Laplace-type equations, which are obtained from the derivatives taken with respect to the fourteen variables on vertices of the face-centered cube.  These equations are listed explicitly here using the notations of Section \ref{sec:YBE}.  The equations here are also referred to as one of types-A, -B, or C, according to which type of face-centered quad equation they correspond to through the method of Section \ref{sec:CAFCCIRF}.

There are four equations obtained from derivatives with respect to the variables at interior vertices on the left hand side of Figure \ref{SSRYBE}
\begin{align}
\label{ey1}
 \text{\scalebox{1.0}{$\frac{\partial}{\partial x_i}\Bigl(\lag_{u_2-v_1}(x_f,x_i)+\ol_{u_2-v_2}(x_h,x_i)+\ol_{u_1-v_1}(x_a,x_i)+\lag_{u_1-v_2}(x_b,x_i)\Bigr)=0,$}} \\
\label{ey2}
 \text{\scalebox{1.0}{$\frac{\partial}{\partial x_j}\Bigl(\lam_{u_2-w_1}(x_h,x_j)+\olam_{u_2-w_2}(x_d,x_j)+\olam_{u_1-w_1}(x_b,x_j)+\lam_{u_1-w_2}(x_c,x_j)\Bigr)=0,$}} \\
\label{ey3}
 \text{\scalebox{1.0}{$\frac{\partial}{\partial x_k}\Bigl(\lam_{v_2-w_1}(x_f,x_k)+\olam_{v_2-w_2}(x_e,x_k)+\olam_{v_1-w_1}(x_h,x_k)+\lam_{v_1-w_2}(x_d,x_k)\Bigr)=0,$}} \\
\label{ey}
 \text{\scalebox{1.0}{$\frac{\partial}{\partial x_h}\Bigl(\lag_{v_1-v_2}(x_f,x_h)+\ol_{u_2-v_2}(x_h,x_i)+\lam_{u_2-w_1}(x_h,x_j)+\olam_{v_1-w_1}(x_h,x_k)\Bigr)=0.$}}
\end{align}
The equation \eqref{ey1} is of type-A, the equations \eqref{ey2}, \eqref{ey3} are of type-B, and the equation \eqref{ey} is of type-C.

There are four equations obtained from derivatives with respect to the variables at interior vertices on the right hand side of Figure \ref{SSRYBE}
\begin{align}
\label{ez1}
 \text{\scalebox{1.0}{$\frac{\partial}{\partial x_i'}\Bigl(\lag_{u_2-v_1}(x_e,x_i')+\ol_{u_2-v_2}(x_d,x_i')+\ol_{u_1-v_1}(x_h',x_i')+\lag_{u_1-v_2}(x_c,x_i')\Bigr)=0,$}} \\
\label{ez2}
 \text{\scalebox{1.0}{$\frac{\partial}{\partial x_j'}\Bigl(\lam_{u_2-w_1}(x_f,x_j')+\olam_{u_2-w_2}(x_e,x_j')+\olam_{u_1-w_1}(x_a,x_j')+\lam_{u_1-w_2}(x_h',x_j')\Bigr)=0,$}} \\
\label{ez3}
 \text{\scalebox{1.0}{$\frac{\partial}{\partial x_k'}\Bigl(\lam_{v_2-w_1}(x_a,x_k')+\olam_{v_2-w_2}(x_h',x_k')+\olam_{v_1-w_1}(x_b,x_k')+\lam_{v_1-w_2}(x_c,x_k')\Bigr)=0,$}}  \\
\label{ez}
 \text{\scalebox{1.0}{$\frac{\partial}{\partial x_h'}\Bigl(\lag_{v_2-v_1}(x_h',x_c)+\ol_{u_1-v_1}(x_h',x_i')+\lam_{u_1-w_2}(x_h',x_j')+\olam_{v_2-w_2}(x_h',x_k')\Bigr)=0.$}}
\end{align}
The equation \eqref{ez1} is of type-A, the equations \eqref{ez2}, \eqref{ez3} are of type-B, and the equation \eqref{ez} is of type-C.

Finally, there are six equations obtained  from derivatives with respect to the variables at boundary vertices of Figure \ref{SSRYBE} (these are all equations of type-C)
\begin{align}
\label{ea}
\frac{\partial}{\partial x_a}\Bigl(\ol_{u_1-v_1}(x_a,x_i)+\lag_{v_2-v_1}(x_a,x_b)-\olam_{u_1-w_1}(x_a,x_j')-\lam_{v_2-w_1}(x_a,x_k')\Bigr)=0, \\
\label{eb}
\frac{\partial}{\partial x_b}\Bigl(\lag_{u_1-v_2}(x_b,x_i)+\lag_{v_2-v_1}(x_a,x_b)+\olam_{u_1-w_1}(x_b,x_j)-\olam_{v_1-w_1}(x_b,x_k')\Bigr)=0, \\
\label{ec}
\frac{\partial}{\partial x_c}\Bigl(\lag_{u_1-v_2}(x_c,x_i')+\lag_{v_2-v_1}(x_h',x_c)+\lam_{v_1-w_2}(x_c,x_k')-\lam_{u_1-w_2}(x_c,x_j)\Bigr)=0, \\
\label{ed}
\frac{\partial}{\partial x_d}\Bigl(\ol_{u_2-v_2}(x_d,x_i')+\lag_{v_1-v_2}(x_d,x_e)-\lam_{v_1-w_2}(x_d,x_k)-\olam_{u_2-w_2}(x_d,x_j)\Bigr)=0, \\
\label{ee}
\frac{\partial}{\partial x_e}\Bigl(\lag_{u_2-v_1}(x_e,x_i')+\lag_{v_1-v_2}(x_d,x_e)+\olam_{u_2-w_2}(x_e,x_j')-\olam_{v_2-w_2}(x_e,x_k)\Bigr)=0, \\
\label{ef}
\frac{\partial}{\partial x_e}\Bigl(\lag_{u_2-v_1}(x_f,x_i)+\lag_{v_1-v_2}(x_f,x_h)+\lam_{v_2-w_1}(x_f,x_k)-\lam_{u_2-w_1}(x_f,x_j')\Bigr)=0.
\end{align}

\end{appendices}

\bibliography{MComp}
\bibliographystyle{utphys}

\end{document}